\documentclass[12pt,twoside]{article} 
\usepackage{graphicx}
\usepackage{subeqnarray}

\begin{document}
%
%
%
%
\def\astrobj#1{#1}
\newenvironment{lefteqnarray}{\arraycolsep=0pt\begin{eqnarray}}
{\end{eqnarray}\protect\aftergroup\ignorespaces}
\newenvironment{lefteqnarray*}{\arraycolsep=0pt\begin{eqnarray*}}
{\end{eqnarray*}\protect\aftergroup\ignorespaces}
\newenvironment{leftsubeqnarray}{\arraycolsep=0pt\begin{subeqnarray}}
{\end{subeqnarray}\protect\aftergroup\ignorespaces}
\newcommand{\diff}{{\rm\,d}}
\newcommand{\pprime}{{\prime\prime}}
\newcommand{\szeta}{\mskip 3mu /\mskip-10mu \zeta}
\newcommand{\srho}{\mskip 3mu /\mskip-10mu \rho}
\newcommand{\sP}{\mskip 3mu /\mskip-10mu p}
\newcommand{\FC}{\mskip 0mu {\rm F}\mskip-10mu{\rm C}}
\newcommand{\appleq}{\stackrel{<}{\sim}}
\newcommand{\appgeq}{\stackrel{>}{\sim}}
\newcommand{\Int}{\mathop{\rm Int}\nolimits}
\newcommand{\Nint}{\mathop{\rm Nint}\nolimits}
\newcommand{\arcsinh}{\mathop{\rm arcsinh}\nolimits}
\newcommand{\range}{{\rm -}}
\newcommand{\displayfrac}[2]{\frac{\displaystyle #1}{\displaystyle #2}}

\newtheorem{Theorem}{\quad Theorem}[section] 

\newtheorem{Definition}[Theorem]{\quad Definition} 

\newtheorem{Corollary}[Theorem]{\quad Corollary} 

\newtheorem{Lemma}[Theorem]{\quad Lemma} 

\newtheorem{Example}[Theorem]{\quad Example} 

\centerline{} 

\centerline{\bf APPLIED MATHEMATICAL SCIENCES, Vol. x, 20xx, no. xx, xxx - xxx} 

\centerline{\bf HIKARI Ltd, \ www.m-hikari.com}

\centerline{\bf http://dx.doi.org/10.12988/}

\centerline{} 

\centerline{} 

\centerline {\Large{\bf  Multi-component, concentric, copolar,}} 

\centerline{} 

\centerline{\Large{\bf  axisymmetric, rigidly rotating polytropes:}} 

\centerline{} 

\centerline{\Large{\bf  an improved and extended theory}} 

\centerline{} 

\centerline{\bf {R. Caimmi}
\footnote
{Affiliated up to September 30th 2014. Current status: Studioso Senior.
Current position: in retirement due to age limits.}
} 

\centerline{} 

\centerline{ Physics and Astronomy Department, Padua University} 

\centerline{Vicolo Osservatorio 3/2, I-35122 Padova, Italy} 

\centerline{email: roberto.caimmi@unipd.it} 









\centerline{}

{\footnotesize Copyright $\copyright$ 20xx R. Caimmi. This is an open access 
article distributed under the Creative Commons Attribution License, which permits 
unrestricted use, distribution, and reproduction in any medium, provided the original 
work is properly cited.}

\begin{abstract} 

\noindent\noindent
With respect to earlier investigations, the theory of multi-component,
concentric, copolar, axisymmetric, rigidly rotating polytropes is improved and
extended, including subsystems with nonzero density on the boundary and
subsystems with intersecting boundaries.   The formulation is restricted to
two subsystems for simplicity but, in principle, can be extended to $N$
subsystems.   Equilibrium configurations are independent of the nature of the
fluid i.e. collisional or collisionless, provided the polytropic index lies
within the range, $1/2\le n\le5$, as in one-component systems.   The solution
of the equilibrium equations is expanded in power series, which can be
continued up to the boundary and outside via starting points placed at
increasingly larger distance from the centre of mass.   A detailed analysis is
devoted to special cases where the solution of the equilibrium equations can
be expressed analytically.   Finally a guidance example is shown, involving
homogeneous subsystems with intersecting boundaries, where a substantially
flattened component extends outside a slightly flattened one.

\end{abstract} 

{\bf Subject Classification:} Stellar and Galactic Astrophysics \\ 

{\bf Keywords:} stars: equilibrium,  galaxies: equilibrium,
polytropes: rigid rotation, tidal effects.

\section{Introduction} \label{intro}

Special attention has been devoted to polytropes since more than a century,
and is continuing to be devoted at present.  In fact, polytropic models are
useful not only for the rough estimates of some processes in real stars, but
also in the precise investigation of some matters of principle, such as the
effect of increasing central condensation on nonradial pulsation models
[26], the structure of supermassive and superdense stars [42],
the collapse of stellar cores [17],
the fission and the equatorial shedding of matter due to rotation [24],
[23], [22], [30], [33], [28], [19], the
oscillation and stability of rapidly rotating stellar models [38], 
[2], [3], [32], the structure of neutron stars and quark stars [16], [25],
the expression of the total mass of a
rotating configuration as a function of the central density [35],
equilibrium configurations and triaxiality in stars and stellar systems
[39], [40].   For exhaustive review and
complete references, an interested reader is addressed to classical textbooks
concerning stellar structure [14] and, in addition,
applications in astrophysics and related fields [21].   Among others,
polytropes can provide a continuous transition from null to infinite matter
concentration i.e. homogeneous and Roche (mass points surrounded by vanishing
atmospheres) models, distorted by rotation and/or tidal interaction [24],
Chap.\,IX, \S235.

On the other hand, large-scale stellar systems (galaxies and galaxy clusters)
appear to be made of at least two subsystems interacting only via gravitation
e.g., bulge+disk, core+halo, visible baryonic (including leptons)+dark
nonbaryonic matter.   In most applications, each component is treated
separately and the model is a simple superposition of the two matter
distributions e.g., [12].   To this respect, models involving
two-component polytropes make a further step in that each subsystem readjusts
itself in presence of tidal interaction [10], [5], [7]
and, in addition, both collisional and collisionless fluids can
be described within the range of polytropic index, $1/2\le n\le5$, [39] where
$n=0,5,$ relates to homogeneous and Roche or Plummer models, respectively.

The current paper aims to improve and extend earlier investigation on
two-component polytropes [10], [5], [7], mainly focusing on the following
points: (i) formulation of the theory where due effort is devoted to the
connection with one-component polytropes; (ii) extension of the theory to
subsystems with nonzero density on the boundary; (iii) extension of the theory
to subsystems with intersecting boundaries; (iv) extension of the theory to
collisionless fluids; (v) series expansion of the solution of equilibrium
equations, which can be continued up to the boundary and outside, via starting
points placed at increasingly larger distance from the centre of mass; (vi)
detailed analysis of special cases where the solution of equilibrium equations
can be expressed analytically; (vii) a guidance example involving homogeneous
subsystems with intersecting boundaries, where a substantially flattened
component extends outside a slightly flattened one.

The paper is organized as follows.   Points (i)-(vi) mentioned above make the
subject of Section \ref{bath}, where each argument is discussed in one or more
Subsections.   The guidance example stated on point (vii) is shown in Section
\ref{guex}.   The discussion is performed in Section \ref{disc} and the
conclusion is drawn in Section \ref{conc}.   Further details on the
formulation of the theory can be seen in the Appendix.

\section{Basic theory}
\label{bath}
\subsection{Collisional fluids}
\label{coaf}

The theory of multi-component polytropes has been developed in earlier
attempts [10], [5], [7] and shall not be
repeated, unless extensions and improvements are involved.   An interested
reader is addressed to the above quoted parent papers.   In any case,
attention is restricted to concentric, copolar, axisymmetric, rigidly rotating
polytropes which interact only via gravitation.

Accordingly, the pressure and the density within a generic subsystem, $w$, on
a total of $N$, read:
\begin{lefteqnarray}
\label{eq:pwc}
&& p_w=K_w(\rho_w^{1+1/n_w}-\rho_{{\rm b},w}^{1+1/n_w})~~;\quad
w=i_1,i_2,...,i_N~~; \\
\label{eq:rhow}
&& \rho_w=\lambda_w\theta_w^{n_w}~~;\quad w=i_1,i_2,...,i_N~~;
\end{lefteqnarray}
where $K_w$ and the polytropic index, $n_w$, are two constants,
$\rho_{{\rm b},w}$
is the density on the boundary, $\lambda_w$ is the central density,
$\theta_w^{n_w}$ is a reduced density, and
the general case relates to nonzero density on the boundary e.g., [24],
Chap.\,IX, \S235, [41], [35].

Strictly speaking, $K_w$ has no physical meaning as related dimensions would
involve density to a real power.   The combination of Eq.\,(\ref{eq:pwc})
with its counterpart particularized to the centre of mass yields:
\begin{lefteqnarray}
\label{eq:rpw}
&& \sP_w=\frac{\rho_w^{1+1/n_w}-
\rho_{{\rm b},w}^{1+1/n_w}}{\lambda_w^{1+1/n_w}-\rho_{{\rm b},w}^{1+1/n_w}}=
\frac{\srho_w^{1+1/n_w}-\srho_{{\rm b},w}^{1+1/n_w}}
{1-\srho_{{\rm b},w}^{1+1/n_w}}=\frac{\theta_w^{n_w+1}-\theta_{{\rm b},w}^
{n_w+1}}{1-\theta_{{\rm b},w}^{n_w+1}}~~; \\
\label{eq:srhow}
&& \sP_w=\frac{p_w}{\pi_w}~~;\qquad \srho_w=\frac{\rho_w}{\lambda_w}=
\theta_w^{n_w}~~;
\end{lefteqnarray}
where $\pi_w$ is the central pressure and $\sP$, $\srho$, are a reduced
pressure and a reduced density, respectively.

Let a Cartesian reference frame have origin coinciding with the
centre of mass of the whole system (which, in turn, coincides with the centre
of mass of
each subsystem) and coordinate axes coinciding with the principal axes of
inertia of the whole system (which, in turn, coincide with the principal axes
of inertia of each subsystem).   For reasons of simplicity, the centre of mass
shall be hereafter quoted as the centre or the origin (of the reference
frame), unless otherwise specified.   Without loss of generality, let the
polar axis coincide with the $x_3$ coordinate axis.

A necessary and sufficient condition for the $w$ subsystem to be in
equilibrium is [5], [7]:
\begin{equation}
\label{eq:cnse}
\Delta{\cal V}_w=\Delta{\cal V}_{\rm G}+2\Omega_w^2=-4\pi G\rho+2\Omega_w^2~~;
\quad w=i_1,i_2,...,i_N~~;
\end{equation}
where ${\cal V}_w$ is the total (gravitational + centrifugal) potential on the
$w$ subsystem,
${\cal V}_{\rm G}$ is the gravitational potential induced by the whole
system, $\Omega_w$ is the angular velocity of the $w$ subsystem and
$\rho=\rho_{i_1}+\rho_{i_2}+...+\rho_{i_N}$ is the local density of the
whole system.

From this point on, attention shall be restricted to the simple case of two
concentric, copolar, axisymmetric, rigidly rotating polytropes, hereafter
quoted as EC2 polytropes, which may exhibit non intersecting or intersecting
boundaries [5], [7].   Accordingly, the two subsystems may or may not lie
one completely within the other.   Let the two subsystems be denoted as
$i_1=i$; $i_2=j$;
respectively, and $u=i,j$; $v=j,i$; respectively, whenever they remain
unspecified.   Let a generic unspecified subsystem be denoted as $w=u,v$.

Let the volume where both subsystems coexist be defined as the common region,
and the volume where only one is present be defined as the noncommon region.
Without loss of generality, let the subsystem, $i$, exhibit the pole closer to
the origin i.e. the shorter polar axis, and let $i$, $j$, be defined as the
inner and the outer subsystem, respectively.   Strictly speaking, the inner
subsystem should be defined in connection with the undistorted configuration,
$\Omega_i=\Omega_j=0$, inferred from the knowledge of the initial conditions
which, on the other hand, is not possible to ascertain for astrophysical
systems.

The gravitational potential induced by the whole system on a generic internal
point, ${\sf P}(x_1,x_2,x_3)$, is [5], [7]:
\begin{equation}
\label{eq:VG}
{\cal V}_G=K_w(n_w+1)\lambda_w^{1/n_w}(\theta_w-\theta_{{\rm b},w})-\frac12
\Omega_w^2(x_1^2+x_2^2)+{\cal V}_{{\rm b},w}~~;
\end{equation}
where $\theta_{{\rm b},w}^{n_w}$ is the reduced density,
$\theta_w^{n_w}$, on the $w$ boundary and ${\cal V}_{{\rm b},w}$ is the
potential on the $w$ boundary.

The second term on the right-hand side of Eq.\,(\ref{eq:VG})
relates to the centrifugal potential, ${\cal V}_{C,w}$, which implies the
remaining part relates to the total potential,
${\cal V}_w={\cal V}_G+{\cal V}_{C,w}$, as:
\begin{equation}
\label{eq:V}
{\cal V}_w=K_w(n_w+1)\lambda_w^{1/n_w}(\theta_w-\theta_{{\rm b},w})+
{\cal V}_{{\rm b},w}~~;
\end{equation}
and the equipotential surfaces coincide with both the isopycnic i.e. constant
density surfaces and the isobaric surfaces [24], Chap.\,IX, \S224.

The condition of centrifugal support on a generic point on the equatorial
plane, $(\varpi,0)$, reads:
\begin{lefteqnarray}
\label{eq:Fw0}
&& \left(\frac{\partial{\cal V}_w}{\partial r}\right)_{\varpi,0}=
\left(\frac{\partial{\cal V}_G}{\partial r}\right)_{\varpi,0}+
\left(\frac{\partial{\cal V}_{C,w}}{\partial r}\right)_{\varpi,0}=0~~;
\end{lefteqnarray}
where $\varpi=(x_1^2+x_2^2)^{1/2}$.  According to the above considerations,
Eq.\,(\ref{eq:Fw0}) via (\ref{eq:VG}) reduces to:
\begin{lefteqnarray}
\label{eq:thw0}
&& \left(\frac{\partial\theta_w}{\partial r}\right)_{\varpi,0}=0~~;
\end{lefteqnarray}
where positive values on the left-hand side cause instability in that
centrifugal support is exceeded e.g., [10], [5], [7].

A necessary and sufficient condition for the $w$ subsystem to be in
equilibrium, Eq.\,(\ref{eq:cnse}), in the case under discussion, via
Eqs.\,(\ref{eq:rhow}) and (\ref{eq:VG}) reduces to:
\begin{equation}
\label{eq:PEC}
K_w(n_w+1)\lambda_w^{1/n_w}\Delta\theta_w-2\Omega_w^2=-4\pi G\sum[\lambda_w
\theta_w^{n_w}]~~;
\end{equation}
where, in general, $\sum(F_w)=F_u+F_v$ and $\theta_v^{n_v}=0$ in the noncommon
region filled by $u$ subsystem.

Let the following parameters be defined as [5], [7]:
\begin{lefteqnarray}
\label{eq:dimq}
&& \alpha_w=\Lambda_w^{1/2}\alpha_{w1}~~;\quad\upsilon_w=\Lambda_w
\upsilon_{w1}~~;\quad\xi_w=\Lambda_w^{-1/2}\xi_{w1}~~;\quad\Lambda_w
=\frac{\lambda_w}{\sum(\lambda_w)}~~;\qquad
\end{lefteqnarray}
where $\alpha$ is a scaling radius, $\upsilon$ a rotation parameter,
$\Lambda_w$
a central density ratio, $\xi$ a scaled radial coordinate, and the
index, $w1$, denotes a one-component polytrope, hereafter quoted as EC1
polytrope, defined by the $w$ subsystem
i.e. the limit of a vanishing (other than $w$) subsystem.   Accordingly,
related explicit expressions read e.g., [4]:
\begin{lefteqnarray}
\label{eq:diEC}
&& \alpha_{w1}=\left[\frac{(n_w+1)K_w\lambda_w^{1/n_w}}{4\pi G\lambda_w}
\right]^{1/2};\quad
\upsilon_{w1}=\frac{\Omega_w^2}{2\pi G\lambda_w}~~;
\quad\xi_{w1}=\frac r{\alpha_{w1}}~~;\qquad
\end{lefteqnarray}
where $K_w\lambda_w^{1+1/n_w}$ is dimensioned as a pressure and $r$ is the
usual radial coordinate.

The additional relations:
\begin{lefteqnarray}
\label{eq:csir}
&& \alpha_{v1}\xi_{v1}=\alpha_v\xi_v=\alpha_u\xi_u=\alpha_{u1}\xi_{u1}~~; \\
\label{eq:bsir}
&& \alpha_{v1}\Xi_{v1}=\alpha_v\Xi_v=\alpha_u\xi_u^\ast=\alpha_{u1}\xi_{u1}^
\ast~~; \\
\label{eq:Ltot}
&& \sum(\Lambda_w)=1~~; \\
\label{eq:lalfa}
&& \sum(\lambda_w)\alpha_w^2=\lambda_w\alpha_{w1}^2~~; \\
\label{eq:upcsi}
&& \upsilon_w\xi_w^2=\upsilon_{w1}\xi_{w1}^2~~;
\end{lefteqnarray}
follow from Eq.\,(\ref{eq:dimq}) via (\ref{eq:diEC}), where $\Xi_v$,
$\xi_u^\ast$, define the interface provided
$\alpha_v\Xi_v(\mu)<\alpha_u\Xi_u(\mu)$ along the
direction considered.   It is worth noticing $\xi_u^\ast=\xi_u^\ast(\mu)$ does
not imply $\theta_u(\xi_u^\ast)={\rm const}$, unless the two subsystems rotate
to the same extent, $\upsilon_u=\upsilon_v$.

The substitution of Eq.\,(\ref{eq:dimq}), (\ref{eq:diEC}), into (\ref{eq:VG})
and (\ref{eq:PEC}), after replacing Cartesian with polar coordinates, yields
[5], [7]:
\begin{lefteqnarray}
\label{eq:VGd}
&& {\cal V}_G=4\pi G\sum(\lambda_w)\alpha_w^2\left\{\theta_w-\theta_{{\rm b},
w}-\frac16\upsilon_w\xi_w^2[1-P_2(\mu)]\right\}+{\cal V}_{{\rm b},w}~~; \\
\label{eq:PEd}
&& \Delta\theta_w-\upsilon_w=-\sum(\Lambda_w\theta_w^{n_w})~~;
\end{lefteqnarray}
where $(r,\mu)=(\alpha_w\xi_w,\cos\delta)$ are polar coordinates ($\delta$
polar angle), implying $x_1^2+x_2^2=\alpha_w^2\xi_w^2(2/3)[1-P_2(\mu)]$,
$P_2(\mu)=(3\mu^2-1)/2$ is
the Legendre polynomial of degree, 2, and a vanishing subsystem in the non
common region implies a null density, $\theta_v^{n_v}(\xi_v,\mu)=0$,
$\xi_v\ge\Xi_v$, along the direction considered, which makes the sum on the
right-hand side of Eq.\,(\ref{eq:PEd}) reduce to a single term,
$\Lambda_u\theta_u^{n_u}(\xi_u,\mu)$, along the direction considered.

In terms of the parameters, $\alpha_{w1}$, $\upsilon_{w1}$, and the variable,
$\xi_{w1}$, Eq.\,(\ref{eq:VGd}) via (\ref{eq:lalfa}) and (\ref{eq:upcsi})
translates into:
\begin{lefteqnarray}
\label{eq:VGd1}
&& {\cal V}_G=4\pi G\lambda_w\alpha_{w1}^2\left\{\theta_w-\theta_{{\rm b},w}-
\frac16\upsilon_{w1}\xi_{w1}^2[1-P_2(\mu)]\right\}+{\cal V}_{{\rm b},w}~~;
\end{lefteqnarray}
which is the formal expression of the gravitational potential of EC1
polytropes related to
central density, $\lambda_w$, scaling radius, $\alpha_{w1}$, and rotation
parameter, $\upsilon_{w1}$.

In dimensionless polar coordinates, the Laplace operator translates into:
\begin{equation}
\label{eq:Dep}
\Delta=\frac1{\xi_w^2}\frac\partial{\partial\xi_w}\left(\xi_w^2\frac\partial
{\partial\xi_w}\right)+\frac1{\xi_w^2}\frac\partial{\partial\mu}\left\{\frac23
[1-P_2(\mu)]\frac\partial{\partial\mu}\right\}~~;
\end{equation}
accordingly, Eq.\,(\ref{eq:PEd}) takes the explicit form:
\begin{equation}
\label{eq:EC2}
\frac1{\xi_w^2}\frac\partial{\partial\xi_w}\left(\xi_w^2\frac{\partial
\theta_w}{\partial\xi_w}\right)+
\frac1{\xi_w^2}\frac\partial{\partial\mu}\left[(1-\mu^2)\frac{\partial
\theta_w}{\partial\mu}\right]-\upsilon_w=-\sum(\Lambda_w\theta_w^{n_w})~~;
\end{equation}
which shall be hereafter quoted as the EC2 equation.
The related boundary conditions read [10], [5], [7]:
\begin{lefteqnarray}
\label{eq:EC2b}
&& \theta_w(0,\mu)=1~~;\qquad\left(\frac{\partial\theta_w}
{\partial\xi_w}\right)_{0,\mu}=0~~;\qquad\left(\frac{\partial\theta_w}
{\partial\mu}\right)_{0,\mu}=0~~; \\
\label{eq:EC2c}
&& \theta_w(\xi_w^{\ast-},\mu)=\theta_w(\xi_w^{\ast+},\mu)~~; \\
\label{eq:EC2d}
&& \left(\frac{\partial\theta_w}{\partial\xi_w}\right)_{\xi_w^{\ast-},
\mu}=\left(\frac{\partial\theta_w}{\partial\xi_w}\right)_{\xi_w^{\ast+},\mu}
~~;\qquad\left(\frac{\partial\theta_w}{\partial\mu}\right)_{\xi_w^{\ast-},\mu}
=\left(\frac{\partial\theta_w}{\partial\mu}\right)_{\xi_w^{\ast+},\mu}~~;
\end{lefteqnarray}
where $\xi_w^{\ast-}<\xi_w^\ast$ and $\xi_w^{\ast+}>\xi_w^\ast$ are infinitely
close to the interface, $\xi_w^\ast=\Xi_v,\xi_u^\ast$.

In terms of the parameters defined by Eq.\,(\ref{eq:dimq}), the
EC2 equation, Eq.\,(\ref{eq:EC2}), via Eq.\,(\ref{eq:diEC}) translates into:
\begin{lefteqnarray}
\label{eq:ECE}
&&
\frac1{\xi_{u1}^2}\frac\partial{\partial\xi_{u1}}\left(\xi_{u1}^2
\frac{\partial\theta_u}{\partial\xi_{u1}}\right)+
\frac1{\xi_{u1}^2}\frac\partial{\partial\mu}\left[(1-\mu^2)\frac{\partial
\theta_u}{\partial\mu}\right]-\upsilon_{u1}=
-\theta_u^{n_u}-\Lambda_{vu}\theta_v^{n_v}~~;\qquad \\
\label{eq:Lvu}
&& \Lambda_{vu}=\frac{\Lambda_v}{\Lambda_u}=\frac{\lambda_v}{\lambda_u}~~;
\end{lefteqnarray}
where the last term on the right-hand side of Eq.\,(\ref{eq:ECE}) could be
interpreted as a tidal interaction due to the $v$ subsystem.

Within the noncommon region, $\theta_v^{n_v}(\xi_v,\mu)=0$, the EC2 equation
reduces to its counterpart related to a single polytrope with polytropic
index, $n_u$, central density, $\lambda_u$, central pressure,
$\pi_u=K_u(\lambda_u^{1+1/n_u}-\rho_{{\rm b},u}^{1+1/n_u})$, scaling radius,
$\alpha_{u1}$, rotation
parameter, $\upsilon_{u1}$, and related solutions formally coincide even if
the boundary conditions are different.   A
special integral is $\theta_u^{(p)}(\xi_{u1},\mu)=\upsilon_{u1}^{1/n_u}$.

In the limit of a vanishing inner (along the direction considered) subsystem,
$\lambda_v\to0$, $\Lambda_v\to0$, $\Lambda_u\to1$, $\alpha_{u1}\to\alpha_u$,
$\upsilon_{u1}\to\upsilon_u$, $\xi_{u1}\to\xi_u$, and Eq.\,(\ref{eq:ECE})
reduces to its counterpart related to EC1 systems.   On the other
hand, $K_v\to+\infty$, $\Omega_v\to0$, to ensure finite and nonzero $\alpha_v$
(implying the same for $\xi_v$), $\upsilon_{v1}$, respectively.

In the limit of a vanishing outer (along the direction considered) subsystem,
$\lambda_u\to0$, $\Lambda_u\to0$, $\Lambda_v\to1$, $\alpha_{v1}\to\alpha_v$,
$\upsilon_{v1}\to\upsilon_v$, $\xi_{v1}\to\xi_v$, and Eq.\,(\ref{eq:ECE})
(where the indexes, $u$ and $v$, are mutually exchanged)
reduces to its counterpart related to EC1 systems.   On the other
hand, $K_u\to+\infty$, $\Omega_u\to0$, to ensure finite and nonzero $\alpha_u$
(implying the same for $\xi_u$), $\upsilon_{u1}$, respectively.
In addition, the gravitational potential within the noncommon region tends to
the gravitational potential induced by the inner subsystem outside the
interface, as:
\begin{lefteqnarray}
\label{eq:VGd2}
&& 4\pi G\lambda_v\alpha_u^2\left\{\theta_u(\xi_u,\mu)-\frac16\upsilon_u
\xi_u^2[1-P_2(\mu)]\right\}+{\cal V}_{{\rm b},u}=
{\cal V}_v^{({\rm ext})}(\xi_v,\mu)+{\cal V}_{{\rm b},v}~~;\qquad
\end{lefteqnarray}
where ${\cal V}_v^{({\rm ext})}$ can be expanded in Legendre polynomials
e.g., [27], Chap.\,VII, \S 193.

If one subsystem is completely lying within the other, a vanishing outer
component resembles generalized Roche models [24], Chap.\,IX, \S 234,
keeping in mind the following differences: (i) the massive body is a polytrope
instead of a homogeneous spheroid, and (ii) the vanishing atmosphere extends
down to the centre instead of branching off from the boundary of the massive
body.

The comparison between the alternative expressions of the gravitational
potential, Eq.\,(\ref{eq:VGd}), yields:
\begin{eqnarray}
\label{eq:Pij}
&& \phantom{=}
4\pi G\sum(\lambda_w)\alpha_u^2\left\{\theta_u-\theta_{{\rm b},u}-\frac
16\upsilon_u\xi_u^2[1-P_2(\mu)]\right\}+{\cal V}_{{\rm b},u} \nonumber \\
&& =
4\pi G\sum(\lambda_w)\alpha_v^2\left\{\theta_v-\theta_{{\rm b},v}-\frac16
\upsilon_v\xi_v^2[1-P_2(\mu)]\right\}+{\cal V}_{{\rm b},v}~~;\quad\qquad
\end{eqnarray}
which holds within and on the interface, $\xi_w\le\xi_w^\ast$.

At the centre, $(\xi_w,\mu)=(0,\mu)$, Eq.\,(\ref{eq:Pij})
by use of (\ref{eq:EC2b}) reduces to:
\begin{eqnarray}
\label{eq:P0}
&& 4\pi G\sum(\lambda_w)\alpha_u^2(1-\theta_{{\rm b},u})+{\cal V}_{{\rm b},u}=
   4\pi G\sum(\lambda_w)\alpha_v^2(1-\theta_{{\rm b},v})+{\cal V}_{{\rm b},v}
   ~~;\qquad
\end{eqnarray}
and the substitution of Eq.\,(\ref{eq:P0}) into (\ref{eq:Pij}) after
some algebra yields:
\begin{eqnarray}
\label{eq:P2ij}
&& \alpha_u^2\left\{1-\theta_u+\frac16\upsilon_u\xi_u^2[1-P_2(\mu)]
\right\}=\alpha_v^2\left\{1-\theta_v+\frac16\upsilon_v\xi_v^2[1-
P_2(\mu)]\right\};\qquad \nonumber \\
&& \xi_w\le\xi_w^\ast~~;
\end{eqnarray}
with no explicit boundary dependence.

The combination of Eqs.\,(\ref{eq:csir}) and (\ref{eq:P2ij}) after
little algebra yields:
\begin{lefteqnarray}
\label{eq:ispuv}
&& \theta_v(\xi_v,\mu)=\theta_u(\xi_u,\mu)+(1-\Gamma_{uv})
[1-\theta_u(\xi_u,\mu)] \nonumber \\
&& \phantom{\theta_v(\xi_v,\mu)=\theta_u(\xi_u,\mu)}+
\Gamma_{uv}\frac{\upsilon_v-\upsilon_u}6
\xi_u^2[1-P_2(\mu)]~~;\qquad\xi_w\le\xi_w^\ast~~;\qquad \\
\label{eq:Gauv}
&& \Gamma_{uv}=\frac{\alpha_u^2}{\alpha_v^2}~~;
\end{lefteqnarray}
where the third term on the right-hand side of Eq.\,(\ref{eq:ispuv}) is null
for subsystems rotating to the same extent $(\upsilon_u=\upsilon_v)$ and the
second is null for subsystems with coinciding scaling radii, $\Gamma_{uv}=1$
or $\alpha_u=\alpha_v$.

Accordingly, four different situations can be considered concerning
subsystems, namely:
\begin{description}
\item[$\bullet$]
rotating to different extents and showing different scaling radii;
\item[$\bullet$]
rotating to different extents but showing equal scaling radii;
\item[$\bullet$]
rotating to the same extent but showing different scaling radii;
\item[$\bullet$]
rotating to the same extent and showing equal scaling radii.
\end{description}

With regard to the last case,
the system reduces to a single matter distribution with central density,
$\lambda=\lambda_u+\lambda_v$, and central pressure,
$\pi=K_u(\lambda_u^{1+1/n_u}-\rho_{{\rm b},u}^{1+1/n_u})+K_v
(\lambda_v^{1+1/n_v}-\rho_{{\rm b},v}^{1+1/n_v})$.
The additional restriction of equal polytropic indexes, $n_u=n_v=n$, implies a
single reduced density profile, $\theta_u^{n_u}=\theta_v^{n_v}=\theta^n$ and
hence $\rho=\rho_u+\rho_v=\lambda_u
\theta_u^{n_u}+\lambda_v\theta_v^{n_v}=\lambda\theta^n$, where the central
pressure reads $\pi=K(\lambda^{1+1/n}-\rho_{\rm b}^{1+1/n})$,
$K=K_u(\lambda_u^{1+1/n}-\rho_{{\rm b},u}^{1+1/n})/
(\lambda^{1+1/n}-\rho_{\rm b}^{1+1/n})+
K_v(\lambda_v^{1+1/n}-\rho_{{\rm b},v}^{1+1/n})/
(\lambda^{1+1/n}-\rho_{\rm b}^{1+1/n})$, conform to Dalton's law on gas
partial pressure, and $\rho_{\rm b}=\rho_{{\rm b},u}+\rho_{{\rm b},v}$ is the
boundary density.

The particularization of Eq.\,(\ref{eq:P2ij}) to the polar axis, $P_2(\mu)=1$,
shows that $\theta_u=1-\zeta$ implies $\theta_v=1-\Gamma_{uv}\zeta$
and, in particular, $\theta_v=1-\zeta=\theta_u$ for
$\Gamma_{uv}=1$ i.e. related isopycnic surfaces are tangent on the polar axis
within the common region, $\xi_w\le\xi_w^\ast$.

At the pole of the interface, $(\xi_i,\mu)=(\Xi_i,1)$, $(\xi_j,\mu)=
(\xi_j^\ast,1)$, $\alpha_i\Xi_i=\alpha_j\xi_j^\ast$, $\theta_i=\theta_{{\rm b},i}$,
$\theta_j=\theta_{{\rm b},j}^\ast$, and Eq.\,(\ref{eq:Pij}) reduces to:
\begin{eqnarray}
\label{eq:PC}
&& {\cal V}_{{\rm b},i}=4\pi G\sum(\lambda_w)\alpha_j^2(\theta_{{\rm b},j}^
\ast-\theta_{{\rm b},j})+{\cal V}_{{\rm b},j}~~;
\end{eqnarray}
where $\Xi_i(\mu)$, $\xi_j^\ast(\mu)$, are the equations of the inner
boundary in terms of related subsystems and $\theta_{{\rm b},j}^\ast=
\theta_{{\rm b},j}(\xi_j^\ast,1)$ denotes the isopycnic surface of the outer
subsystem which is tangent to the interface on the pole, according to the
above considerations.

The particularization of Eq.\,(\ref{eq:ispuv}), $u=j$, $v=i$, to the pole of
the interface, yields after some
algebra:
%
\begin{equation}
\label{eq:raij}
\Gamma_{ji}=\frac{1-\theta_{{\rm b},i}}{1-\theta_{{\rm b},j}^\ast}~~;
\end{equation}
where $\Gamma_{ji}\ge1$ implies $\theta_{{\rm b},j}^\ast\ge\theta_{{\rm b},i}$
and vice versa.

The substitution of Eq.\,(\ref{eq:raij}) into (\ref{eq:P2ij}), by use
of (\ref{eq:Gauv}), after some algebra yields:
\begin{equation}
\label{eq:thic}
\alpha_i^2\left\{\theta_i-\theta_{{\rm b},i}-\frac16\upsilon_i\xi_i^2[1-P_2
(\mu)]\right\}=\alpha_j^2\left\{\theta_j-\theta_{{\rm b},j}^\ast-\frac16
\upsilon_j\xi_j^2[1-P_2(\mu)]\right\}~~;
\end{equation}
which depends only on scaling radii, boundary densities, and rotation
parameters.

On the polar axis, $P_2(\mu)=1$, Eq.\,(\ref{eq:thic}) reduces to:
\begin{equation}
\label{eq:ratc}
\Gamma_{ji}=\frac{\theta_i-\theta_{{\rm b},i}}{\theta_j-
\theta_{{\rm b},j}^\ast}~~;
\end{equation}
which includes Eq.\,(\ref{eq:raij}) and also holds, for any direction via
Eq.\,(\ref{eq:thic}), in the special case of subsystems rotating to the same
extent, $\upsilon_i=\upsilon_j$, where the equipotential surface of the $j$
subsystem, $\theta_j=\ell_j$, and the equipotential surface of the $i$
subsystem, $\theta_i=\theta_{{\rm b},i}+\Gamma_{ji}(\ell_j-\theta_{{\rm b},j}^
\ast)=\ell_i$, are coincident in the physical space, as expected.

Turning to the general case, equilibrium configurations may be determined to a
selected order of
approximation, using a method outlined in earlier attempts [10], [5], [7].

\subsection{Collisionless fluids}
\label{colf}

The results of Subsection \ref{coaf}, related to multi-component (in
particular, two-component) polytropes, can be extended from the collisional to
the collisionless case following the same kind of procedure used for
EC1 polytropes.    To this respect, it has been established an
exact collisionless dynamical counterpart exists for collisional polytropes
within the polytropic index range, $1/2\le n\le5$, [39], [21],
Chap.\,6, \S 6.1.9, [1], Chap.\,4, \S 4.3.3.   The
same holds for the isothermal sphere [1], Chap.\,4, \S4.3.3.   A similar
method can be used in dealing with multi-component (in
particular, two-component) polytropes.   For detailed calculations, an
interested reader is addressed to the above mentioned references.

Accordingly, the distribution function of the $w$ subsystem reads:
\begin{equation}
\label{eq:fw}
f_w(x_1,x_2,x_3,v_1,v_2,v_3)=\cases{
C_w(H_w-H)^{n_w-3/2}~~; & $H\le H_w<0~~;$ \cr
0~~;                          & $H>H_w~~;$      \cr
}
\end{equation}
where $C_w$ is a positive constant dimensioned as [L$^{-2n_w-3}$T$^{2n_w}$]
and $H_w$ a negative constant dimensioned as [L$^2$T$^{-2}$].   In addition,
the requirement of a finite mass implies $n_w>1/2$.   In the limiting case,
$n_w\to1/2$, Eq.\,(\ref{eq:fw}) reduces to:
\begin{equation}
\label{eq:fwdd}
f_w(x_1,x_2,x_3,v_1,v_2,v_3)=
C_w\delta(H_w-H)~~;
\end{equation}
where $\delta(H_w-H)$ is the Dirac's delta dimensioned as [L$^{-2}$T$^2$].

The integration of the distribution function, expressed by Eq.\,(\ref{eq:fw})
and (\ref{eq:fwdd}), over the coordinate space, yields the density
distribution:
\begin{lefteqnarray}
\label{eq:roww}
&& \rho_w(x_1,x_2,x_3)=\mu_w\psi_w^{n_w}(x_1,x_2,x_3)~~;\quad n_w\ge\frac12~~;
 \\
\label{eq:muw}
&& \mu_w=2^{5/2}\pi\overline{m}_wC_wB\left(\frac32,n_w-\frac12\right)~~; \\
\label{eq:psiw}
&& \psi_w(x_1,x_2,x_3)=H_w+{\cal V}_{\rm G}(x_1,x_2,x_3)+\frac12\Omega_w^2
(x_1^2+x_2^2)~~;
\end{lefteqnarray}
where $\overline{m}_w$ is the mean particle mass of the $w$ subsystem,
${\cal V}_{\rm G}$ is the total gravitational potential and $B(p,q)$
coincides with the Euler's complete beta function for $p>0$, $q>0$, and equals
unity for $q=0$ i.e. $n_w=1/2$.

The comparison between Eqs.\,(\ref{eq:rhow}) and (\ref{eq:roww}) shows that
collisional and collisionless polytropic subsystems with equal density
profiles and polytropic indexes, subjected to equal tidal and centrifugal
potential, are related as:
\begin{equation}
\label{eq:laps}
\rho_w=\lambda_w\theta_w^{n_w}=\mu_w\psi_w^{n_w}~~;
\end{equation}
which is equivalent to:
\begin{equation}
\label{eq:lap2}
\psi_w=\left(\frac{\lambda_w}{\mu_w}\right)^{1/n_w}\theta_w~~;
\end{equation}
where the ratio, $(\lambda_w/\mu_w)^{1/n_w}$, is dimensioned as a potential,
[L$^2$T$^{-2}$].

The Poisson equation is readily derived from Eq.\,(\ref{eq:psiw}) as:
\begin{equation}
\label{eq:Poiss}
\Delta{\cal V}_{\rm G}+2\Omega_w^2=\Delta\psi_w=-4\pi G\rho+2\Omega_w^2~~;
\end{equation}
where $\rho=\sum(\rho_w)=\sum(\mu_w\psi_w^{n_w})$ is the total density and
$\psi_w=0$ outside the $w$ subsystem.

The mean square velocity of a collisionless fluid, $\overline{(v_w^2)}$, may
be calculated using the theorem of the mean with regard to the phase space.
In the case under discussion, the result is:
\begin{equation}
\label{eq:v2m}
\overline{(v_w^2)}=\frac3{n_w+1}\psi_w~~;\quad n_w\ge\frac12~~;
\end{equation}
which, under the restriction of isotropic stress tensor and nonrelativistic
velocities, relates to the pressure as e.g., [21], Chap.\,1, \S 1.7:
\begin{equation}
\label{eq:pwn}
p_w=\frac13\rho_w\overline{(v_w^2)}=\frac1{n_w+1}\rho_w\psi_w~~;\quad v_w\ll c
~~;\quad n_w\ge\frac12~~;
\end{equation}
and the substitution of Eqs.\,(\ref{eq:roww}) and (\ref{eq:laps}) into
(\ref{eq:pwn}) yields:
\begin{lefteqnarray}
\label{eq:pw2}
&& p_w=K_w\rho_w^{1+1/n_w}~~; \\
\label{eq:Kw}
&& K_w=\frac1{n_w+1}\frac1{\mu_w^{1/n_w}}~~;\quad n_w\ge\frac12~~;
\end{lefteqnarray}
a comparison between Eqs.\,(\ref{eq:pwc}) and (\ref{eq:pwn}) shows that the
pressure of collisionless polytropic subsystems has the same formal expression
of their dynamical collisional counterparts where the density on the boundary
is null, $\rho_{{\rm b},w}=0$, provided $n_w\ge1/2$.

The pressure has necessarily to be null on the boundary, $p_{{\rm b},w}=0$,
which implies $\rho_{{\rm b},w}=0$, $\psi_{{\rm b},w}=0$.   Accordingly, the
particularization of Eq.\,(\ref{eq:psiw}) to the surface of the $w$
subsystem yields:
\begin{equation}
\label{eq:H0}
H_w+{\cal V}_{{\rm b},w}=0~~;
\end{equation}
or $H_w=-{\cal V}_{{\rm b},w}$ i.e. the opposite of the potential
induced by the whole system on the surface of the $w$ subsystem.

The above results for collisionless polytropic subsystems may be formulated in
terms of dimensionless variables and parameters, defined by
Eq.\,(\ref{eq:dimq}), by use of Eq.\,(\ref{eq:laps}), yielding the same formal
expression of their dynamical collisional counterparts, provided $n_w\ge1/2$.

The existence of
collisionless polytropes with nonzero boundary densities, exhibiting equal
configurations with respect to their collisional counterparts, remains (to the
knowledge of the author) an open  question.

\subsection{The general problem}
\label{gepr}

For axisymmetric configurations, the exact solutions of the EC2 equation,
Eq.\,(\ref{eq:EC2}), may be expanded in series of even
Legendre polynomials multiplied by functions of the radial coordinate,
$\theta_{2\ell,w}(\xi_w)$, and coefficients, $A_{2\ell,w}$, which depend on
the rotation parameters provided $2\ell>0$.   Odd Legendre polynomials in the
series expansion are ruled out by symmetry with respect to the equatorial
plane.   For further details, an interested reader is addressed to an earlier
attempt [4].

\subsubsection{The noncommon region}
\label{ncr}

The presence of either subsystem within the noncommon region allows the
notation be shortened for sake of simplicity, throughout the current
Subsection, as:
\begin{equation}
\label{eq:snnr}
\xi=\xi_{w1}~~;\quad\upsilon=\upsilon_{w1}~~;\quad n=n_w~~;\quad
A_{2\ell}=A_{2\ell,w}^{({\rm ncm})}~~;\quad
\theta_{2\ell}=\theta_{2\ell,w}^{({\rm ncm})}~~;
\end{equation}
where the apex, (ncm), denotes the noncommon region.   Accordingly,
the solution of Eq.\,(\ref{eq:EC2}) via
(\ref{eq:ECE}) can be expanded in Legendre polynomials as [10], [7]:
\begin{lefteqnarray}
\label{eq:thn}
&& \theta(\xi,\mu)=\sum_{\ell=0}^{+\infty}A_{2\ell}\theta_{2\ell}(\xi)P_
{2\ell}(\mu)~~; \\
\label{eq:Lp}
&& P_\ell(\mu)=\frac1{2^\ell}\frac1{\ell!}\frac{\diff^\ell}{\diff\mu^\ell}
[(\mu^2-1)^\ell]~~;\qquad\vert P_\ell(\mu)\vert\le1~~;\qquad\ell=0,1,2,...~~;
\qquad \\
\label{eq:Le}
&& \frac{\diff}{\diff\mu}\left[(1-\mu^2)\frac{\diff P_\ell}{\diff\mu}\right]=
-\ell(\ell+1)P_\ell(\mu)~~; \\
\label{eq:A2ln}
&& 
\lim_{\upsilon_w\to0}A_{2\ell}=\delta_{2\ell,0}~~;
\end{lefteqnarray}
where $\delta$ is the Kronekher symbol; $A_{2\ell}$, $2\ell>0$, are
coefficients which depend on the rotation
parameters; $\theta_{2\ell}$ are the EC2 associated function of degree,
$2\ell$; and the Legendre polynomials, defined by Eq.\,(\ref{eq:Lp}), obey
the Legendre equation, Eq.\,(\ref{eq:Le}).

In absence of rotation,
$\upsilon_u=\upsilon_v=0$, the system attains the related undistorted
spherical configuration and the EC2 associated function, $\theta_0(\xi)$, via
Eq.\,(\ref{eq:EC2b})
coincides with the EC2 function, $\theta(\xi,\mu)$, hence $A_0=1$ according
to Eq.\,(\ref{eq:A2ln}).
%
%
Conversely, the coefficients, $A_{2\ell}$, $2\ell>0$, depend on
the rotation parameters and vanish for spherical configurations, according to
Eq.\,(\ref{eq:A2ln}).

If the distorsion due to rigid rotation may be considered as a small
perturbation on the spherical shape, then the first term of the series
expansion on the right-hand side of Eq.\,(\ref{eq:thn}) is dominant, which
implies the following inequality:
\begin{lefteqnarray}
\label{eq:S2n}
&& \vert R_1(\xi,\mu)\vert\ll\vert\theta_0(\xi)\vert~~; \\
\label{eq:R2n}
&& R_1(\xi,\mu)=\sum_{\ell=1}^{+\infty}A_{2\ell}\theta_{2\ell}(\xi)P_{2\ell}
(\mu)~~;
\end{lefteqnarray}
and the power on the right-hand side of Eq.\,(\ref{eq:ECE}) can safely be
approximated as:
\begin{lefteqnarray}
\label{eq:thn2l}
&& \theta^n(\xi,\mu)=\theta_0^n(\xi)+n\theta_0^{n-1}(\xi)R_1(\xi,\mu)~~;
\end{lefteqnarray}
which is a series expansion in Legendre polynomials.   With regard to the
special cases, $n=0,1$, Eq.\,(\ref{eq:thn2l}) is exact.

The series on the right-hand side of Eq.\,(\ref{eq:thn}) describes the
expansion of the nonrotating noncommon region as a whole, via $\theta_0$, and
superimposed on this an oblateness, via $R_1$.   More specifically, $\theta_0$
relates to an expanded spherical shell where the radial contribution of rigid
rotation adds to the undistorted configuration, and $R_1$ quantifies the
meridional distortion.

Let $\xi_{\rm ex}$ define the (fictitious) spherical isopycnic surface of the
expanded spherical shell, as:
\begin{lefteqnarray}
\label{eq:isp0}
&& \theta_0(\xi_{\rm ex})=\theta(\xi_{\rm un})=\kappa~~;
\end{lefteqnarray}
where the index, ex, un, denotes the expanded and the undistorted spherical
shell, respectively, and radial expansion implies
$\xi_{\rm un}\le\xi_{\rm ex}$.

The substitution of Eqs.\,(\ref{eq:R2n}) and (\ref{eq:isp0}) into
(\ref{eq:thn}) yields:
\begin{lefteqnarray}
\label{eq:R1z}
&& R_1(\xi_{\rm ex},\mp\mu_{\rm ex})=0~~;
\end{lefteqnarray}
and the locus, $(\xi_{\rm ex},\mp\mu_{\rm ex})$, defines the intersection
between the oblate isopycnic surface, $\theta(\xi,\mu)=\kappa$, and the
(fictitious) spherical isopycnic surface, $\theta_0(\xi_{\rm ex})=\kappa$.
In the
nonrotating limit, $\upsilon\to0$, $\xi_{\rm ex}\to\xi_{\rm un}$, and the term
containing $P_2(\mu)$ is expected to be dominant with respect to the others in
Eq.\,(\ref{eq:R2n}).   Accordingly, $\mu_{\rm ex}$ relates to $P_2(\mu)=0$,
hence $\mu_{\rm ex}\to1/\sqrt3$.   For further details, an interested reader
is addressed to the parent paper [9].

The substitution of Eqs.\,(\ref{eq:thn}) and (\ref{eq:thn2l}) into
(\ref{eq:ECE}), after equating
separately the terms of the same degree in Legendre polynomials, shows
Eq.\,(\ref{eq:ECE}) is equivalent to the set of EC2 associated equations
e.g., [6]:
\begin{lefteqnarray}
\label{eq:t0n}
&& \frac1{\xi^2}\frac{\diff}{\diff\xi}\left(\xi^2\frac{\diff\theta_0}{\diff
\xi}\right)-\upsilon=-\vert\theta_0\vert^n\cos(n\pi\zeta)~~; \\
\label{eq:t2n}
&& \frac1{\xi^2}\frac{\diff}{\diff\xi}\left(\xi^2\frac{\diff\theta_2}{\diff
\xi}\right)-\frac6{\xi^2}\theta_2=-n\vert\theta_0\vert^{n-1}\cos[(n-1)\pi
\zeta]\theta_2~~; \\
\label{eq:t2ln}
&& \frac1{\xi^2}\frac{\diff}{\diff\xi}\left(\xi^2\frac{\diff\theta_{2\ell}}
{\diff\xi}\right)-\frac{(2\ell+1)2\ell}{\xi^2}\theta_{2\ell}=-n\vert\theta_0
\vert^{n-1}\cos[(n-1)\pi\zeta]\theta_{2\ell}~~;\quad \\
\label{eq:zin}
&& \zeta=\cases{
0~~; & $\theta_0\ge0~~;$ \cr
1~~; & $\theta_0<0~~;$   \cr
}
\end{lefteqnarray}
where the occurrence of absolute values and cosines on the
right-hand side of Eqs.\,(\ref{eq:t0n})-(\ref{eq:t2ln}), makes $\theta_0^n$
and $\theta_0^{n-1}$, $\theta_0<0$, be evaluated as the real part of the
principal value of complex powers; the boundary conditions relate to the
interface.

The solutions of the EC2 associated equations,
Eqs.\,(\ref{eq:t0n})-(\ref{eq:t2ln}), may be expanded in Taylor series as:
\begin{lefteqnarray}
\label{eq:sersn}
&& \theta_{2\ell}(\xi)=\sum_{k=0}^{+\infty}a_{2\ell,k}(\xi_0)(\xi-\xi_0)^k~~;
 \\
\label{eq:cersn}
&& a_{2\ell,0}(\xi_0)=\theta_{2\ell}(\xi_0)~~;\qquad a_{2\ell,k}(\xi_0)=\frac1
{k!}\left(\frac{\diff^k\theta_{2\ell}}{\diff\xi^k}\right)_{\xi_0}~~;
\end{lefteqnarray}
where, in particular, $a_{2\ell,1}(\xi_0)=\theta_{2\ell}^\prime(\xi_0)$ and
$a_{2\ell,2}(\xi_0)=\theta_{2\ell}^\pprime(\xi_0)/2$.  It can be seen the
convergence radius tends to be null if the starting point, $\xi_0$, tends to a
singular point outside the origin, $\xi_0^\dagger$,
$\theta_0(\xi_0^\dagger)=0$, which implies
$\theta_0(\xi)$ and $\theta_0(\xi_0)$ are both positive or negative.   The
first and the second derivative of the EC2 associated functions, $\theta_
{2\ell}^\prime(\xi)$ and $\theta_{2\ell}^\pprime(\xi)$, can readily be
expanded in Taylor series via Eq.\,(\ref{eq:sersn}).   For further details, an
interested reader is addressed to the parent paper [6].

The powers of $\vert\theta_0\vert$ appearing on the right-hand side of
Eqs.\,(\ref{eq:t0n})-(\ref{eq:t2ln}) can be expanded in series as e.g., [18]:
\begin{lefteqnarray}
\label{eq:serpn}
&& \vert\theta_0(\xi)\vert^x=\vert\theta_0(\xi_0)\vert^x\sum_{k=0}^{+\infty}
C_k^{(x)}(\xi-\xi_0)^k~~; \\
\label{eq:cerpn}
&& C_k^{(x)}=\frac1{\theta_0(\xi_0)}\frac1k\sum_{i=1}^k(ix-k+i)a_{0,i}C_{k-i}^
{(x)}~~;\qquad C_0^{(x)}=1~~;
\end{lefteqnarray}
where $a_{0,i}$ are defined by Eq.\,(\ref{eq:cersn}).

On the other hand, using Eqs.\,(\ref{eq:sersn}) and (\ref{eq:cersn}) yields:
\begin{equation}
\label{eq:sern0}
\vert\theta_0(\xi)\vert^x=\vert\theta_0(\xi_0)\vert^x\left[1+\frac1{\theta_0
(\xi_0)}\sum_{k=1}^{+\infty}a_{0,k}(\xi-\xi_0)^k\right]^x~~;
\end{equation}
provided $\xi_0$, $\xi$, both precede or exceed a singular
point, $\xi_0^\dagger$, $\theta_0(\xi_0^\dagger)=0$, i.e.
$\xi_0<\xi<\xi_0^\dagger$ or $\xi_0^\dagger<\xi_0<\xi$.   More specifically,
$\theta_0(\xi)/\theta_0(\xi_0)>0$ in the case under
discussion, which implies the quantity within square brackets on the
right-hand side of Eq.\,(\ref{eq:sern0}) is always positive.

The
following identity:
\begin{equation}
\label{eq:LEi}
\frac1{\xi^2}\frac{\diff}{\diff\xi}\left(\xi^2\frac{\diff\theta_{2\ell}}{\diff
\xi}\right)=\theta_{2\ell}^\pprime+\frac2\xi\theta_{2\ell}^\prime~~;
\end{equation}
implies use of the Taylor series expansion:
\begin{equation}
\label{eq:cn1s}
\frac1\xi=\frac1{\xi_0}\sum_{k=0}^{+\infty}(-1)^k\left(\frac{\xi-\xi_0}{\xi_0}
\right)^k~~;\qquad\vert\xi-\xi_0\vert<\xi_0~~;
\end{equation}
where the starting point, $\xi_0$, has to be replaced by $\xi_0+\Delta\xi<2\xi
_0$ whenever $\xi\ge2\xi_0$, to ensure convergence.

The substitution of the series expansions for $\theta_0^\pprime$, $\theta_0^
\prime$, $\vert\theta_0\vert^n$, $1/\xi$, via
Eqs.\,(\ref{eq:sersn})-(\ref{eq:cn1s}), into
(\ref{eq:t0n}), keeping in mind the coefficients of $(\xi-\xi_0)^k$
on both sides must necessarily be equal, yields after a lot of algebra [6]:
\begin{leftsubeqnarray}
\slabel{eq:a0ka}
&& a_{0,k+2}=-\frac1{(k+1)(k+2)}\frac1{\xi_0}\left[\left(C_k^{(n)}\xi_0+
C_{k-1}^{(n)}\right)\vert\theta_0(\xi_0)\vert^n\cos(n\pi\zeta)\right.
\nonumber \\
&& \phantom{a_{0,k+2}=-\frac1{(k+1)(k+2)}\frac1{\xi_0}\left[\right.}+\left.
(k+1)(k+2)a_{0,k+1}^{\phantom{|}}\right]~~;\qquad k>1~~;\qquad \\
\slabel{eq:a0kb}
&& a_{0,0}=\theta_0(\xi_0)~~;\qquad a_{0,1}=\theta_0^\prime(\xi_0)~~; \\
\slabel{eq:a0kc}
&& a_{0,2}=-\frac1{1\cdot2}\frac1{\xi_0}\left\{C_0^{(n)}\xi_0\left[\vert
\theta_0(\xi_0)\vert^n\cos(n\pi\zeta)-\upsilon\right]+1\cdot2~a_{0,1}\right\}
~~; \\
\slabel{eq:a0kd}
&& a_{0,3}=-\frac1{2\cdot3}\frac1{\xi_0}\left[\left(C_1^{(n)}\xi_0+C_0^{(n)}
\right)\vert\theta_0(\xi_0)\vert^n\cos(n\pi\zeta)-\upsilon+2\cdot3~a_{0,2}
\right]~~;
\label{seq:a0k}
\end{leftsubeqnarray}
which, together with Eqs.\,(\ref{eq:serpn}) and (\ref{eq:cerpn}), make the
series expansion, Eq.\,(\ref{eq:sersn}), $2\ell=0$, be a solution of
Eq.\,(\ref{eq:t0n})
for nonsingular starting points, $\xi_0$, provided values of $\theta_0(\xi_0)$
and $\theta_0^\prime(\xi_0)$ are known.   For further details, an interested
reader is addressed to the parent paper [6].

The factor, $6/\xi^2$, appearing on the left-hand side of Eq.\,(\ref{eq:t2n}),
implies use of the Taylor series expansion:
\begin{equation}
\label{eq:cn2s}
\frac1{\xi^2}=\frac1{\xi_0^2}\sum_{k=0}^{+\infty}(-1)^k(k+1)\left(\frac{\xi-
\xi_0}{\xi_0}\right)^k~~;\qquad\vert\xi-\xi_0\vert<\xi_0~~;
\end{equation}
where the starting point, $\xi_0$, has to be replaced by $\xi_0+\Delta\xi<2\xi
_0$ whenever $\xi\ge2\xi_0$, to ensure convergence.

The substitution of the series expansions for $\theta_2^\pprime$, $\theta_2^
\prime$, $\theta_2$, $\vert\theta_0\vert^{n-1}$, $1/\xi$, $1/\xi^2$, via
Eqs.\,(\ref{eq:sersn})-(\ref{eq:cn2s}), into
(\ref{eq:t2n}), keeping in mind the coefficients of $(\xi-\xi_0)^k$ on both
sides must necessarily be equal, yields after a lot of algebra [6]:
\begin{leftsubeqnarray}
\slabel{eq:a2ka}
&& a_{2,k+2}=-\frac{a_{2,k+1}}{\xi_0}+\frac1{(k+1)(k+2)}\left\{\frac6
{\xi_0^2}\sum_{i=0}^k\frac{(-1)^i}{\xi_0^i}a_{2,k-i}-n\vert\theta_0(\xi_0)
\vert^{n-1}\right. \nonumber \\
&& \phantom{}\left.\times
\cos[(n-1)\pi\zeta]
\left[\sum_{i=0}^kC_i^{(n-1)}a_{2,k-i}+\frac1{\xi_0}\sum_{i=0}^{k-1}C_i^
{(n-1)}a_{2,k-i-1}\right]\right\}~;~~k>1~;\qquad \\
\slabel{eq:a2kb}
&& a_{2,0}=\theta_2(\xi_0)~~;\qquad a_{2,1}=\theta_2^\prime(\xi_0)~~; \\
\slabel{eq:a2kc}
&& a_{2,2}=-\frac{a_{2,1}}{\xi_0}+\frac1{1\cdot2}\left\{\frac6{\xi_0^2}
a_{2,0}-n\vert\theta_0(\xi_0)\vert^{n-1}\cos[(n-1)\pi\zeta]a_{2,0}\right\}~~; \\
\slabel{eq:a2kd}
&& a_{2,3}=-\frac{a_{2,2}}{\xi_0}+\frac1{2\cdot3}\left\{\frac6{\xi_0^2}
\left[a_{2,1}-\frac{a_{2,0}}{\xi_0}\right]-n\vert\theta_0(\xi_0)\vert^{n-1}
\cos[(n-1)\pi\zeta]\right. \nonumber \\
&& \phantom{a_{2,3}=-\frac{a_{2,2}}{\xi_0}+}\times\left.
\left[a_{2,1}+\left(C_1^{(n-1)}+\frac{C_0^{(n-1)}}{\xi_0}\right)a_{2,0}
\right]\right\}~~;
\label{seq:a2k}
\end{leftsubeqnarray}
which, together with Eqs.\,(\ref{eq:serpn}) and (\ref{eq:cerpn}), make the
series expansion, Eq.\,(\ref{eq:sersn}), $2\ell=2$, be a solution of
Eq.\,(\ref{eq:t2n})
for nonsingular starting points, $\xi_0$, provided values of
$\theta_2(\xi_0)$, $\theta_2^\prime(\xi_0)$, $\theta_0(\xi_0)$
and $\theta_0^\prime(\xi_0)$ are known.   For further details, an interested
reader is addressed to the parent paper [6].

The above procedure can be extended to the EC2 associated equations,
Eqs.\,(\ref{eq:t2ln}), $2\ell>2$, to express related solutions as series
expansions
starting from nonsingular points.   Nevertheless, the EC2 associated functions
of order, $2\ell>2$, are of little practical interest in that $A_{2\ell}=0$
[10], [4] and for this reason they shall not be considered in the following.

In terms of the variable, $\xi_w$, the EC2 associated functions in the 
noncommon region, $\theta_{2\ell,w}(\xi_w)$, via Eqs.\,(\ref{eq:dimq}),
(\ref{eq:snnr}), (\ref{eq:sersn}), can be expanded in Taylor series as:
\begin{lefteqnarray}
\label{eq:sncn1}
&& \theta_{2\ell,w}(\xi_w)=\theta_{2\ell,w}(\xi_{0,w})+\sum_{k=1}^{+\infty}
a_{2\ell,k}^{\rm(ncm)}(\xi_{0,w})(\xi_w-\xi_{0,w})^k~;~~\xi_w^\ast\le
\xi_w\le\Xi_w~~;\qquad \\
\label{eq:sncn2}
&& a_{2\ell,k}^{\rm(ncm)}(\xi_{0,w})=\Lambda_w^{k/2}a_{2\ell,k}(\xi_{0,w1})~~;
\end{lefteqnarray}
where $n=n_w$, $\upsilon=\upsilon_{w1}=\upsilon_w/\Lambda_w$,
$\xi=\xi_{w1}=\Lambda_w^{1/2}\xi_w$, when using
Eqs.\,(\ref{seq:a0k}) and (\ref{seq:a2k}).   The first starting point relates
to the interface, $\xi_{0,w}=\xi_w^\ast$, which can be determined from the
knowledge of the EC2 associated functions within the common region.

\subsubsection{The common region}
\label{crg}

The presence of both subsystems within the common region allows the
notation be shortened for sake of simplicity, throughout the current
Subsection, as:
\begin{equation}
\label{eq:sncr}
A_{2\ell,w}=A_{2\ell,w}^{({\rm com})}~~;\quad
\theta_{2\ell,w}=\theta_{2\ell,w}^{({\rm com})}~~;
\end{equation}
where the apex, (com), denotes the common region.   Accordingly, the solution
of Eq.\,(\ref{eq:EC2}) can be expanded in Legendre polynomials as:
\begin{lefteqnarray}
\label{eq:thc}
&& \theta_w(\xi_w,\mu)=\sum_{\ell=0}^{+\infty}A_{2\ell,w}\theta_{2\ell,w}
(\xi_w)P_{2\ell}(\mu)~~; \\
\label{eq:A2lc}
&& 
\lim_{\upsilon_w\to0}A_{2\ell,w}=\delta_{2\ell,0}~~;
\end{lefteqnarray}
where $\delta$ is the Kronecker symbol; $A_{2\ell,w}$ are coefficients which
depend on the rotation
parameters; $\theta_{2\ell,w}$ are the EC2 associated function of degree,
$2\ell$; and the Legendre polynomials, defined by Eq.\,(\ref{eq:Lp}), obey
the Legendre equation, Eq.\,(\ref{eq:Le}).

In absence of rotation,
$\upsilon_u=\upsilon_v=0$, the system attains the related undistorted
spherical configuration and the EC2 associated function, $\theta_{0,w}
(\xi_w)$, via Eq.\,(\ref{eq:EC2b})
coincides with the EC2 function, $\theta_w(\xi_w,\mu)$, hence $A_{0,w}=1$
according to Eq.\,(\ref{eq:A2lc}).
%
%
Conversely, the coefficients, $A_{2\ell,w}$, $2\ell>0$, depend on
the rotation parameters and vanish for spherical configurations, according to
Eq.\,(\ref{eq:A2lc}).

If the distorsion due to rigid rotation may be considered as a small
perturbation on the spherical shape, then the first term of the series
expansion on the right-hand side of Eq.\,(\ref{eq:thc}) is dominant, which
implies the following inequality:
\begin{lefteqnarray}
\label{eq:S2c}
&& \vert R_{1,w}(\xi_w,\mu)\vert\ll\vert\theta_{0,w}(\xi_w)\vert~~; \\
\label{eq:R2c}
&& R_{1,w}(\xi_w,\mu)=\sum_{\ell=1}^{+\infty}A_{2\ell,w}\theta_{2\ell,w}
(\xi_w)P_{2\ell}(\mu)~~;
\end{lefteqnarray}
and the powers on the right-hand side of Eq.\,(\ref{eq:EC2}) can safely be
approximated as:
\begin{lefteqnarray}
\label{eq:thc2l}
&& \theta_w^{n_w}(\xi_w,\mu)=\theta_{0,w}^{n_w}(\xi_w)+n_w\theta_{0,w}^{n_w-1}
(\xi_w)R_{1,w}(\xi_w,\mu)~~;
\end{lefteqnarray}
which is a series expansion in Legendre polynomials.   With regard to the
special cases, $n_w=0,1$, Eq.\,(\ref{eq:thc2l}) is exact.

The series on the right-hand side of Eq.\,(\ref{eq:thc}) describes the
expansion of the nonrotating common region as a whole, via $\theta_{0,w}$, and
superimposed on this an oblateness, via $R_{1,w}$.
More specifically, $\theta_{0,w}$
relates to an expanded sphere where the radial contribution of rigid
rotation adds to the undistorted configuration, and $R_{1,w}$ quantifies the
meridional distortion.

Let $\xi_{{\rm ex},w}$ define the (fictitious) spherical isopycnic surface of
the expanded sphere, as:
\begin{lefteqnarray}
\label{eq:is0w}
&& \theta_{0,w}(\xi_{{\rm ex},w})=\theta(\xi_{{\rm un},w})=\kappa_w~~;
\end{lefteqnarray}
where the index, ex, un, denotes the expanded and the undistorted sphere,
respectively, and radial expansion implies
$\xi_{{\rm un},w}\le\xi_{{\rm ex},w}$.

The substitution of Eqs.\,(\ref{eq:R2c}) and (\ref{eq:is0w}) into
(\ref{eq:thc}) yields:
\begin{lefteqnarray}
\label{eq:Rzw}
&& R_{1,w}(\xi_{{\rm ex},w},\mp\mu_{{\rm ex},w})=0~~;
\end{lefteqnarray}
and the locus, $(\xi_{{\rm ex},w},\mp\mu_{{\rm ex},w})$, defines the
intersection
between the oblate isopycnic surface, $\theta_w(\xi_w,\mu)=\kappa_w$, and the
(fictitious) spherical isopycnic surface,
$\theta_{0,w}(\xi_{{\rm ex},w})=\kappa_w$.
In the nonrotating limit, $\upsilon_w\to0$,
$\xi_{{\rm ex},w}\to\xi_{{\rm un},w}$, and the term
containing $P_2(\mu)$ is expected to be dominant with respect to the others in
Eq.\,(\ref{eq:R2c}).   Accordingly, $\mu_{{\rm ex},w}$ relates to
$P_2(\mu)=0$, hence $\mu_{{\rm ex},w}\to1/\sqrt3$.   For further details, an
interested reader is addressed to the parent paper [9].

The substitution of Eqs.\,(\ref{eq:thc}) and (\ref{eq:thc2l}) into
(\ref{eq:EC2}), after equating
separately the terms of the same degree in Legendre polynomials, shows
Eq.\,(\ref{eq:EC2}) is equivalent to the set of EC2 associated equations
e.g., [6]:
\begin{lefteqnarray}
\label{eq:At0c}
&& \frac1{\xi_w^2}\frac\diff{\diff\xi_w}\left[\xi_w^2\frac{\diff(A_{0,w}
\theta_{0,w})}{\diff\xi_w}\right]-\upsilon_w=-\sum\left[\Lambda_w\vert A_{0,w}
\theta_{0,w}\vert^{n_w}\cos(n_w\pi\zeta_w)\right]~~;\qquad \\
\label{eq:At2c}
&& \frac1{\xi_w^2}\frac\diff{\diff\xi_w}\left[\xi_w^2\frac{\diff(A_{2,w}
\theta_{2,w})}{\diff\xi_w}\right]-\frac6{\xi_w^2}A_{2,w}\theta_{2,w}
\nonumber \\
&& =-\sum\left\{\Lambda_wn_w\vert A_{0,w}\theta_{0,w}\vert^{n_w-1}\cos
[(n_w-1)\pi\zeta_w]A_{2,w}\theta_{2,w}\right\}~; \\
\label{eq:At2lc}
&& \frac1{\xi_w^2}\frac\diff{\diff\xi_w}\left[\xi_w^2\frac{\diff(A_{2\ell,w}
\theta_{2\ell,w})}{\diff\xi_w}\right]-\frac{2\ell(2\ell+1)}{\xi_w^2}
A_{2\ell,w}\theta_{2\ell,w} \nonumber \\
&& =-\sum\left\{\Lambda_wn_w\vert A_{0,w}\theta_{0,w}\vert^{n_w-1}
\cos[(n_w-1)\pi\zeta_w]A_{2\ell,w}\theta_{2\ell,w}\right\}~; \\
\label{eq:t02ci}
&& \theta_{2\ell,w}(0)=\delta_{2\ell,0}~~;\qquad\theta_{2\ell,w}^\prime(0)=0
~~; \\
\label{eq:zic}
&& \zeta_w=\cases{
0~~; & $\theta_{0,w}\ge0~~;$ \cr
1~~; & $\theta_{0,w}<0~~;$   \cr
}
\end{lefteqnarray}
where the occurrence of absolute values and cosines on the
right-hand side of Eqs.\,(\ref{eq:At0c})-(\ref{eq:At2lc}), makes
$\theta_{0,w}^{n_w}$ and $\theta_{0,w}^{n_w-1}$, $\theta_{0,w}<0$, be
evaluated as the real part of the principal value of complex powers; the
boundary conditions relate to the centre.

The right-hand side of Eqs.\,(\ref{eq:At0c})-(\ref{eq:At2lc}) is independent
of the subsystem under consideration, which implies
$A_{2\ell,u}=A_{2\ell,v}=A_{2\ell}$, as shown in Appendix \ref{a:seco}.
Accordingly, Eqs.\,(\ref{eq:At0c})-(\ref{eq:At2lc}) reduce to:
\begin{lefteqnarray}
\label{eq:t0c}
&& \frac1{\xi_w^2}\frac\diff{\diff\xi_w}\left(\xi_w^2\frac{\diff\theta_{0,w}}
{\diff\xi_w}\right)-\upsilon_w=-\sum\left[\Lambda_w\vert\theta_{0,w}\vert^
{n_w}\cos(n_w\pi\zeta_w)\right]~~; \\
\label{eq:t2c}
&& \frac1{\xi_w^2}\frac\diff{\diff\xi_w}\left(\xi_w^2\frac{\diff\theta_{2,w}}
{\diff\xi_w}\right)-\frac6{\xi_w^2}\theta_{2,w} \nonumber \\
&& =-\sum\left\{\Lambda_wn_w\vert
\theta_{0,w}\vert^{n_w-1}\cos[(n_w-1)\pi\zeta_w]\theta_{2,w}\right\}~; \\
\label{eq:t2lc}
&& \frac1{\xi_w^2}\frac\diff{\diff\xi_w}\left(\xi_w^2\frac{\diff\theta_
{2\ell,w}}{\diff\xi_w}\right)-\frac{2\ell(2\ell+1)}{\xi_w^2}\theta_{2\ell,w}
\nonumber \\
&& =-\sum\left\{\Lambda_wn_w\vert\theta_{0,w}\vert^{n_w-1}
\cos[(n_w-1)\pi\zeta_w]\theta_{2\ell,w}\right\}~;
\end{lefteqnarray}
where the explicit dependence on $A_{2\ell,w}$ has been erased.

Related solutions of the EC2 associated equations,
Eqs.\,(\ref{eq:t0c})-(\ref{eq:t2lc}), may be expanded in Taylor series as:
\begin{lefteqnarray}
\label{eq:sersc}
&& \theta_{2\ell,w}(\xi_w)=\sum_{k=0}^{+\infty}a_{2\ell,k}^{(w,w)}(\xi_{0,w})
(\xi_w-\xi_{0,w})^k~~; \\
\label{eq:cersc}
&& a_{2\ell,0}^{(w,w)}(\xi_{0,w})=\theta_{2\ell,w}(\xi_{0,w});~~
a_{2\ell,k}^{(w,w)}(\xi_{0,w})=\frac1{k!}\left(\frac{\diff^k\theta_{2\ell,w}}
{\diff\xi_w^k}\right)_{\xi_{0,w}};
\end{lefteqnarray}
where, in particular, $a_{2\ell,1}^{(w,w)}(\xi_{0,w})=\theta_{2\ell,w}^\prime
(\xi_{0,w})$ and $a_{2\ell,2}^{(w,w)}(\xi_{0,w})=\theta_{2\ell,w}^\pprime
(\xi_{0,w})/2$.  It can be seen the convergence radius tends to be null if the
starting point, $\xi_{0,w}$, tends to a singular point outside the origin,
$\xi_{0,w}^\dagger$, $\theta_{0,w}(\xi_{0,w}^\dagger)=0$, which implies
$\theta_{0,w}(\xi_w)$ and $\theta_{0,w}(\xi_{0,w})$ are both positive or
negative.   The first and the second derivative of the EC2 associated
functions, $\theta_{2\ell,w}^\prime(\xi_w)$ and
$\theta_{2\ell,w}^\pprime(\xi_w)$, can readily be
expanded in Taylor series via Eq.\,(\ref{eq:sersc}).   For further details, an
interested reader is addressed to the parent paper [6].

The solutions of the EC2 associated equations related to the other subsystem,
let it be $v$, $w=u$, along the direction considered, may be expanded in
Taylor series as shown in Appendix \ref{a:seco}.

The powers of $\vert\theta_{0,w}\vert$ appearing on the right-hand side of
Eqs.\,(\ref{eq:t0c})-(\ref{eq:t2lc}) can be expanded in series as e.g., [18]:
\begin{lefteqnarray}
\label{eq:serpc}
&& \vert\theta_{0,w}(\xi_w)\vert^x=\vert\theta_{0,w}(\xi_{0,w})\vert^x
\sum_{k=0}^{+\infty}C_{k,w}^{(x)}(\xi_w-\xi_{0,w})^k~~; \\
\label{eq:cerpc}
&& C_{k,w}^{(x)}=\frac1{\theta_{0,w}(\xi_{0,w})}\frac1k\sum_{i=1}^k(ix-k+i)
a_{0,i}^{(w,u)}C_{k-i,w}^{(x)}~~;\qquad C_{0,w}^{(x)}=1~~;
\end{lefteqnarray}
where $w=u,v$ and $a_{0,i}^{(w,u)}$ are defined by Eqs.\,(\ref{eq:cersc}) and
(\ref{eq:cersv}) written in Appendix \ref{a:seco}.

On the other hand, the combination of Eqs.\,(\ref{eq:sersc}) and
(\ref{eq:cersc}) yields:
\begin{equation}
\label{eq:serc0}
\vert\theta_{0,w}(\xi_u)\vert^x=\vert\theta_{0,w}(\xi_{0,u})\vert^x\left[1+
\frac1{\theta_{0,w}(\xi_{0,u})}\sum_{k=1}^{+\infty}a_{0,k}^{(w,u)}(\xi_u-
\xi_{0,u})^k\right]^x~~;
\end{equation}
provided $\xi_{0,u}$, $\xi_u$, both precede or exceed a singular
point, $\xi_{0,u}^\dagger$, $\theta_{0,u}(\xi_{0,u}^\dagger)=0$, i.e.
$\xi_{0,u}<\xi_u<\xi_{0,u}^\dagger$ or $\xi_{0,u}^\dagger<\xi_{0,u}<\xi_u$.
More specifically,
$\theta_{0,w}(\xi_u)/\theta_{0,w}(\xi_{0,u})>0$ in the case under
discussion, which implies the quantity within square brackets on the
right-hand side of Eq.\,(\ref{eq:serc0}) is always positive.

The following identity:
\begin{equation}
\label{eq:LEc}
\frac1{\xi_w^2}\frac{\diff}{\diff\xi_w}\left(\xi_w^2\frac{\diff
\theta_{2\ell,w}}{\diff\xi}\right)=\theta_{2\ell,w}^\pprime+\frac2\xi_w
\theta_{2\ell,w}^\prime~~;
\end{equation}
implies use of the Taylor series expansion:
\begin{equation}
\label{eq:cc1s}
\frac1{\xi_w}=\frac1{\xi_{0,w}}\sum_{k=0}^{+\infty}(-1)^k\left(\frac{\xi_w-
\xi_{0,w}}{\xi_{0,w}}\right)^k~~;\qquad\vert\xi_w-\xi_{0,w}\vert<\xi_{0,w}~~;
\end{equation}
where the starting point, $\xi_{0,w}$, has to be replaced by $\xi_{0,w}+\Delta
\xi_w<2\xi_{0,w}$ whenever $\xi_w\ge2\xi_{0,w}$, to ensure convergence.

The substitution of the series expansions for $\theta_{0,w}^\pprime$,
$\theta_{0,w}^\prime$, $\vert\theta_{0,w}\vert^{n_w}$, $1/\xi_w$, via
Eqs.\,(\ref{eq:sersc})-(\ref{eq:cc1s}), into
(\ref{eq:t0c}), keeping in mind the coefficients of $(\xi_u-\xi_{0,u})^k$ on
both sides must necessarily be equal, yields after a lot of algebra [6]:
\begin{leftsubeqnarray}
\slabel{eq:a0kca}
&& a_{0,k+2}^{(u,u)}=-\frac1{(k+1)(k+2)}\frac1{\xi_{0,u}}\left\{\sum\left\{
\left[C_{k,w}^{(n_w)}\xi_{0,u}+C_{k-1,w}^{(n_w)}\right]\Lambda_w\vert
\theta_{0,w}(\xi_{0,u})\vert^{n_w}\right.\right. \nonumber \\
&& \phantom{a_{k+2,0}}\left.\left.
\phantom{\sum}\times
\cos(n_w\pi\zeta_w)\right\}+(k+1)(k+2)a_{0,k+1}^{(u,u)}\right\}~~;\qquad k>1
~~;\qquad \\
\slabel{eq:a0kcb}
&& a_{0,0}^{(u,u)}=\theta_{0,u}(\xi_{0,u})~~;\qquad a_{0,1}^{(u,u)}=
\theta_{0,u}^\prime(\xi_{0,u})~~; \\
\slabel{eq:a0kcc}
&& a_{0,2}^{(u,u)}=-\frac1{1\cdot2}\frac1{\xi_{0,u}}\left\{\sum\left[C_{0,w}^
{(n_w)}\xi_{0,u}\Lambda_w\vert\theta_{0,w}(\xi_{0,u})\vert^{n_w}\cos(n_w\pi
\zeta_w)\right]\right. \nonumber \\
&& \phantom{a_{2,0}^{(u,u)}=-\frac1{1\cdot2}\frac1{\xi_{u,0}}\left\{\right.}
\left.-C_{0,u}^{(n_u)}\xi_{0,u}\upsilon_u+1\cdot2~a_{0,1}^{(u,u)}\right\};
\qquad \\
\slabel{eq:a0kcd}
&& a_{0,3}^{(u,u)}=-\frac1{2\cdot3}\frac1{\xi_{0,u}}\left\{\sum\left\{\left[
C_{1,w}^{(n_w)}\xi_{0,u}+C_{0,w}^{(n_w)}\right]\Lambda_w\vert\theta_{0,w}
(\xi_{0,u})\vert^{n_w}\cos(n_w\pi\zeta_w)\right\}\right. \nonumber \\
&& \phantom{a_{3,0}^{(u,u)}=-\frac1{2\cdot3}\frac1{\xi_{0,u}}\left\{\right.}
-\left.
\upsilon_u+2\cdot3~a_{0,2}^{(u,u)}\right\};
\label{seq:a0kc}
\end{leftsubeqnarray}
which, together with Eqs.\,(\ref{eq:serpc}) and (\ref{eq:cerpc}), make the
series expansion, Eq.\,(\ref{eq:sersc}), $2\ell=0$, be a solution of
Eq.\,(\ref{eq:t0c})
for nonsingular starting points, $\xi_{0,u}$, provided values of $\theta_{0,u}
(\xi_{0,u})$ and $\theta_{0,u}^\prime(\xi_{u,0})$ are known.   For further
details, an interested reader is addressed to the parent paper [6].    The
coefficients, $a_{0,k+2}^{(v,u)}$, needed for the calculation of
$\theta_{0,v}(\xi_u)$ via Eq.\,(\ref{eq:sersv}), are related to their
counterparts, $a_{0,k+2}^{(u,u)}$, via Eq.\,(\ref{eq:alkuv}), as shown in
Appendix \ref{a:seco}.

The series, defined by Eq.\,(\ref{eq:sersc}), where the coefficients are
expressed by Eq.\,(\ref{seq:a0kc}), represents a solution of the associated
EC2 equation of degree 0 for nonsingular starting points, $\xi_{0,u}$, with
regard to assigned input parameters and boundary conditions.   For the central
singular starting point, $\xi_{0,u}=0$, Eq.\,(\ref{eq:sersc}) reduces to the
MacLaurin series expansion:
\begin{lefteqnarray}
\label{eq:sers0}
&& \theta_{2\ell,w}(\xi_w)=\sum_{k=0}^{+\infty}a_{2\ell,k}^{(w,w)}(0)\xi_w^k
~~; \\
\label{eq:cers0}
&& a_{2\ell,0}^{(w,w)}(0)=\theta_{2\ell,w}(0)=\delta_{2\ell,0};~~
a_{2\ell,k}^{(w,w)}(0)=\frac1{k!}\left(\frac{\diff^k\theta_{2\ell,w}}
{\diff\xi_w^k}\right)_0;~~
\end{lefteqnarray}
which implies the series expansion of the ratio, $\theta_{2\ell,w}^\prime
(\xi_w)/\xi_w$, as:
\begin{lefteqnarray}
\label{eq:ders0}
&& \frac{\theta_{2\ell,w}^\prime(\xi_w)}{\xi_w}=\sum_{k=0}^{+\infty}ka_
{2\ell,k}^{(w,w)}(0)\xi_w^{k-2}~~;
\end{lefteqnarray}
where $a_{2\ell,1}^{(w,w)}(0)=0$ according to Eqs.\,(\ref{eq:t02ci}) and
(\ref{eq:ders0}).   In addition, Eqs.\,(\ref{eq:serpc}) and (\ref{eq:cerpc})
reduce to:
\begin{lefteqnarray}
\label{eq:serp0}
&& \vert\theta_{0,w}(\xi_w)\vert^x=1+\sum_{k=1}^{+\infty}C_{k,w}^{(x)}\xi_w^k
~~; \\
\label{eq:cerp0}
&& C_{k,w}^{(x)}=\frac1k\sum_{i=1}^k(ix-k+i)a_{0,i}^{(w,u)}C_{k-i,w}^{(x)}~~;
\qquad C_{0,w}^{(x)}=1~~;
\end{lefteqnarray}
where $w=u,v$ and $a_{0,i}^{(w,u)}$ are defined by Eqs.\,(\ref{eq:cers0}) and
(\ref{eq:cersv}) in Appendix \ref{a:seco}, where $\xi_{0,u}=0$.

In the case under discussion, $2\ell=0$, following the procedure used for
nonsingular starting points yields:
\begin{leftsubeqnarray}
\slabel{eq:a0kc0a}
&& a_{0,2k+2}^{(u,u)}=-\frac1{2k(2k+2)(2k+3)} \nonumber \\
&& \phantom{a} \times
\sum\left\{\Lambda_w\sum_{i=1}^k
\frac{(2in_w-2k+2i)(2k-2i+2)(2k-2i+3)}{\Gamma_{uw}(1-\delta_{ik}\upsilon_w)}
a_{0,2i}^{(w,u)}a_{0,2k-2i+2}^{(w,u)}\right\}; \nonumber \\
&& \phantom{a~}
k>0~~; \\  
\slabel{eq:a0kc0b}
&& a_{0,0}^{(u,u)}=1~~;\qquad a_{0,1}^{(u,u)}=0~~;\qquad a_{0,2}^{(u,u)}=
-\frac{1-\upsilon_u}6~~; \\
\slabel{eq:a0kc0d}
&& a_{0,2k+1}^{(u,u)}=0~~;\qquad C_{2k+1,u}^{(n_u)}=0~~;\qquad k>0~~;\qquad
C_{1,u}^{(n_u)}=a_{0,1}^{(u,u)}=0~~;
\label{seq:a0kc0}
\end{leftsubeqnarray}
where the coefficients, $a_{0,2i}^{(v,u)}$, needed for the calculation of
$a_{0,2k+2}^{(u,u)}$, are related to their counterparts, $a_{0,2i}^{(u,u)}$,
via Eq.\,(\ref{eq:lkuv0}) as shown in Appendix \ref{a:seco}.

The series, defined by Eq.\,(\ref{eq:sers0}), where the coefficients are
expressed by Eq.\,(\ref{seq:a0kc0}), represents a solution of the associated
EC2 equation of degree, 0, for the singular starting point, $\xi_{0,u}=0$, and
the boundary conditions, $\theta_{0,u}(0)=1$, $\theta_{0,u}^\prime(0)=0$.

The factor, $6/\xi_w^2$, appearing on the left-hand side of
Eq.\,(\ref{eq:t2c}), implies use of the Taylor series expansion:
\begin{equation}
\label{eq:cc2s}
\frac1{\xi_w^2}=\frac1{\xi_{0,w}^2}\sum_{k=0}^{+\infty}(-1)^k(k+1)\left(\frac
{\xi_w-\xi_{0,w}}{\xi_{0,w}}\right)^k~~;\qquad\vert\xi_w-\xi_{0,w}\vert<
\xi_{0,w}~~;
\end{equation}
where the starting point, $\xi_{0,w}$, has to be replaced by $\xi_{0,w}+
\Delta\xi_w<2\xi_{0,w}$ whenever $\xi_w\ge2\xi_{0,w}$, to ensure convergence.

The substitution of the series expansions for $\theta_{2,w}^\pprime$,
$\theta_{2,w}^\prime$, $\theta_{2,w}$, $\vert\theta_{0,w}\vert^{n_w-1}$,
$1/\xi_w$, $1/\xi_w^2$, by use of
Eqs.\,(\ref{eq:sersc})-(\ref{eq:cc1s}), 
(\ref{eq:cc2s}), into
(\ref{eq:t2c}), keeping in mind the coefficients of $(\xi_w-\xi_{0,w})^k$ on
both sides must necessarily be equal, yields after a lot of algebra [6]:
\begin{leftsubeqnarray}
\slabel{eq:a2kca}
&& a_{2,k+2}^{(u,u)}=-\frac{a_{2,k+1}^{(u,u)}}{\xi_{0,u}}+\frac1{(k+1)(k+2)}
\left\{\frac6{\xi_{0,u}^2}\sum_{i=0}^k\frac{(-1)^i}{\xi_{0,u}^i}a_{2,k-i}^
{(u,u)}\right. \nonumber \\
&& \phantom{a_{2,k+2}^{(u,u)}=}\left.-
\sum\left\{\Lambda_wn_w
\vert\theta_{0,w}(\xi_{0,w})\vert^{n_w-1}\cos[(n_w-1)\pi\zeta_w]
\phantom{\frac||}\right.\right. \nonumber \\
&& \phantom{a_{2,k+2}^{(u,u)}=}\left.\left.\times
\left[\sum_{i=0}^kC_{i,w}^{(n_w-1)}a_{2,k-i}^{(w,u)}+\frac1{\xi_{0,u}}
\sum_{i=0}^{k-1}C_{i,w}^{(n_w-1)}a_{2,k-i-1}^{(w,u)}\right]\right\}\right\}~;
~~k>1~;~\qquad \\
\slabel{eq:a2kcb}
&& a_{2,0}^{(u,u)}=\theta_{2,u}(\xi_{0,u})~~;\qquad a_{2,1}^{(u,u)}=\theta_
{2,u}^\prime(\xi_{0,u})~~; \\
\slabel{eq:a2kcc}
&& a_{2,2}^{(u,u)}=-\frac{a_{2,1}^{(u,u)}}{\xi_{0,u}}+\frac1{1\cdot2}\left\{
\frac6{\xi_{0,u}^2}a_{2,0}^{(u,u)}-\sum\left\{\Lambda_wn_w\vert\theta_{0,w}
(\xi_{0,w})\vert^{n_w-1}\phantom{\frac||}\right.\right. \nonumber \\
&& \phantom{a_{2,2}^{(u,u)}=-\frac{a_{2,1}^{(u,u)}}{\Xi_{0,u}}+\frac1{1\cdot2}
\left\{\right.}\left.\left.\times
\cos[(n_w-1)\pi\zeta_w]C_{0,w}^{(n_w-1)}a_{2,0}^{(w,u)}\phantom{\frac||}\right
\}\right\}~~; \\
\slabel{eq:a2kcd}
&& a_{2,3}^{(u,u)}=-\frac{a_{2,2}^{(u,u)}}{\xi_{0,u}}+\frac1{2\cdot3}\left\{
\frac6{\xi_{0,u}^2}\left[a_{2,1}^{(u,u)}-\frac{a_{2,0}^{(u,u)}}{\xi_{0,u}}
\right]-\sum\left\{\Lambda_wn_w\vert\theta_{0,w}(\xi_{0,w})\vert^{n_w-1}
\right.\right. \nonumber \\
&& \phantom{a_{2,3}=}\times\left.\left.
\cos[(n_w-1)\pi\zeta_w]
\left[a_{2,1}^{(w,u)}+\left(C_{1,w}^{(n_w-1)}+\frac{C_{0,w}^{(n_w-1)}}
{\xi_{0,w}}\right)a_{2,0}^{(w,u)}\right]\right\}\right\};~
\label{seq:a2kc}
\end{leftsubeqnarray}
which, together with Eqs.\,(\ref{eq:serpc}) and (\ref{eq:cerpc}), make the
series expansion, Eq.\,(\ref{eq:sersc}), $2\ell=2$, be a solution of
Eq.\,(\ref{eq:t2c})
for nonsingular starting points, $\xi_{0,u}$, provided values of
$\theta_{2,w}(\xi_{0,u})$, $\theta_{2,w}^\prime(\xi_{0,u})$, $\theta_{0,w}
(\xi_{0,w})$ and $\theta_{0,w}^\prime(\xi_{0,w})$ are known.   For further
details, an interested reader is addressed to the parent paper [6].   The
coefficients, $a_{2,k+2}^{(v,u)}$, needed for the calculation of
$\theta_{2,v}(\xi_{0,v})$ via Eq.\,(\ref{eq:sersv}), are related to their
counterparts, $a_{2,k+2}^{(u,u)}$, via Eq.\,(\ref{eq:alkuv}), as shown in
Appendix \ref{a:seco}.

For the central singular starting point, $\xi_{0,u}=0$, following the
procedure used for nonsingular starting points
yields:
\begin{leftsubeqnarray}
\slabel{eq:a2kc0a}
&& a_{2,k+2}^{(u,u)}=-\frac1{(2k+2)(2k+3)-6}\sum\left\{\Lambda_wn_w\sum_{i=0}^
kC_{i,w}^{(n_w-1)}a_{2,2k-2i}^{(w,u)}\right\};k>1;~~\qquad \\
\slabel{eq:a2kc0b}
&& a_{2,0}^{(u,u)}=0~~;\qquad a_{2,1}^{(u,u)}=0~~; \\
\slabel{eq:a2kc0c}
&& a_{2,2}^{(u,u)}(2\cdot3-6)=-\sum\left\{\Lambda_wn_wC_{0,w}^{(n_w-1)}a_{2,0}
^{(w,u)}\right\}~~; \\
&& a_{2,2k+1}^{(u,u)}=0~~;\qquad C_{2k+1,w}^{(n_w-1)}=0~~;
\label{seq:a2kc0}
\end{leftsubeqnarray}
where the coefficients, $a_{2,2k-2i}^{(v,u)}(\xi_{0,v})$, needed for the
calculation of $a_{2,2k+2}^{(u,u)}(\xi_{0,u})$, are related to their
counterparts, $a_{2,2k-2i}^{(u,u)}(\xi_{0,u})$, via Eq.\,(\ref{eq:lkuv0}) as
shown in Appendix \ref{a:seco}.

The coefficient, $a_{2,2}^{(u,u)}(0)$, is left undetermined according to
Eq.\,(\ref{eq:a2kc0c}).   On the other hand, the EC2 associated function,
$\theta_{2,w}(\xi_w)$, remains undefined by a constant factor, $A_2$,
according to Eq.\,(\ref{eq:t2c}).   Without loss of generality,
$\theta_{2,w}(\xi_w)$ may be fixed as:
\begin{equation}
\label{eq:a2kc0d}
a_{2,2}^{(u,u)}(0)=1~~;
\end{equation}
where $A_2$ has still to be determined.   The series, defined by
Eq.\,(\ref{eq:sers0}), $2\ell=2$, when the coefficients are expressed by
Eqs.\,(\ref{seq:a2kc0})-(\ref{eq:a2kc0d}), represents a solution of the EC2
associated equation
of degree, $2\ell=2$, for the singular starting point, $\xi_{0,u}=0$, and the
boundary conditions, $\theta_{2,w}(0)=0$, $\theta_{2,w}^\prime(0)=0$.

\subsection{Behaviour of EC2 associated functions near
singularities outside the origin}
\label{sing}

Owing to uniform convergence of the power series, defined by
Eqs.\,(\ref{eq:sncn1}), (\ref{eq:sersc}), differentiation or integration can
be performed term by term within the convergence radius,
$\vert\xi_w-\xi_{0,w}\vert<\Delta_{\rm C}\xi_w$.
Unfortunately, related coefficients are expressed in a rather cumbersome form
via the recursion formulae, Eqs.\,(\ref{seq:a0k}), (\ref{seq:a2k}),
(\ref{seq:a0kc}), (\ref{seq:a0kc0}), (\ref{seq:a2kc}), (\ref{seq:a2kc0}),
which implies no simple means for analysing series convergence, leaving aside
a few special cases.   It can be seen a lower limit to the convergence radius
satisfies $\Delta_{\rm C}\xi_w>1$ for the singular starting point,
$\xi_{0,w}=0$, and an upper limit satisfies
$\Delta_{\rm C}\xi_w<\min(\xi_{0,w},\vert\xi_{0,w}^\dagger-\xi_{0,w}\vert)$,
where $\xi_{0,w}^\dagger$ is the zero of $\theta_{0,w}$ which is nearest to
$\xi_{0,w}$.   For further details, an interested reader is addressed to the
parent papers [36], [29], [6].

The power series, expressed by Eqs.\,(\ref{eq:sncn1}), (\ref{eq:sersc}), do
not converge as $\xi_{0,w}\to\xi_{0,w}^\dagger$, [6] which implies the
convergence radius tends to be null as the starting point approaches a zero of
$\theta_{0,w}$, $\lim_{\xi_{0,w}\to\xi_{0,w}^\dagger}\Delta_{\rm C}\xi_w=0$.
Accordingly, the
starting point, $\xi_{0,w}$, and the ending point on the convergence radius,
$\xi_{0,w}+\Delta_{\rm C}\xi_w$, lie on the same side with respect to the
nearest singular point, $\xi_{0,w}^\dagger$.   More specifically,
$\vert\xi_w-\xi_{0,w}\vert<\Delta_{\rm C}\xi_w$, $\xi_w<\xi_{0,w}^\dagger$,
and an approximation is needed for evaluating the EC2 associated function on
points, $\xi_w>\xi_{0,w}^\dagger$, via series expansions expressed by
Eqs.\,(\ref{eq:sncn1}), (\ref{eq:sersc}).

With regard to the EC2 equation, Eq.\,(\ref{eq:EC2}), related to the common
region, it may safely be assumed the term containing $\vert\theta_v(\xi_v,\mu)
\vert^{n_v}$, where $v$ relates to the inner boundary along the direction
considered, is negligible in
comparison with the other ones in the neighbourhood of a singular point,
$\vert\xi_{0,v}^\dagger-\xi_{0,v}\vert<\epsilon_v$, where $\epsilon_v$ is an
assigned tolerance.   Accordingly, the EC2 associated equations,
Eqs.\,(\ref{eq:t0c})-(\ref{eq:t2lc}), $w=u$, reduce to
Eqs.\,(\ref{eq:t0n})-(\ref{eq:t2ln}), respectively, via Eqs.\,(\ref{eq:dimq})
and (\ref{eq:snnr}).
Then power-series solutions of the EC2 associated equations
are expressed by Eq.\,(\ref{eq:sersn}) and related coefficients, for
$2\ell=0,2$, by Eqs.\,(\ref{seq:a0k}), (\ref{seq:a2k}),  particularized to the
case under discussion, including the singular starting point, $\xi_{0,w}=0$.
Finally, the EC2 associated functions, $w=v$,
%
%
can be determined within the common region, via Eq.\,(\ref{eq:isq2l}), and
related coefficients appearing in series expansions via
Eqs.\,(\ref{eq:alkuv})-(\ref{eq:lkuv0}), as shown in Appendix \ref{a:seco}.

With regard to the EC2 equation, Eq.\,(\ref{eq:EC2}), related to the noncommon
region, it may safely be assumed the term containing $\vert\theta_u(\xi_u,\mu)
\vert^{n_u}$, where $u$ relates to the outer boundary along the direction
considered, is negligible in
comparison with the other ones in the neighbourhood of a singular point,
$\vert\xi_{0,u}^\dagger-\xi_{0,u}\vert<\epsilon_u$, where $\epsilon_u$ is an
assigned
tolerance.   Accordingly, the EC2 equation, Eq.\,(\ref{eq:EC2}), reduces to:
\begin{equation}
\label{eq:EC20}
\frac1{\xi_u^2}\frac\partial{\partial\xi_u}\left(\xi_u^2\frac{\partial
\theta_u}{\partial\xi_u}\right)+
\frac1{\xi_u^2}\frac\partial{\partial\mu}\left[(1-\mu^2)\frac{\partial
\theta_u}{\partial\mu}\right]-\upsilon_u=0~~;
\end{equation}
and it can be verified via Eq.\,(\ref{eq:Le}) a solution is:
\begin{equation}
\label{eq:EC20s}
\theta_u(\xi_u,\mu)=\sum_{\ell=0}^{+\infty}\left[D_{2\ell}\xi_u^{2\ell}+
\frac{\overline D_{2\ell}}{\xi_u^{2\ell+1}}\right]P_{2\ell}(\mu)+\frac16
\upsilon_u\xi_u^2\left[1-P_2(\mu)\right]~~;
\end{equation}
where $D_{2\ell}$, $\overline D_{2\ell}$, are constants to be determined.

With regard to the first nonzero singular point,
$\xi_{0,u}^\dagger=\Xi_{{\rm ex},u}$,
the comparison between Eqs.\,(\ref{eq:thn}) and (\ref{eq:EC20s}) at a
specified transition point, $\xi_u=\hat\xi_u<\Xi_{{\rm ex},u}$, equating the
terms of equal degree in Legendre polynomials, yields:
\begin{lefteqnarray}
\label{eq:g00}
&& A_0\theta_{0,u}(\hat\xi_u)=D_0+\frac{\overline{D}_0}{\hat\xi_u}+\frac16
\upsilon_u\hat\xi_u^2~~;\qquad A_0=1~~; \\
\label{eq:g20}
&& A_2\theta_{2,u}(\hat\xi_u)=D_2\hat\xi_u^2+\frac{\overline{D}_2}
{\hat\xi_u^3}-\frac16\upsilon_u\hat\xi_u^2~~; \\
\label{eq:g2l0}
&& A_{2\ell}\theta_{2\ell,u}(\hat\xi_u)=D_{2\ell}\hat\xi_u^{2\ell}+\frac
{\overline{D}_{2\ell}}{\hat\xi_u^{2\ell+1}}~~;\qquad2\ell>2~~;
\end{lefteqnarray}
where the constants, $D_0$, $\overline{D}_0$; $D_2$,
$\overline{D}_2$; $D_{2\ell}$, $\overline{D}_{2\ell}$; can be inferred from
the continuity of the gravitational potential and the radial component of the
gravitational force,
which implies the continuity of the EC2 associated functions and their first
derivatives on the transition point, $\hat\xi_u$.

To this respect, it suffices solving the system of 
Eqs.\,(\ref{eq:g00}), (\ref{eq:g20}), (\ref{eq:g2l0}), and related radial
first derivatives on both sides:
\begin{lefteqnarray}
\label{eq:d00}
&& A_0\theta_{0,u}^\prime(\hat\xi_u)=-\frac{\overline{D}_0}{\hat\xi_u^2}+
\frac13\upsilon_u\hat\xi_u~~;\qquad A_0=1~~; \\
\label{eq:d20}
&& A_2\theta_{2,u}^\prime(\hat\xi_u)=2D_2\hat\xi_u-\frac{3\overline{D}_2}
{\hat\xi_u^4}-\frac13\upsilon_u\hat\xi_u~~; \\
\label{eq:d2l0}
&& A_{2\ell}\theta_{2\ell,u}^\prime(\hat\xi_u)=2\ell D_{2\ell}\hat\xi_u^
{2\ell-1}-\frac{(2\ell+1)\overline{D}_{2\ell}}{\hat\xi_u^{2\ell+2}}~~;\qquad
2\ell>2~~;
\end{lefteqnarray}
respectively.   After some algebra, the result is:
\begin{lefteqnarray}
\label{eq:D2l0}
&& D_{2\ell}=\frac1{4\ell+1}\frac1{\hat\xi_u^{2\ell}} \nonumber \\
&& \phantom{D_{2\ell}}\times
\left\{A_{2\ell}[(2\ell+1)\theta_{2\ell,u}(\hat\xi_u)+\hat\xi_u
\theta_{2\ell,u}^\prime(\hat\xi_u)]-(2\ell+3)
\frac{\delta_{2\ell,0}-\delta_{2\ell,2}}6\upsilon_u\hat\xi_u^2\right\};
\qquad \\
\label{eq:bD2l0}
&& \overline{D}_{2\ell}=\frac{\hat\xi_u^{2\ell+1}}{4\ell+1} \nonumber \\
&& \phantom{\overline{D}_{2\ell}}\times
\left\{A_{2\ell}[2\ell\theta_{2\ell,u}(\hat\xi_u)-\hat\xi_u\theta_{2\ell,u}^
\prime(\hat\xi_u)]-(2\ell-2)\frac{\delta_{2\ell,0}-\delta_{2\ell,2}}6
\upsilon_u\hat\xi_u^2\right\};
\end{lefteqnarray}
where the transition point may be related to the first nonzero singular point,
as $\hat\xi_u=\zeta_u\Xi_{{\rm ex},u}$, $0<1-\zeta_u\ll1$.

It is worth noticing the counterpart of Eq.\,(\ref{eq:g20}) in the parent
paper [6], Eq.\,(24) therein, lacks the last
quadratic term for the following reason.   The EC2 associated function,
$\theta_{2,u}$, has been inferred from a solution of the EC2 equation in the
current paper and from a solution of the EC2 associated equation of degree,
$2\ell=2$, in the parent paper, where the former expression is exact and the
latter is approximate, in the case under discussion.

On the basis of the above results, the behaviour of the EC2 associated
functions, $\theta_{0,w}$, $\theta_{2,w}$, in the neighbourhood of first
nonzero singular point, $\xi_{0,w}^\dagger=\Xi_{{\rm ex},w}$, can be
ascertained via Eqs.\,(\ref{eq:sersn}),
(\ref{seq:a0k}), (\ref{seq:a2k}), for the inner boundary, $w=v$ along the
direction considered, and Eqs.\,(\ref{eq:g00})-(\ref{eq:bD2l0}) for the outer
boundary, $w=u$ along the direction considered.

Similar results can be inferred for other singular points via
Eqs.\,(\ref{eq:sersc}), (\ref{seq:a0kc}), (\ref{seq:a2kc}),
outside the boundary in connection with the analytical
continuation of $\theta_{2\ell,u}(\xi_u)$.

\subsection{Special cases}\label{speca}

For a restricted number of special cases, the solution of both EC2 and
associated EC2 equations may be expressed analytically.   To save space and
simplify the notation, the common and noncommon region shall be dealt with in
different Subsubsections.   Accordingly, it shall be intended that unlabelled 
symbols
in each Subsubsection are related to the region under discussion therein.

\subsubsection{The noncommon region}\label{ncm5}

With regard to the noncommon region, both the EC2 equation,
Eq.\,(\ref{eq:ECE}), and the EC2 associated equations,
Eqs.\,(\ref{eq:t0n})-(\ref{eq:t2ln}), formally coincide with their
counterparts related to EC1 polytropes.   On the other hand, the
boundary conditions are related to the interface instead of the origin, which
implies general solutions with doubled integration constants.

Concerning the special case, $(n_v,n_u)=(n_v,0)$, the EC2 equation,
Eq.\,(\ref{eq:EC2}), reduces to:
\begin{equation}
\label{eq:ECn0}
\frac1{\xi_u^2}\frac\partial{\partial\xi_u}\left(\xi_u^2\frac{\partial
\theta_u}{\partial\xi_u}\right)+
\frac1{\xi_u^2}\frac\partial{\partial\mu}\left[(1-\mu^2)\frac{\partial
\theta_u}{\partial\mu}\right]-\upsilon_u=-\Lambda_u~~;
\end{equation}
which can be turned into Eq.\,(\ref{eq:EC20}) provided the constant,
$\upsilon_u$, is replaced therein by $\upsilon_u-\Lambda_u$, which preserves
related results.   In particular, a solution of Eq.\,(\ref{eq:ECn0}) is:
\begin{equation}
\label{eq:ECn0s}
\theta_u(\xi_u,\mu)=\sum_{\ell=0}^{+\infty}\left[D_{2\ell}\xi_u^{2\ell}+
\frac{\overline D_{2\ell}}{\xi_u^{2\ell+1}}\right]P_{2\ell}(\mu)+\frac16
(\upsilon_u-\Lambda_u)\xi_u^2\left[1-P_2(\mu)\right]~~;
\end{equation}
and the substitution of Eq.\,(\ref{eq:ECn0s}) into (\ref{eq:VGd}), after some
algebra, yields an explicit expression of the gravitational potential as:
\begin{lefteqnarray}
\label{eq:VGn0}
&& {\cal V}_G=4\pi G\sum(\lambda_w)\alpha_u^2\left\{\sum_{\ell=0}^{+\infty}
\left[D_{2\ell}\xi_u^{2\ell}+\frac{\overline D_{2\ell}}{\xi_u^{2\ell+1}}
\right]P_{2\ell}(\mu)-\frac16\Lambda_u\xi_u^2\left[1-P_2(\mu)\right]\right.
\nonumber \\
&& \phantom{{\cal V}_G=4\pi G\sum(\lambda_w)\alpha_u^2\left\{\right.}+\left.
c_{{\rm b},u}^\dagger\right\}~~;
\end{lefteqnarray}
where $c_{{\rm b},u}^\dagger$ is defined by Eq.\,(\ref{eq:cccC}) as shown in
Appendix \ref{a:A2}.

The radial component of the gravitational force via
Eqs.\,(\ref{eq:diEC})-(\ref{eq:csir}) reads:
\begin{lefteqnarray}
\label{eq:FGn0}
&& \frac{\partial{\cal V}_G}{\partial r}=\frac1{\alpha_u}\frac
{\partial{\cal V}_G}{\partial\xi_u}=4\pi G\sum(\lambda_w)\alpha_u\left\{
\sum_{\ell=0}^{+\infty}\left[2\ell D_{2\ell}\xi_u^{2\ell-1}-\frac{(2\ell+1)
\overline D_{2\ell}}{\xi_u^{2\ell+2}}\right]P_{2\ell}(\mu)\right.
\nonumber \\
&& \phantom{\frac{\partial{\cal V}_G}{\partial r}=\frac1{\alpha_u}\frac
{\partial{\cal V}_G}{\partial\xi_u}=4\pi G\sum(\lambda_w)\alpha_u\left\}
\sum_{\ell=0}^{+\infty}\right.}-\left.
\frac13\Lambda_u\xi_u\left[1-P_2(\mu)\right]\right\}~~;
\end{lefteqnarray}
on the other hand, the gravitational potential and the radial component of the
gravitational force outside the boundary may be approximated by
Eqs.\,(\ref{eq:grpe}) and (\ref{eq:frpe}), respectively, as shown in Appendix
\ref{a:A2}.

The continuity of the gravitational potential and the radial component of the
gravitational force on a selected point of the boundary, $\Xi_u=\Xi_u(\mu)$,
makes the terms of same order in Legendre polynomials necessarily be equal in
each expression, which implies the following systems of equations:
\begin{lefteqnarray}
\label{eq:sib0}
&& \cases {
D_{0}+\displayfrac{\overline{D}_{0}}{\Xi_{u}^{\phantom{2}}}-
\displayfrac1{6^{\phantom{}}}\Lambda_u\Xi_{u}^2+c_{{\rm b},u}^\dagger=
\displayfrac{c_{0,u}}{\Xi_{u}^{\phantom{2}}}~~;
 & \cr
 & \cr
-\displayfrac{\overline{D}_{0}}{\Xi_{u}^2}-\displayfrac13\Lambda_u\Xi_{u}=
-\displayfrac{c_{0,u}}{\Xi_{u}^2}~~; & \cr
} \\
\label{eq:sib2}
&& \cases {
D_{2}\Xi_{u}^2+\displayfrac{\overline{D}_{2}}{\Xi_{u}^3}+\displayfrac16
\Lambda_u\Xi_u^2=\displayfrac{c_{2,u}}{\Xi_{u}^3}~~;
 & \cr
 & \cr
2D_{2}\Xi_{u}-\displayfrac{3\overline{D}_{2}}{\Xi_{u}^4}+\displayfrac13
\Lambda_u\Xi_u=-\displayfrac{3c_{2,u}}{\Xi_{u}^4}~~; & \cr
} \\
\label{eq:sib2l}
&& \cases {
D_{2\ell}\Xi_{u}^{2\ell}+\displayfrac{\overline{D}_{2\ell}}
{\Xi_{u}^{2\ell+1}}=\displayfrac{c_{2\ell,u}}{\Xi_{u}^{2\ell+1}}~~;
 & \cr
 & \cr
2\ell D_{2\ell}\Xi_{u}^{2\ell-1}-\displayfrac{(2\ell+1)
\overline{D}_{2\ell}}{\Xi_{u}^{2\ell+2}}
=-\displayfrac{(2\ell+1)c_{2\ell,u}}{\Xi_{u}^{2\ell+2}}~~; & \cr
}
\end{lefteqnarray}
and related solutions read:
\begin{lefteqnarray}
\label{eq:Dw00}
&& D_{0}=\frac12\Lambda_u\Xi_u^2-c_{{\rm b},u}^\dagger~~;\qquad\overline{D}_0=
c_{0,u}-\frac13\Lambda_u\Xi_u^3~~; \\
\label{eq:Dw20}
&& D_2=-\frac16\Lambda_u~~;\qquad\overline{D}_2=c_{2,u}~~; \\
\label{eq:Dw2l0}
&& D_{2\ell}=0~~;\qquad\overline{D}_{2\ell}=c_{2\ell,u}~~;\qquad2\ell>2
\end{lefteqnarray}
accordingly, Eq.\,(\ref{eq:ECn0s}) reduces to:
\begin{equation}
\label{eq:ECn0t}
\theta_u(\xi_u,\mu)=D_0+\sum_{\ell=0}^{+\infty}\frac{\overline D_{2\ell}}
{\xi_u^{2\ell+1}}P_{2\ell}(\mu)+\frac16\upsilon_u\xi_u^2\left[1-P_2(\mu)
\right]-\frac16\Lambda_u\xi_u^2~~;
\end{equation}
where the constants, $\overline D_{2\ell}$, remain to be explicitly
determined.

The comparison of Eq.\,(\ref{eq:ECn0s}) with its counterpart expressed in
terms of the EC2 associated functions, Eq.\,(\ref{eq:thn}), by equating the
terms of same degree in Legendre polynomials, yields:
\begin{lefteqnarray}
\label{eq:h00}
&& A_0\theta_{0,u}(\xi_u)=D_0+\frac{\overline{D}_0}{\xi_u}+\frac16
(\upsilon_u-\Lambda_u)\xi_u^2~~;\qquad A_0=1~~; \\
\label{eq:h20}
&& A_2\theta_{2,u}(\xi_u)=D_2\xi_u^2+\frac{\overline{D}_2}{\xi_u^3}-
\frac16(\upsilon_u-\Lambda_u)\xi_u^2~~; \\
\label{eq:h2l0}
&& A_{2\ell}\theta_{2\ell,u}(\xi_u)=D_{2\ell}\xi_u^{2\ell}+
\frac{\overline{D}_{2\ell}}{\xi_u^{2\ell+1}}~~;\qquad2\ell>2~~;
\end{lefteqnarray}
and additional relations can be inferred from the continuity of the
gravitational potential and the radial component of the gravitational force on
a selected point of the interface, $\xi_u^\ast=\xi_u^\ast(\mu)$.   To this
respect, it is sufficient solving the systems of Eqs.\,(\ref{eq:h00}),
(\ref{eq:h20}), (\ref{eq:h2l0}), and related radial first derivatives on both
sides:
\begin{lefteqnarray}
\label{eq:e00}
&& A_0\theta_{0,u}^\prime(\xi_u)=-\frac{\overline{D}_0}{\xi_u^2}+
\frac13(\upsilon_u-\Lambda_u)\xi_u~~;\qquad A_0=1~~; \\
\label{eq:e20}
&& A_2\theta_{2,u}^\prime(\xi_u)=2D_2\xi_u-\frac{3\overline{D}_2}
{\hat\xi_u^4}-\frac13(\upsilon_u-\Lambda_u)\xi_u~~; \\
\label{eq:e2l0}
&& A_{2\ell}\theta_{2\ell,u}^\prime(\xi_u)=2\ell D_{2\ell}\xi_u^{2\ell-1}-
\frac{(2\ell+1)\overline{D}_{2\ell}}{\xi_u^{2\ell+2}}~~;\qquad2\ell>2~~;
\end{lefteqnarray}
particularized to $\xi_u=\xi_u^\ast$,
where values on the left-hand side relate to the common region and the
coefficients, $A_{2\ell}$, attain the same value for both subsystems, as shown
in Appendix \ref{a:seco}.   After some algebra, the result is:
\begin{leftsubeqnarray}
\slabel{eq:nD02la}
&& D_{2\ell}=\frac1{4\ell+1}\frac1{(\xi_{u}^\ast)^{2\ell}}\left\{A_{2\ell}
\left[(2\ell+1)\theta_{2\ell,u}(\xi_{u}^\ast)+\xi_{u}^\ast\theta_{2\ell,u}^
\prime(\xi_{u}^\ast)\right]\right. \nonumber \\
&& \phantom{D_{2\ell}=\frac1{4\ell+1}\frac1{(\xi_{u}^\ast)^{2\ell}}\left\{
\right.}\left.
-(2\ell+3)\frac{\delta_{2\ell,0}-\delta_{2\ell,2}}6(\upsilon_u-\Lambda_u)
(\xi_{u}^\ast)^2\right\}~~; \\
\slabel{eq:nD02lb}
&& \overline{D}_{2\ell}=\frac{(\xi_{u}^\ast)^{2\ell+1}}{4\ell+1}\left\{
A_{2\ell}\left[2\ell\theta_{2\ell,u}(\xi_{u}^\ast)-\xi_{u}^\ast
\theta_{2\ell,u}^\prime(\xi_{u}^\ast)\right]\right. \nonumber \\
&& \phantom{\overline{D}_{2\ell}=\frac{(\xi_{u}^\ast)^{2\ell+1}}{4\ell+1}
\left\{\right.}
\left.-(2\ell-2)\frac{\delta_{2\ell,0}-\delta_{2\ell,2}}6(\upsilon_u-
\Lambda_u)(\xi_{u}^\ast)^2\right\}~~;
\label{seq:nD02l}
\end{leftsubeqnarray}
where $D_{2\ell}=0$, $2\ell>2$, according to Eq.\,(\ref{eq:Dw2l0}), which
implies $A_{2\ell}=0$, $2\ell>2$, via Eq.\,(\ref{eq:nD02la}) and, in turn,
$\overline D_{2\ell}=0$, $2\ell>2$, via Eq.\,(\ref{eq:nD02lb}).   Accordingly,
Eq.\,(\ref{eq:ECn0t}) reduces to:
\begin{equation}
\label{eq:ECn0u}
\theta_u(\xi_u,\mu)=D_0+\frac{\overline D_0}{\xi_u}+\frac{\overline D_2}
{\xi_u^3}P_{2}(\mu)-\frac16\Lambda_u\xi_u^2+\frac16\upsilon_u\xi_u^2\left[1-
P_2(\mu)\right]~~;
\end{equation}
where $D_0$, $\overline D_0$, $\overline D_2$, are expressed by
Eq.\,(\ref{seq:nD02l}) particularized to $2\ell=0,2$, respectively.

Finally, the substitution of Eq.\,(\ref{eq:Dw00}) into (\ref{eq:nD02la})
yields:
\begin{equation}
\label{eq:cbu00c}
c_{{\rm b},u}^\dagger=-\theta_{0,u}(\xi_u^\ast)-\xi_{u}^\ast
\theta_{0,u}^\prime(\xi_{u}^\ast)+\frac12\upsilon_u(\xi_u^\ast)^2+
\frac12\Lambda_u[\Xi_u^2-(\xi_u^\ast)^2]~~;
\end{equation}
and the substitution of Eq.\,(\ref{eq:Dw20}) into
(\ref{eq:nD02la}) after some algebra produces:
\begin{equation}
\label{eq:A200}
A_2=-\frac56\frac{\upsilon_u(\xi_{u}^\ast)^2}{3\theta_{2,u}(\xi_u^\ast)+
\xi_{u}^\ast\theta_{2\ell,u}^\prime(\xi_{u}^\ast)}~~;
\end{equation}
to be compared with
their counterparts related to the boundary, Eqs.\,(\ref{eq:ccCs}) and
(\ref{eq:AC2l}) written in Appendix \ref{a:A2}, respectively.

The EC2 associated functions, $\theta_{0,u}$, $\theta_{2,u}$, expressed by
Eqs.\,(\ref{eq:h00}), (\ref{eq:h20}), respectively, can be expanded in Taylor
series of starting point, $\xi_{0,u}$.   To this aim, the substitution of
Eqs.\,(\ref{eq:idcs}) and (\ref{eq:Tfm}), $m=1,3$, written in Appendix
\ref{a:seco}, into (\ref{eq:h00}) and (\ref{eq:h20}), after some algebra
yields:
\begin{lefteqnarray}
\label{eq:s0u}
&& A_0\theta_{0,u}(\xi_u)=D_0+\frac{\overline{D}_0}{\xi_{0,u}}\sum_{k=0}^
{+\infty}(-1)^k\left(\frac{\xi_{u}-\xi_{0,u}}{\xi_{0,u}}\right)^k+\frac16
(\upsilon_u-\Lambda_u) \nonumber \\
&& \phantom{A_0\theta_{0,u}(\xi_u)=}\times
\left[\xi_{0,u}^2+2\xi_{0,u}(\xi_u-\xi_{0,u})+
(\xi_u-\xi_{0,u})^2\right]~~;\qquad A_0=1~~;\qquad \\
\label{eq:s2u}
&& A_2\theta_{2,u}(\xi_u)=\frac{\overline{D}_2}{\xi_{0,u}^3}\sum_{k=0}^
{+\infty}(-1)^k\frac{(k+1)(k+2)}2\left(\frac{\xi_u-\xi_{0,u}}{\xi_{0,u}}
\right)^k-\frac16\upsilon_u \nonumber \\
&& \phantom{A_2\theta_{2,u}(\xi_u)=}\times
\left[\xi_{0,u}^2+2\xi_{0,u}(\xi_u-\xi_{0,u})+
(\xi_u-\xi_{0,u})^2\right]~~;
\end{lefteqnarray}
provided $\vert\xi_u-\xi_{0,u}\vert<\xi_{0,u}$.

The comparison of Eqs.\,(\ref{eq:s0u}) and (\ref{eq:s2u}) with related Taylor
series expansions, Eq.\,(\ref{eq:sersn}), yields the explicit expression of
the coefficients as:
\begin{leftsubeqnarray}
\slabel{eq:a0knu0a}
&& A_0a_{0,1}=-\frac{\overline{D}_0}{\xi_{0,u}^2}+\frac16
(\upsilon_u-\Lambda_u)2\xi_{0,u}~~;\qquad A_0=1~~; \\
\slabel{eq:a0knu0b}
&& A_0a_{0,2}=\frac{\overline{D}_0}{\xi_{0,u}^3}+\frac16(\upsilon_u-\Lambda_u)
~~;\qquad A_0=1~~; \\
\slabel{eq:a0knu0c}
&& A_0a_{0,k}=(-1)^k\frac{\overline{D}_0}{\xi_{0,u}^{k+1}}~~;\qquad k>2~~;
\qquad A_0=1~~;  
\label{seq:a0knu0}
\end{leftsubeqnarray}
\begin{leftsubeqnarray}
\slabel{eq:a2knu0a}
&& A_2a_{2,1}=-\frac{3\overline{D}_2}{\xi_{0,u}^4}-\frac16\upsilon_u2\xi_{0,u}
~~; \\
\slabel{eq:a2knu0b}
&& A_2a_{2,2}=\frac{6\overline{D}_2}{\xi_{0,u}^5}-\frac16\upsilon_u~~; \\
\slabel{eq:a2knu0c}
&& A_2a_{2,k}=(-1)^k\frac{(k+1)(k+2)}2\frac{\overline{D}_2}{\xi_{0,u}^{k+3}}
~~;\qquad k>2~~;
\label{seq:a2knu0}
\end{leftsubeqnarray}
according to Eq.\,(\ref{eq:cersn}), where
$a_{2\ell,0}=\theta_{2\ell,u}(\xi_{0,u})$ is also included.

Let $\theta_{0,u}(\Xi_{{\rm ex},u})=\theta_u(\Xi_u,\mu)=\theta_{{\rm b},u}$ be
the (fictitious) spherical isopycnic surface of the expanded sphere, related to the
boundary.   By use of Eq.\,(\ref{eq:h00}), an explicit expression reads:
\begin{equation}
\label{eq:tn0ex}
D_0+\frac{\overline D_0}{\Xi_u}+\frac16(\upsilon_u-\Lambda_u)\Xi_u^2=
\theta_{{\rm b},u}~~;
\end{equation}
which is a third-degree equation where the scaled radius,
$\Xi_{{\rm ex},u}$, is the lowest positive solution.   In the special case,
$\theta_{{\rm b},u}=0$, Eq.\,(\ref{eq:tn0ex}) via (\ref{eq:g00}) reduces to:
\begin{equation}
\label{eq:tn00ex}
D_0+\frac{\overline D_0}{\Xi_u}+\frac16\upsilon_u\Xi_u^2=0~~;
\end{equation}
which, in the nonrotating limit, $\upsilon_u\to0$, has a single solution,
$\Xi_{{\rm ex},u}=-\overline D_0/D_0$, that has necessarily to be positive,
hence $\theta_{0,u}(\xi_u)<0$ as $\xi_u>\Xi_{{\rm ex},u}$.

Concerning the special case, $(n_v,n_u)=(n_v,1)$, the EC2 equation,
Eq.\,(\ref{eq:ECE}), takes the form:
\begin{lefteqnarray}
\label{eq:ECn1}
&& \frac1{\xi^2}\frac\partial{\partial\xi}\left(\xi^2
\frac{\partial\theta}{\partial\xi}\right)+
\frac1{\xi^2}\frac\partial{\partial\mu}\left[(1-\mu^2)\frac{\partial
\theta}{\partial\mu}\right]-\upsilon=-\theta~~;
\end{lefteqnarray}
where $\theta=\theta_u$, $\xi=\Lambda_u^{1/2}\xi_u$, $\upsilon=\Lambda_u^{-1}
\upsilon_u$, via Eqs.\,(\ref{eq:dimq}), (\ref{eq:snnr}).
A special integral of Eq.\,(\ref{eq:ECn1}) is
$\theta^{(p)}(\xi,\mu)=\upsilon$.

Similarly, the associated EC2 equations, Eqs.\,(\ref{eq:t0n})-(\ref{eq:t2ln}),
reduce to:
\begin{lefteqnarray}
\label{eq:t2ln1}
&& \frac1{\xi^2}\frac{\diff}{\diff\xi}\left(\xi^2\frac{\diff\theta_{2\ell}}
{\diff\xi}\right)-\frac{(2\ell+1)2\ell}{\xi^2}\theta_{2\ell}-\delta_{2\ell,0}
\upsilon=-\theta_{2\ell}~~;
\end{lefteqnarray}
where $\theta_{2\ell}=\theta_{2\ell,u}$ via Eq.\,(\ref{eq:snnr}).   A special
integral of Eq.\,(\ref{eq:t2ln1}) is
$\theta_{2\ell}^{(p)}(\xi)=\delta_{2\ell,0}\upsilon$.

In the case under discussion, the set of associated EC2 equations is exactly
equivalent to the EC2 equation. Accordingly, if
$A_{2\ell}\theta_{2\ell}(\xi)+\delta_{2\ell,0}\upsilon$ are solutions of
Eqs.\,(\ref{eq:t2ln1}), the solution of Eq.\,(\ref{eq:ECn1}) reads:
\begin{lefteqnarray}
\label{eq:thn1}
&& \theta(\xi,\mu)=\sum_{\ell=0}^{+\infty}A_{2\ell}\theta_{2\ell}(\xi)
P_{2\ell}(\mu)+\upsilon~~; \qquad A_0=1~~;
\end{lefteqnarray}
where the sum is the general integral of the associated homogeneous equation.

The solutions of Eqs.\,(\ref{eq:t2ln1}) are e.g., [4], [20]:
\begin{lefteqnarray}
\label{eq:ten12l}
&& A_{2\ell}\theta_{2\ell}(\xi)=D_{2\ell}\hat{J}_{+2\ell+1/2}(\xi)+
\overline{D}_{2\ell}\hat{J}_{-2\ell-1/2}(\xi)+\delta_{2\ell,0}\upsilon~~; \\
\label{eq:Jt}
&& \hat{J}_{\mp k\mp1/2}(\xi)=(2k+1)!!\sqrt{\frac\pi{2\xi}}J_{\mp k\mp1/2}
(\xi)~~; \\
\label{eq:facd}
&& (2k+1)!!=(2k+1)\cdot(2k-1)\cdot...\cdot5\cdot3\cdot1~~;\qquad k\ge0~~;
\end{lefteqnarray}
where $J_{\mp k\mp1/2}(\xi)$ are the Bessel functions of half-integer degree,
$\mp k\mp1/2$ e.g., [37], Chap.\,24.

The substitution of Eq.\,(\ref{eq:ten12l}) into (\ref{eq:thn1}) yields:
\begin{lefteqnarray}
\label{eq:thn2}
&& \theta(\xi,\mu)=\sum_{\ell=0}^{+\infty}\left[D_{2\ell}\hat J_{+2\ell+1/2}
(\xi)+\overline D_{2\ell}\hat J_{-2\ell-1/2}(\xi)\right]P_{2\ell}(\mu)+
\upsilon~~;
\end{lefteqnarray}
which is an exact solution of Eq.\,(\ref{eq:ECn1}).

Related expressions of the gravitational potential and radial component of the
gravitational force, via Eqs.\,(\ref{eq:diEC}), (\ref{eq:csir}),
(\ref{eq:VGd}), (\ref{eq:cccC}) written in Appendix \ref{a:A2}, read:
\begin{lefteqnarray}
\label{eq:VGdn1}
&& {\cal V}_G(\xi_u,\mu)=4\pi G\sum(\lambda_w)\alpha_u^2\left\{\sum_{\ell=0}^
{+\infty}\left[D_{2\ell}\hat{J}_{+2\ell+1/2}(\Lambda_u^{1/2}\xi_u)
\right.\right.\nonumber \\
&& \phantom{{\cal V}_G(\xi_u,\mu)=}
\left.\left.+
\overline{D}_{2\ell}\hat{J}_{-2\ell-1/2}(\Lambda_u^{1/2}\xi_u)\right]+
\frac{\upsilon_u}{\Lambda_u}-\frac16\upsilon_u\xi_u^2[1-
P_2(\mu)]+c_{{\rm b},u}^\dagger\right\};\qquad \\
\label{eq:FGdn1}
&& \frac{\partial{\cal V}_G}{\partial r}=\frac1{\alpha_u}\frac
{\partial{\cal V}_G}{\partial\xi_u}=4\pi G\sum(\lambda_w)\alpha_u\left\{
\sum_{\ell=0}^{+\infty}\left[D_{2\ell}\hat{J}_{+2\ell+1/2}^\prime
(\Lambda_u^{1/2}\xi_u)\right.\right.\nonumber \\
&& \phantom{\frac{\partial{\cal V}_G}{\partial r}=\frac1{\alpha_u}\frac
{\partial{\cal V}_G}{\partial\xi_u}=}\left.\left.+
\overline{D}_{2\ell}\hat{J}_{-2\ell-1/2}^\prime(\Lambda_u^{1/2}\xi_u)\right]-
\frac13\upsilon_u\xi_u[1-P_2(\mu)]\right\};
\end{lefteqnarray}
where the prime denotes derivation with respect to $\xi_u$.   On the other
hand, the gravitational potential and the radial component of the
gravitational force outside the boundary may be approximated by
Eqs.\,(\ref{eq:grpe}) and (\ref{eq:frpe}), respectively, as shown in Appendix
\ref{a:A2}.

The continuity of the gravitational potential and the radial component of the
gravitational force on the interface along a selected direction,
$\xi_u^\ast=\xi_u^\ast(\mu)$, yields the following relations:
\begin{lefteqnarray}
\label{eq:tin12l}
&& A_{2\ell}\theta_{2\ell}(\xi_u^\ast)=D_{2\ell}\hat{J}_{+2\ell+1/2}
(\Lambda_u^{1/2}\xi_u^\ast)+\overline{D}_{2\ell}\hat{J}_{-2\ell-1/2}
(\Lambda_u^{1/2}\xi_u^\ast)+\delta_{2\ell,0}\Lambda_u^{-1}\upsilon_u~~;\qquad
\\
\label{eq:din12l}
&& A_{2\ell}\theta_{2\ell}^\prime(\xi_u^\ast)=D_{2\ell}\hat{J}_{+2\ell+1/2}
^\prime(\Lambda_u^{1/2}\xi_u^\ast)+\overline{D}_{2\ell}\hat{J}_{-2\ell-1/2}
^\prime(\Lambda_u^{1/2}\xi_u^\ast)~~;
\end{lefteqnarray}
where $A_0=1$.   In terms of the constants, $A_{2\ell}$, $D_{2\ell}$,
$\overline{D}_{2\ell}$, Eqs.\,(\ref{eq:tin12l}) and (\ref{eq:din12l}) make a
system of two equations in three unknowns, implying an infinity of solutions.

Conversely, the system of equations:
\begin{lefteqnarray}
\label{eq:sis2l}
&& \cases {
D_{2\ell}^\ast\hat{J}_{+2\ell+1/2}(\Lambda_u^{1/2}\xi_u^\ast)+
\overline{D}_{2\ell}^\ast\hat{J}_{-2\ell-1/2}(\Lambda_u^{1/2}\xi_u^\ast)
=\theta_{2\ell}(\xi_u^\ast)-\delta_{2\ell,0}\Lambda_u^{-1}\upsilon_u~~;
 & \cr
 & \cr
D_{2\ell}^\ast\hat{J}_{+2\ell+1/2}^{\prime}(\Lambda_u^{1/2}\xi_u^\ast)+
\overline{D}_{2\ell}^\ast\hat{J}_{-2\ell-1/2}^{\prime}(\Lambda_u^{1/2}
\xi_u^\ast)=\theta_{2\ell}^\prime(\xi_u^\ast)~~; & \cr
} \\
\label{eq:D2la}
&& D_{2\ell}^\ast=\frac{D_{2\ell}}{A_{2\ell}}~~;\qquad
\overline D_{2\ell}^\ast=\frac{\overline D_{2\ell}}{A_{2\ell}}~~;\qquad
A_0=1~~;
\end{lefteqnarray}
exhibits a unique solution as:
\begin{lefteqnarray}
\label{eq:sol2l}
&& D_{2\ell}^\ast=\frac{{\cal D}_{2\ell}^\ast}{{\cal D}^{\ast(2\ell)}}~~;
\qquad\overline{D}_{2\ell}^\ast=\frac{\overline{\cal D}_{2\ell}^\ast}{{\cal D}
^{\ast(2\ell)}}~~; \\
\label{eq:det2l1}
&& {\cal D}_{2\ell}^\ast=\hat{J}_{-2\ell-1/2}^\prime(\Lambda_u^{1/2}\xi_u^
\ast)\left[\theta_{2\ell}(\xi_u^\ast)-\delta_{2\ell,0}\Lambda_u^{-1}\upsilon_u
\right] \nonumber \\
&& \phantom{{\cal D}_{2\ell}^\ast=}
-\hat{J}_{-2\ell-1/2}(\Lambda_u^{1/2}\xi_u^\ast)\theta_{2\ell}^\prime
(\xi_u^\ast)~~; \\
\label{eq:adt2l1}
&& \overline{\cal D}_{2\ell}^\ast=\hat{J}_{+2\ell+1/2}(\Lambda_u^{1/2}\xi_u^
\ast)
\theta_{2\ell}^\prime(\xi_u^\ast) \nonumber \\
&& \phantom{\overline{\cal D}_{2\ell}^\ast=}
-\hat{J}_{+2\ell+1/2}^\prime(\Lambda_u^{1/2}
\xi_u^\ast)\left[\theta_{2\ell}(\xi_u^\ast)-\delta_{2\ell,0}\Lambda_u^{-1}
\upsilon_u\right]~~; \\
\label{eq:dts2l}
&& {\cal D}^{\ast(2\ell)}=\hat{J}_{+2\ell+1/2}(\Lambda_u^{1/2}\xi_u^\ast)
\hat{J}_{-2\ell-1/2}^\prime(\Lambda_u^{1/2}\xi_u^\ast) \nonumber \\
&& \phantom{{\cal D}^{\ast(2\ell)}=}
-\hat{J}_{+2\ell+1/2}^\prime
(\Lambda_u^{1/2}\xi_u^\ast)\hat{J}_{-2\ell-1/2}(\Lambda_u^{1/2}\xi_u^\ast)~~;
\end{lefteqnarray}
where ${\cal D}_{2\ell}^\ast$, $\overline{\cal D}_{2\ell}^\ast$,
${\cal D}^{\ast(2\ell)}$, are determinants appearing in the solutions of the
systems, expressed by Eq.\,(\ref{eq:sis2l}).

In general, $A_{2\ell}=0$, $2\ell>2$, according to Eq.\,(\ref{eq:ACsi})
as shown in
Appendix \ref{a:A2}, which implies $D_{2\ell}=0$, $\overline D_{2\ell}=0$,
$2\ell>2$, via Eqs.\,(\ref{eq:tin12l})-(\ref{eq:din12l}).   Then the cases of
interest are $2\ell=0,2$, and using explicit expressions of Bessel functions
of half-integer degree, Eqs.\,(\ref{eq:Jt}) can be rewritten as:
\begin{lefteqnarray}
\label{eq:Jt10}
&& \hat J_{+1/2}(\xi)=\frac{\sin\xi}\xi~~;\qquad\hat J_{-1/2}(\xi)=\frac
{\cos\xi}\xi~~; \\
\label{eq:Jt12p}
&& \hat J_{+5/2}(\xi)=15\left[\left(\frac3{\xi^2}-1\right)\frac{\sin\xi}\xi-
\frac3\xi\frac{\cos\xi}\xi\right]~~; \\
\label{eq:Jt12m}
&& \hat J_{-5/2}(\xi)=15\left[\left(\frac3{\xi^2}-1\right)\frac{\cos\xi}\xi+
\frac3\xi\frac{\sin\xi}\xi\right]~~;
\end{lefteqnarray}
and the substitution of Eqs.\,(\ref{eq:Jt10})-(\ref{eq:Jt12m}) into
(\ref{eq:sis2l})-(\ref{eq:dts2l}), after a lot of algebra yields:
\begin{lefteqnarray}
\label{eq:det01}
&& {\cal D}_0^\ast=-\xi^\ast\{[\theta_0(\xi_u^\ast)-\upsilon]
[\xi^\ast\sin\xi^\ast+\cos\xi^\ast]
+\xi^\ast\theta_0^\prime(\xi_u^\ast)\cos\xi^\ast\}~~; \\
\label{eq:adt01}
&& \overline{\cal D}_0^\ast=-\xi^\ast\{[\theta_0(\xi_u^\ast)-\upsilon]
[\xi^\ast\cos\xi^\ast-\sin\xi^\ast]
-\xi^\ast\theta_0^\prime(\xi_u^\ast)\sin\xi^\ast\}~~; \\
\label{eq:dts0}
&& {\cal D}^{\ast(0)}=-\xi^\ast~~; \\
\label{eq:det21}
&& {\cal D}_2^\ast=-15\left\{\left[\frac9{(\xi^\ast)^2}-4\right]\theta_2
(\xi_u^\ast)+\left[\frac3{(\xi^\ast)^2}-1\right]\xi^\ast\theta_2^\prime
(\xi_u^\ast)\right\}\frac{\cos\xi^\ast}{\xi^\ast} \nonumber \\
&& \phantom{{\cal D}_2=}-
15\left\{\left[\frac9{\xi^\ast}-\xi^\ast\right]\theta_2(\xi_u^\ast)
+3\theta_2^\prime(\xi_u^\ast)\right\}\frac{\sin\xi^\ast}
{\xi^\ast}~~; \\
\label{eq:adt21}
&& \overline{\cal D}_2^\ast=+15\left\{\left[\frac9{(\xi^\ast)^2}-4\right]
\theta_2(\xi_u^\ast)+\left[\frac3{(\xi^\ast)^2}-1\right]\xi^\ast\theta_2^
\prime(\xi_u^\ast)\right\}\frac{\sin\xi^\ast}{\xi^\ast} \nonumber \\
&& \phantom{{\cal D}_2=}-
15\left\{\left[\frac9{\xi^\ast}-\xi^\ast\right]\theta_2(\xi_u^\ast)
+3\theta_2^\prime(\xi_u^\ast)\right\}\frac{\cos\xi^\ast}
{\xi^\ast}~~; \\
\label{eq:dts2}
&& {\cal D}^{\ast(2)}=-\frac{225}{\xi^\ast}~~;
\end{lefteqnarray}
where $\xi^\ast=\Lambda_u^{1/2}\xi_u^\ast$,
$\upsilon=\Lambda_u^{-1}\upsilon_u$,
via Eqs.\,(\ref{eq:dimq}) and (\ref{eq:snnr}).            
The substitution of Eqs.\,(\ref{eq:det01})-(\ref{eq:dts2}) into
(\ref{eq:sol2l}) yields
the explicit expression of the constants, $D_0^\ast$, $\overline D_0^\ast$,
$D_2^\ast$, $\overline D_2^\ast$.

In general, Eq.\,(\ref{eq:ten12l}) via (\ref{eq:D2la}) may be cast under the
form:
\begin{equation}
\label{eq:t12lb}
\theta_{2\ell}(\xi)=D_{2\ell}^\ast\hat{J}_{+2\ell+1/2}(\xi)+\overline
D_{2\ell}^\ast\hat{J}_{-2\ell-1/2}(\xi)+\delta_{2\ell,0}\upsilon~~;
\end{equation}
which, in particular, can be determined on a point of the boundary along a
selected direction, $\Xi=\Xi(\mu)$, $\Xi=\Lambda_u^{1/2}\Xi_u$, together with
related first derivative.   Then the substitution of $\theta_{2\ell}(\Xi_u)$,
$\theta_{2\ell}^\prime(\Xi_u)$, into Eqs.\,(\ref{eq:ACsi})-(\ref{eq:cccC})
written in
Appendix \ref{a:A2}, yields an explicit expression of the constants,
$A_{2\ell}$, $c_{2\ell,u}$, $c_{{\rm b},u}^\dagger$, $c_{{\rm b},u}$.   In
addition, the knowledge of $D_{2\ell}^\ast$, $\overline D_{2\ell}^\ast$,
$A_{2\ell}$, via Eq.\,(\ref{eq:D2la}) implies the knowledge of $D_{2\ell}$,
$\overline D_{2\ell}$.

The EC2 associated functions, $\theta_{2\ell}(\xi_u)$, can be expanded in
Taylor series of starting point, $\xi_{0,u}$, only in the special case of the
singular point, $\xi_{0,u}=0$, where the convergence radius is infinite.
On the other hand, series expansions of the kind considered cannot be used in
the case under discussion, due to the presence of divergent terms as
$\xi_{0,u}\to0$ on the right-hand side of Eq.\,(\ref{eq:t12lb}).  Accordingly,
the coefficients of Taylor series expansions, Eq.\,(\ref{eq:sersn}), cannot be
expressed in simpler form with respect to the regression formulae,
Eqs.\,(\ref{seq:a0k}), (\ref{seq:a2k}).   For further details, an interested
reader is addressed to the parent paper [6].

Let $\theta_{0,u}(\Xi_{{\rm ex},u})=\theta_u(\Xi_u,\mu)=\theta_{{\rm b},u}$ be
the (fictitious) spherical isopycnic surface of the expanded sphere, related
to the
boundary.   By use of Eqs.\,(\ref{eq:D2la}), (\ref{eq:Jt10}) and
(\ref{eq:t12lb}), $2\ell=0$, an explicit expression reads:
\begin{equation}
\label{eq:t10ex}
D_0\frac{\sin(\Lambda_u^{1/2}\Xi_u)}{\Lambda_u^{1/2}\Xi_u}+\overline D_0\frac
{\cos(\Lambda_u^{1/2}\Xi_u)}{\Lambda_u^{1/2}\Xi_u}+\frac{\upsilon_u}
{\Lambda_u}=\theta_{{\rm b},u}~~;
\end{equation}
which is a transcendental equation where the scaled radius,
$\Xi_{{\rm ex},u}$, is the lowest positive solution.
In the nonrotating limit, $\upsilon_u\to0$, restricting to
$\theta_{{\rm b},u}=0$, Eq.\,(\ref{eq:t10ex}) has a single solution,
$\Xi_{{\rm ex},u}=\Lambda_u^{-1/2}\arctan(-\overline D_0/D_0)$, which has
necessarily to be positive, hence $\theta_{0,u}(\xi_u)<0$ as
$\xi_u>\Xi_{{\rm ex},u}$.

Concerning the special case, $(n_v,n_u)=(n_v,5)$, the EC2 equation,
Eq.\,(\ref{eq:ECE}), takes the form:
\begin{lefteqnarray}
\label{eq:ECn5}
&& \frac1{\xi^2}\frac\partial{\partial\xi}\left(\xi^2
\frac{\partial\theta}{\partial\xi}\right)+
\frac1{\xi^2}\frac\partial{\partial\mu}\left[(1-\mu^2)\frac{\partial
\theta}{\partial\mu}\right]-\upsilon=-\theta^5~~;
\end{lefteqnarray}
where $\theta=\theta_u$, $\xi=\Lambda_u^{1/2}\xi_u$, $\upsilon=\Lambda_u^{-1}
\upsilon_u$, via Eqs.\,(\ref{eq:dimq}), (\ref{eq:snnr}).    A
special integral of Eq.\,(\ref{eq:ECn5}) is $\theta^{(p)}(\xi,\mu)=\upsilon^
{1/5}$.   Though both the EC2 equation in the nonrotating limit [14],
Chap.\,IV, \S4
and the associated EC2 equations [8], [20], can be integrated
analytically, related expressions are cumbersome and of little practical
utility, keeping in mind the boundary conditions must be related to the
interface instead of the centre in the case under discussion.   Then
the coefficients of Taylor series expansions, Eq.\,(\ref{eq:sersn}), cannot be
expressed in simpler form with respect to the regression formulae,
Eqs.\,(\ref{seq:a0k}), (\ref{seq:a2k}).   For further details, an interested
reader is addressed to the parent paper [6].

\subsubsection{The common region}\label{cmr5}

With regard to the common region, the system of both the EC2 equations,
Eq.\,(\ref{eq:EC2}), and the EC2 associated equations of the same order,
Eqs.\,(\ref{eq:t0c})-(\ref{eq:t2lc}), for each subsystem, must be integrated
by use of Eqs.\,(\ref{eq:EC2b})-(\ref{eq:EC2d}).
The boundary conditions are related to the origin as in EC1
polytropes.

Concerning the special case, $(n_v,n_u)=(n_v,0)$, the EC2 equation,
Eq.\,(\ref{eq:ECE}), reduces to:
\begin{lefteqnarray}
\label{eq:ECc0}
&& \frac1{\xi^2}\frac\partial{\partial\xi}\left(\xi^2\frac{\partial
\theta}{\partial\xi}\right)+
\frac1{\xi^2}\frac\partial{\partial\mu}\left[(1-\mu^2)\frac{\partial
\theta}{\partial\mu}\right]-\upsilon^\prime=-\theta^n~~; \\
\label{eq:upp}
&& \upsilon^\prime=\upsilon-\Lambda_{uv}=\frac{\upsilon_v-\Lambda_u}
{\Lambda_v}~~;\qquad-\infty<\upsilon^\prime\le\upsilon_v~~;
\end{lefteqnarray}
where $\theta=\theta_v$, $n=n_v$, $\xi=\Lambda_v^{1/2}\xi_v$,
$\upsilon=\upsilon_v/\Lambda_v$, via Eqs.\,(\ref{eq:dimq}), (\ref{eq:snnr}),
and the parameter, $\upsilon^\prime$, can be conceived as a
generalized distortion including both centrifugal and tidal effects.

Accordingly, the EC2 associated equations can be read as in the case of EC1
polytropes [4], [6] where the distortion parameter equals
$\upsilon^\prime$,
which can attain both positive, null, and negative values.   Related solutions
exhibit the same formal expression where, in the case under discussion, the
variable, $\xi$, and the generalized distortion parameter, $\upsilon^\prime$,
are to be formulated in terms of $\xi_v$ and $\upsilon_v$, $\Lambda_{uv}$, via
Eqs.\,(\ref{eq:dimq}) and (\ref{eq:upp}), respectively.

If, in addition, $n_v=0$, Eq.\,(\ref{eq:EC2}) reduces to its counterpart for
EC1 polytropes, as:
\begin{lefteqnarray}
\label{eq:E1c0}
&& \frac1{\xi_w^2}\frac\partial{\partial\xi_w}\left(\xi_w^2\frac{\partial
\theta_w}{\partial\xi_w}\right)+
\frac1{\xi_w^2}\frac\partial{\partial\mu}\left[(1-\mu^2)\frac{\partial
\theta_w}{\partial\mu}\right]-\upsilon_w=-1~~;
\end{lefteqnarray}
which, in turn, coincides with its counterpart related to the noncommon
region, $w=u$,
Eq.\,(\ref{eq:ECn0}), provided $\Lambda_u=1$ therein.   The same holds for
the solution of Eq.\,(\ref{eq:E1c0}) via (\ref{eq:ECn0s}) and for the
expression of the gravitational potential and the radial component of the
gravitational force via Eqs.\,(\ref{eq:VGn0}) and (\ref{eq:FGn0}),
respectively.

Keeping in mind the boundary conditions, Eq.\,(\ref{eq:EC2b}), which implies
absence of divergent terms at the origin, the solution of
Eq.\,(\ref{eq:ECc0}) via (\ref{eq:ECn0s}) reads:
\begin{lefteqnarray}
\label{eq:ECc0s}
&& \theta_w(\xi_w,\mu)=\sum_{\ell=0}^{+\infty}D_{2\ell}\xi_w^{2\ell}P_{2\ell}
(\mu)-\frac16(1-\upsilon_w)\xi_w^2\left[1-P_2(\mu)\right]~~;
\end{lefteqnarray}
where $D_0=1$ via Eq.\,(\ref{eq:EC2b}) and, in general, $D_{2\ell}$ is
denoted as in dealing with the noncommon region, but related values can be
different.   Within the current Subsection, it shall be intended unlabeled
symbols relate to either region, while symbols used for both the common
and noncommon region are denoted by the apex, com, ncm, respectively.

The substitution of Eq.\,(\ref{eq:ECc0s}) into Eq.\,(\ref{eq:VGd}) after some
algebra yields an explicit expression of the gravitational potential, as:
\begin{lefteqnarray}
\label{eq:VGc0}
&& {\cal V}_G=4\pi G\sum(\lambda_w)\alpha_w^2\left\{1+\sum_{\ell=1}^{+\infty}
D_{2\ell}\xi_w^{2\ell}P_{2\ell}(\mu)\right. \nonumber \\
&& \phantom{{\cal V}_G=4\pi G\sum(\lambda_w)\alpha_w^2\left\{\right.}\left.
-\frac16\xi_w^2\left[1-P_2(\mu)\right]+c_{{\rm b},w}^\dagger\right\}
~~;\qquad
\end{lefteqnarray}
where $c_{{\rm b},u}^\dagger$ is defined by Eq.\,(\ref{eq:cccC}) as shown in
Appendix \ref{a:A2}.

The radial component of the gravitational force via
Eqs.\,(\ref{eq:diEC})-(\ref{eq:csir}) reads:
\begin{lefteqnarray}
\label{eq:FGc0}
&& \frac{\partial{\cal V}_G}{\partial r}=\frac1{\alpha_w}\frac
{\partial{\cal V}_G}{\partial\xi_w} \nonumber \\
&& \phantom{\frac{\partial{\cal V}_G}{\partial r}}
=4\pi G\sum(\lambda_w)\alpha_w\left\{
\sum_{\ell=1}^{+\infty}2\ell D_{2\ell}\xi_w^{2\ell-1}P_{2\ell}(\mu)-\frac13
\xi_w\left[1-P_2(\mu)\right]\right\};\qquad
\end{lefteqnarray}
on the other hand, the gravitational potential and the radial component of the
gravitational force within the noncommon region are expressed by
Eqs.\,(\ref{eq:VGn0}) and (\ref{eq:FGn0}), respectively.

The continuity of the gravitational potential and the radial component of the
gravitational force on a selected point of the interface,
$\xi_u^\ast=\xi_u^\ast(\mu)$,
makes the terms of same order in Legendre polynomials necessarily be equal in
each expression, which implies the following systems of equations:
\begin{lefteqnarray}
\label{eq:sic0}
&& \cases {
1-\displayfrac16(\xi_u^\ast)^2+c_{{\rm b},u}^\dagger=
D_{0}^{\rm(ncm)}+\displayfrac{\overline{D}_{0}}{\xi_{u}^\ast}-
\displayfrac1{6^{\phantom{}}}\Lambda_u(\xi_{u}^\ast)^2+c_{{\rm b},u}^\dagger
~~;
 & \cr
 & \cr
-\displayfrac13\xi_u^\ast=-\displayfrac{\overline{D}_{0}}{(\xi_{u}^\ast)^2}-
\displayfrac13\Lambda_u\xi_{u}^\ast~~;
 & \cr
} \\
\label{eq:sic2}
&& \cases {
D_{2}^{\rm(com)}(\xi_{u}^\ast)^2+\displayfrac16(\xi_{u}^\ast)^2=
D_{2}^{\rm(ncm)}(\xi_{u}^\ast)^2+
\displayfrac{\overline{D}_{2}}{(\xi_{u}^\ast)^3}+\displayfrac16
\Lambda_u(\xi_u^\ast)^2~~;
 & \cr
 & \cr
2D_{2}^{\rm(com)}\xi_{u}^\ast+\displayfrac13\xi_u^\ast=
2D_{2}^{\rm(ncm)}\xi_{u}^\ast
-\displayfrac{3\overline{D}_{2}}{(\xi_{u}^\ast)^4}+\displayfrac13
\Lambda_u\xi_u^\ast~~; & \cr
} \\
\label{eq:sic2l}
&& \cases {
D_{2\ell}^{\rm(com)}(\xi_{u}^\ast)^{2\ell}=
D_{2\ell}^{\rm(ncm)}(\xi_{u}^\ast)^{2\ell}+
\displayfrac{\overline{D}_{2\ell}}
{(\xi_{u}^\ast)^{2\ell+1}}~~;
 & \cr
 & \cr
2\ell D_{2\ell}^{\rm(com)}(\xi_{u}^\ast)^{2\ell-1}=
2\ell D_{2\ell}^{\rm(ncm)}(\xi_{u}^\ast)^{2\ell-1}
-\displayfrac{(2\ell+1)
\overline{D}_{2\ell}}{(\xi_{u}^\ast)^{2\ell+2}}~~; & \cr
}
\end{lefteqnarray}
and related solutions via Eqs.\,(\ref{eq:Dw20})-(\ref{eq:Dw2l0})
read:
\begin{lefteqnarray}
\label{eq:Du00}
&& D_{0}^{\rm(ncm)}=1-\frac{1-\Lambda_u}2(\xi_u^\ast)^2~~;\qquad\overline{D}_0
=\frac{1-\Lambda_u}3(\xi_u^\ast)^3~~; \\
\label{eq:Du20}
&& D_2^{\rm(com)}=-\frac16~~;\qquad\overline{D}_2=0~~; \\
\label{eq:Du2l0}
&& D_{2\ell}^{\rm(com)}=0~~;\qquad\overline{D}_{2\ell}=0~~;\qquad2\ell>2
~~;
\end{lefteqnarray}
which, in the case under discussion, yields a simpler formulation of 
Eqs.\,(\ref{eq:ECn0t}), (\ref{eq:h20}), (\ref{eq:h2l0}), as:
\begin{lefteqnarray}
\label{eq:ECn00t}
&& \theta_u^{\rm(ncm)}(\xi_u,\mu)=D_{0}^{\rm(ncm)}+\frac{\overline D_0}{\xi_u}
-\frac16\Lambda_u\xi_u^2+\frac16\upsilon_u\xi_u^2\left[1-P_2(\mu)\right]~~; \\
\label{eq:k20}
&& A_2\theta_{2,u}^{\rm(ncm)}(\xi_u)=-\frac16\upsilon_u\xi_u^2~~; \\
\label{eq:k2l0}
&& A_{2\ell}\theta_{2\ell,u}^{\rm(ncm)}(\xi_u)=0~~;\qquad2\ell>2~~;
\end{lefteqnarray}
and Eq.\,(\ref{eq:h00}) can also be written explicitly by use of
Eq.\,(\ref{eq:Du00}).

The substitution of Eqs.\,(\ref{eq:Du00}), (\ref{eq:Du20})-(\ref{eq:Du2l0}),
into (\ref{eq:Dw00}), (\ref{eq:Dw20})-(\ref{eq:Dw2l0}), respectively, yields:
\begin{lefteqnarray}
\label{eq:cbu00d}
&& c_{{\rm b},u}^\dagger=-1+\frac12(\xi_u^\ast)^2+\frac12\Lambda_u
\left[\Xi_u^2-(\xi_u^\ast)^2\right]~~; \\
\label{eq:c0u00d}
&& c_{0,u}=\frac13(\xi_u^\ast)^3+\frac13\Lambda_u
\left[\Xi_u^3-(\xi_u^\ast)^3\right]~~; \\
\label{eq:c2u00d}
&& c_{2\ell,u}=0~~;\qquad2\ell>0~~;
\end{lefteqnarray}
which completes the explicit formulation of the constants within the
noncommon region.

According to Eq.\,(\ref{eq:k20}), the following choice can be made without
loss of generality:
\begin{lefteqnarray}
\label{eq:tn2u00}
&& \theta_{2,u}^{\rm(ncm)}(\xi_u)=\xi_u^2~~; \\
\label{eq:A220}
&& A_{2}=-\frac16\upsilon_u~~;
\end{lefteqnarray}
as shown by the substitution of Eq.\,(\ref{eq:tn2u00}) into (\ref{eq:A200})
and (\ref{eq:ACsi}) written in Appendix \ref{a:A2}.

The combination of Eqs.\,(\ref{eq:ECc0s}) and
(\ref{eq:Du20})-(\ref{eq:Du2l0}), keeping in mind $D_{0}^{\rm(com)}=1$,
yields:
\begin{equation}
\label{eq:ECc0t}
\theta_w(\xi_w,\mu)=1-\frac{1-\upsilon_w}6\xi_w^2-\frac16\upsilon_w\xi_w^2
P_2(\mu)~~;
\end{equation}
and the comparison of Eq.\,(\ref{eq:ECc0t}), via (\ref{eq:ECc0s}), with its
counterpart expressed in
terms of the EC2 associated functions, Eq.\,(\ref{eq:thc}), by equating the
terms of same degree in Legendre polynomials, yields:
\begin{lefteqnarray}
\label{eq:j00}
&& A_0\theta_{0,w}(\xi_w)=1-\frac{1-\upsilon_w}6\xi_w^2~~;\qquad A_0=1~~; \\
\label{eq:j20}
&& A_2\theta_{2,w}(\xi_w)=-\frac16\upsilon_w\xi_w^2~~; \\
\label{eq:j2l0}
&& \theta_{2\ell,w}(\xi_w)=\xi_w^{2\ell}~~;\qquad2\ell>2~~;
\end{lefteqnarray}
which coincide with their counterparts related to EC1 polytropes [4], [6].
The substitution of Eqs.\,(\ref{eq:j20}) into (\ref{eq:nD02la}),
$2\ell=2$, yields Eq.\,(\ref{eq:Dw20}), as expected.

According to Eq.\,(\ref{eq:j20}), the following choice can be made without
loss of generality:
\begin{lefteqnarray}
\label{eq:tc2u00}
&& \theta_{2,u}(\xi_u)=\xi_u^2~~; \\
\label{eq:A230}
&& A_{2}=-\frac16\upsilon_u~~; \\
\label{eq:tc2v00}
&& \theta_{2,v}(\xi_v)=-\frac16\frac{\upsilon_v}{A_2}\xi_v^2=
\frac{\upsilon_v}{\upsilon_u}\xi_v^2~~;
\end{lefteqnarray}
where Eqs.\,(\ref{eq:tc2u00}) and (\ref{eq:tc2v00}) have the same formal
expression only in the special case of subsystems rotating at the same extent,
$\upsilon_v=\upsilon_u$.   The substitution of Eq.\,(\ref{eq:tc2u00}) and
(\ref{eq:tc2v00}) into (\ref{eq:A2uv}) written in Appendix \ref{a:A2} yields
Eq.\,(\ref{eq:A230}), as expected.

Finally, the substitution of Eq.\,(\ref{eq:ECc0t}) into (\ref{eq:VGd})
discloses the gravitational potential and the
gravitational force are purely radial i.e. independent of both $\mu$ and
$\upsilon_w$ within the common region, in the case under discussion.

The comparison of Eqs.\,(\ref{eq:j00}) and (\ref{eq:j20}) with related
MacLaurin series expansions, Eq.\,(\ref{eq:sers0}), via
Eqs.\,(\ref{eq:A230}) and (\ref{eq:Auv2l}) written in Appendix \ref{a:seco}
yields the explicit expression of the coefficients as:
\begin{lefteqnarray}
\label{eq:a0kcu0a}
&& A_0a_{0,1}^{(w,w)}=0~;\quad A_0a_{0,2}^{(w,w)}=-\frac{1-\upsilon_w}6~;
\quad A_0a_{0,k+2}^{(w,w)}=0~;\quad k>0~; \\
\label{eq:a2kcu0b}
&& A_2a_{2,1}^{(w,w)}=0~;\quad A_2a_{2,2}^{(w,w)}=-\frac16\upsilon_w~;
\quad A_2a_{2,k+2}^{(w,w)}=0~;\quad k>0~;
\end{lefteqnarray}
while in the general case, $\xi_{0,w}>0$, the result via Eq.\,(\ref{eq:idcs})
written in Appendix \ref{a:seco} is:
\begin{lefteqnarray}
\label{eq:a0kcu0c}
&& A_0a_{0,1}^{(w,w)}=-\frac{1-\upsilon_w}62\xi_{0,w}~;\quad
A_0a_{0,2}^{(w,w)}=-\frac{1-\upsilon_w}6~;\quad \nonumber \\
&& A_0a_{0,k+2}^{(w,w)}=0~;\quad k>0~~; \\
\label{eq:a2kcu0d}
&& A_2a_{2,1}^{(w,w)}=-\frac16\upsilon_w2\xi_{0,w}~;~~A_2a_{2,2}^{(w,w)}=
-\frac16\upsilon_w~;~~A_2a_{2,k+2}^{(w,w)}=0~;~~k>0~;\qquad
\end{lefteqnarray}
according to Eq.\,(\ref{eq:cersc}), where
$a_{2\ell,0}^{(w,w)}=\theta_{2\ell,w}(\xi_{0,w})$ is also included and, in any
case, the convergence radius is infinite.   In addition,
Eqs.\,(\ref{seq:a0knu0}) and (\ref{seq:a2knu0}) can be written explicitly via
(\ref{eq:Du00}) and (\ref{eq:Du20}), respectively.

Let $\theta_{0,v}(\Xi_{{\rm ex},v})=\theta_v(\Xi_v,\mu)=\theta_{{\rm b},v}$ be
the (fictitious) spherical isopycnic surface of the expanded sphere, related
to the interface.   By use of Eq.\,(\ref{eq:j00}), an explicit expression
reads:
\begin{equation}
\label{eq:tc0ex}
1-\frac{1-\upsilon_v}6\xi_v^2=\theta_{{\rm b},v}~~;
\end{equation}
which has a unique (acceptable) solution as:
\begin{equation}
\label{eq:Xiexv}
\Xi_{{\rm ex},v}=\left[\frac{6(1-\theta_{{\rm b},v})}{1-\upsilon_v}\right]^
{1/2}~~;
\end{equation}
in the special case, $\theta_{{\rm b},v}=0$, Eq.\,(\ref{eq:Xiexv}) reduces to
its counterpart related to EC1 polytropes [4], [6].

Concerning the special case, $(n_v,n_u)=(1,1)$, the EC2 equation,
Eq.\,(\ref{eq:EC2}), via (\ref{eq:ispuv}) takes the form:
\begin{lefteqnarray}
\label{eq:ECc1}
&& \frac1{\xi_u^2}\frac\partial{\partial\xi_u}\left(\xi_u^2
\frac{\partial\theta_u}{\partial\xi_u}\right)+
\frac1{\xi_u^2}\frac\partial{\partial\mu}\left[(1-\mu^2)\frac{\partial
\theta_u}{\partial\mu}\right]=-(\Lambda_u+\Gamma_{uv}\Lambda_v)\theta_u
\nonumber \\
&& \qquad\qquad-
\Gamma_{uv}\Lambda_v\frac{\upsilon_v-\upsilon_u}6\xi_u^2\left[1-P_2(\mu)
\right]+\upsilon_u-\Lambda_v(1-\Gamma_{uv})~~;
\end{lefteqnarray}
where the indexes, $u$, $v$, can freely be exchanged one with respect to the
other.

Let a new parameter be defined as:
\begin{lefteqnarray}
\label{eq:Bu2}
&& B_u^2=\Lambda_u+\Gamma_{uv}\Lambda_v~~;
\end{lefteqnarray}
where, by use of Eqs.\,(\ref{eq:dimq}) and (\ref{eq:Gauv}), the following
identities can be verified after little algebra:
\begin{lefteqnarray}
\label{eq:Buv1}
&& \frac{B_u^2}{B_v^2}=\frac{\alpha_u^2}{\alpha_v^2}=\Gamma_{uv}~~; \\
\label{eq:Buv2}
&& \frac{\Lambda_u}{B_u^2}+\frac{\Lambda_v}{B_v^2}=1~~; \\
\label{eq:Buv3}
&& (B_u^2-\Lambda_u)(B_v^2-\Lambda_v)=\Lambda_u\Lambda_v~~;
\end{lefteqnarray}
accordingly, Eq.\,(\ref{eq:ECc1}) can be written under the equivalent form:
\begin{lefteqnarray}
\label{eq:ECc2}
&& \frac1{\xi^2}\frac\partial{\partial\xi}\left(\xi^2
\frac{\partial\theta_u}{\partial\xi}\right)+
\frac1{\xi^2}\frac\partial{\partial\mu}\left[(1-\mu^2)\frac{\partial
\theta_u}{\partial\mu}\right]=-\theta_u+1
\nonumber \\
&& \qquad\qquad-\frac{1-\upsilon_u}{B_u^2}-\frac{\Gamma_{uv}\Lambda_v}{B_u^4}
\frac{\upsilon_v-\upsilon_u}6\xi^2\left[1-P_2(\mu)\right]~~; \\
\label{eq:Buv4}
&& \xi=B_u\xi_u=B_v\xi_v~~;
\end{lefteqnarray}
in the limit of a vanishing $v$ subsystem, $\Lambda_v\to0$, $\Lambda_u\to1$,
$B_u\to1$, and Eq.\,(\ref{eq:ECc2}) reduces to its counterpart related to EC1
polytropes [4].

Similarly, the associated EC2 equations, Eqs.\,(\ref{eq:t0c})-(\ref{eq:t2lc}),
via Eqs.\,(\ref{eq:Auv2l})-(\ref{eq:isq2l}) written in Appendix \ref{a:seco}
reduce to:
\begin{lefteqnarray}
\label{eq:t2lc1}
&& \frac1{\xi^2}\frac{\diff}{\diff\xi}\left[\xi^2\frac{\diff(A_{2\ell}
\theta_{2\ell,u})}
{\diff\xi}\right]-\frac{(2\ell+1)2\ell}{\xi^2}A_{2\ell}\theta_{2\ell,u}-
\delta_{2\ell,0}\upsilon^\prime=-A_{2\ell}\theta_{2\ell,u} \nonumber \\
&& \qquad+\delta_{2\ell,0}\left[1-
\frac1{B_u^2}-\frac{\Gamma_{uv}\Lambda_v}{B_u^4}\frac{\upsilon_v-\upsilon_u}6
\xi^2\right]+\delta_{2\ell,2}\frac{\Gamma_{uv}\Lambda_v}{B_u^4}
\frac{\upsilon_v-\upsilon_u}6\xi^2~~;\qquad \\
\label{eq:uppBu}
&& \upsilon^\prime=\frac{\upsilon_u}{B_u^2}=\frac{\upsilon_u}{\Lambda_u}\frac
{\Lambda_u}{\Lambda_u+\Gamma_{uv}\Lambda_v}=\frac{\upsilon\Lambda_u}
{\Lambda_u+\Gamma_{uv}\Lambda_v}~~;
\end{lefteqnarray}
where $\upsilon^\prime$ has to be conceived as a generalized distortion
including both centrifugal and tidal effects.

In the case under discussion, the set of associated EC2 equations is exactly
equivalent to the EC2 equation. Accordingly, if
$A_{2\ell}\theta_{2\ell,u}(\xi)$ 
are solutions of
Eqs.\,(\ref{eq:t2lc1}), the solution of Eq.\,(\ref{eq:ECc2}) reads:
\begin{lefteqnarray}
\label{eq:thc1}
&& \theta_u(\xi,\mu)=\sum_{\ell=0}^{+\infty}A_{2\ell}\theta_{2\ell,u}(\xi)
P_{2\ell}(\mu)~~; \qquad A_0=1~~;
\end{lefteqnarray}
on the other hand, the solutions of Eqs.\,(\ref{eq:t2lc1}) are e.g., [4],
[20]:
\begin{lefteqnarray}
\label{eq:tec12l}
&& A_{2\ell}\theta_{2\ell,u}(\xi)=D_{2\ell}\hat{J}_{+2\ell+1/2}(\xi)+
\overline{D}_{2\ell}\hat{J}_{-2\ell-1/2}(\xi)+(\delta_{2\ell,0}+
\delta_{2\ell,2})\theta_{2\ell,u}^{(p)}(\xi)~~;\qquad
\end{lefteqnarray}
where $\hat J_{\mp k\mp1/2}(\xi)$ are defined by Eq.\,(\ref{eq:Jt}),
$\overline{D}_{2\ell}=0$ via Eq.\,(\ref{eq:t02ci}), and
$\theta_{2\ell,u}^{(p)}$ are special integrals expressed as:
\begin{lefteqnarray}
\label{eq:thp1}
&& \theta_{0,u}^{(p)}(\xi)=1-\frac{1-\upsilon_u}{B_u^2}+\frac
{\Gamma_{uv}\Lambda_v}{B_u^4}(\upsilon_v-\upsilon_u)\left(1-\frac16\xi^2
\right)~~; \\
\label{eq:thp2}
&& \theta_{2,u}^{(p)}(\xi)=\frac{\Gamma_{uv}\Lambda_v}{B_u^4}\frac
{\upsilon_v-\upsilon_u}6\xi^2~~;
\end{lefteqnarray}
finally, the substitution of Eqs.\,(\ref{eq:tec12l})-(\ref{eq:thp2}) into
(\ref{eq:thc1}) yields:
\begin{lefteqnarray}
\label{eq:thc2}
&& \theta_u(\xi_u,\mu)=\sum_{\ell=0}^{+\infty}\left\{D_{2\ell}\hat{J}_
{+2\ell+1/2}(B_u\xi_u)+\delta_{2\ell,0}\left[1-\frac{1-\upsilon_u}{B_u^2}+
\frac{\Gamma_{uv}\Lambda_v}{B_u^4}(\upsilon_v-\upsilon_u)
\right.\right. \nonumber \\
&& \phantom{\theta_u(\xi_u,\mu)=\sum_{\ell=0}^{+\infty}}\left.\left.\times
\left(1-\frac16B_u^2
\xi_u^2\right)\right]+\delta_{2\ell,2}\frac{\Gamma_{uv}\Lambda_v}{B_u^4}\frac
{\upsilon_v-\upsilon_u}6B_u^2\xi_u^2\right\}P_{2\ell}(\mu)~~;
\end{lefteqnarray}
which is an exact solution of Eq.\,(\ref{eq:ECc2}).   The constant, $D_0$, can
be inferred from the boundary conditions at the origin via
Eqs.\,(\ref{eq:EC2b}) and (\ref{eq:Jt}), the last implying
$\hat{J}_{+2\ell+1/2}(0)=\delta_{2\ell,0}$.   After little algebra,
the result is:
\begin{equation}
\label{eq:D0uc}
D_0=\frac{1-\upsilon_u}{B_u^2}-\frac{\Gamma_{uv}\Lambda_v}{B_u^4}
(\upsilon_v-\upsilon_u)~~;
\end{equation}
in the limit of a vanishing $v$ subsystem, $\Lambda_v\to0$, $\Lambda_u\to1$,
$B_u\to1$, which yields $D_0\to1-\upsilon_u$, as expected from EC1 polytropes
[4].

Related expressions of the gravitational potential and radial component of the
gravitational force, via Eqs.\,(\ref{eq:VGd}) 
and (\ref{eq:cccC}) written in Appendix \ref{a:A2}, read:
\begin{lefteqnarray}
\label{eq:VGdc1}
&& {\cal V}_G(\xi_u,\mu)=4\pi G\sum(\lambda_w)\alpha_u^2\left\{\sum_{\ell=0}^
{+\infty}\left[D_{2\ell}\hat{J}_{+2\ell+1/2}(B_u\xi_u)
\right.\right.\nonumber \\
&& \phantom{}\left.\left.+
\delta_{2\ell,0}\left[1-\frac{1-\upsilon_u}{B_u^2}+\frac{\Gamma_{uv}\Lambda_v}
{B_u^4}(\upsilon_v-\upsilon_u)\left(1-\frac16B_u^2\xi_u^2\right)\right]
\right.\right. \nonumber \\
&& \phantom{}\left.\left.+
\delta_{2\ell,2}\frac{\Gamma_{uv}\Lambda_v}{B_u^4}\frac{\upsilon_v-\upsilon_u}
6B_u^2\xi_u^2\right\}P_{2\ell}(\mu)-\frac16\upsilon_u\xi_u^2[1-P_2(\mu)]+
c_{{\rm b},u}^\dagger\right\};\qquad \\
\label{eq:FGdc1}
&& \frac{\partial{\cal V}_G}{\partial r}=\frac1{\alpha_u}\frac
{\partial{\cal V}_G}{\partial\xi_u}=4\pi G\sum(\lambda_w)\alpha_u\left\{
\sum_{\ell=0}^{+\infty}\left[D_{2\ell}\hat{J}_{+2\ell+1/2}^\prime
(B_u\xi_u)\right.\right.\nonumber \\
&& \phantom{}\left.\left.-
\delta_{2\ell,0}\frac{\Gamma_{uv}\Lambda_v}{B_u^4}(\upsilon_v-\upsilon_u)\frac
13B_u^2\xi_u+\delta_{2\ell,2}\frac{\Gamma_{uv}\Lambda_v}{B_u^4}\frac
{\upsilon_v-\upsilon_u}3B_u^2\xi_u\right\}P_{2\ell}(\mu) \right.\nonumber \\
&& \phantom{}\left.-
\frac13\upsilon_u\xi_u[1-P_2(\mu)]\right\};\qquad
\end{lefteqnarray}
where the prime denotes derivation with respect to $\xi_u$.   The counterparts
of Eqs.\,(\ref{eq:VGdc1}) and (\ref{eq:FGdc1}), related to the noncommon
region, are expressed by Eqs.\,(\ref{eq:VGdn1}) and (\ref{eq:FGdn1}),
respectively.

The continuity of the gravitational potential and the radial component of the
gravitational force on
the interface along a selected direction, $\xi_u^\ast=\xi_u^\ast(\mu)$,
via Eqs.\,(\ref{eq:tin12l}), (\ref{eq:din12l}), (\ref{eq:thc2}),
(\ref{eq:D0uc}), implies the following systems of equations:
\begin{lefteqnarray}
\label{eq:sit0}
&& \cases {
D_0^{\rm(com)}\left[\hat{J}_{+1/2}(B_u\xi_u^\ast)-1\right]+1-\frac
{\Gamma_{uv}\Lambda_v}{B_u^2}\frac{\upsilon_v-\upsilon_u}6(\xi_u^\ast)^2
& \cr
=D_0^{\rm(ncm)}\hat{J}_{+1/2}(\Lambda_u^{1/2}\xi_u^\ast)+
\overline{D}_0\hat{J}_{-1/2}(\Lambda_u^{1/2}\xi_u^\ast)+\frac{\upsilon_u}
{\Lambda_u}~~;
 & \cr
 & \cr
D_0^{\rm(com)}\hat{J}_{+1/2}^{\prime}(B_u\xi_u^\ast)-\frac{\Gamma_{uv}\Lambda_v}
{B_u^2}\frac{\upsilon_v-\upsilon_u}3\xi_u^\ast
& \cr
=D_0^{\rm(ncm)}\hat{J}_{+1/2}^
{\prime}(\Lambda_u^{1/2}\xi_u^\ast)+\overline{D}_0\hat{J}_{-1/2}^{\prime}
(\Lambda_u^{1/2}\xi_u^\ast)~~; & \cr
} \\
\label{eq:sit2}
&& \cases {
D_2^{\rm(com)}\hat{J}_{+5/2}(B_u\xi_u^\ast)+\frac
{\Gamma_{uv}\Lambda_v}{B_u^2}\frac{\upsilon_v-\upsilon_u}6(\xi_u^\ast)^2
 & \cr
=D_2^{\rm(ncm)}\hat{J}_{+5/2}(\Lambda_u^{1/2}\xi_u^\ast)+
\overline{D}_2\hat{J}_{-5/2}(\Lambda_u^{1/2}\xi_u^\ast)~~;
 & \cr
 & \cr
D_2^{\rm(com)}\hat{J}_{+5/2}^{\prime}(B_u\xi_u^\ast)+\frac{\Gamma_{uv}\Lambda_v}
{B_u^2}\frac{\upsilon_v-\upsilon_u}3\xi_u^\ast
 & \cr
=D_2^{\rm(ncm)}\hat{J}_{+5/2}^
{\prime}(\Lambda_u^{1/2}\xi_u^\ast)+\overline{D}_2\hat{J}_{-5/2}^{\prime}
(\Lambda_u^{1/2}\xi_u^\ast)~~; & \cr
} \\
\label{eq:sit2l}
&& \cases {
D_{2\ell}^{\rm(com)}\hat{J}_{+2\ell+1/2}(B_u\xi_u^\ast)
 & \cr
=D_{2\ell}^{\rm(ncm)}\hat{J}_{+2\ell+1/2}(\Lambda_u^{1/2}\xi_u^\ast)+
\overline{D}_{2\ell}\hat{J}_{-2\ell-1/2}(\Lambda_u^{1/2}\xi_u^\ast)~~;
 & \cr
 & \cr
D_{2\ell}^{\rm(com)}\hat{J}_{+2\ell+1/2}^{\prime}(B_u\xi_u^\ast)
 & \cr
=D_{2\ell}^{\rm(ncm)}\hat{J}_{+2\ell+1/2}^{\prime}(\Lambda_u^{1/2}\xi_u^\ast)+
\overline{D}_{2\ell}\hat{J}_{-2\ell-1/2}^{\prime}(\Lambda_u^{1/2}\xi_u^\ast)~~;
& \cr
}
\end{lefteqnarray}
which make explicit expressions of
Eqs.\,(\ref{eq:tin12l})-(\ref{eq:din12l}).

The substitution of Eq.\,(\ref{eq:D0uc}) into (\ref{eq:sit0}) yields a system
of two equations in two unknowns, $D_0^{\rm(ncm)}$, $\overline{D}_0$, the
solution of which can be determined using standard methods via 
Eqs.\,(\ref{eq:sis2l})-(\ref{eq:Jt10}) and (\ref{eq:det01})-(\ref{eq:dts0}),
where $\theta_0(\xi_u^\ast)$, $\theta_0^\prime(\xi_u^\ast)$, can be
written explicitly.   With regard to Eq.\,(\ref{eq:sit2l}),
$D_{2\ell}^{\rm(ncm)}=0$, $\overline{D}_{2\ell}=0$, $2\ell>2$, via 
Eqs.\,(\ref{eq:tin12l})-(\ref{eq:din12l}) and (\ref{eq:ACsi}) written in
Appendix
\ref{a:A2}, which implies $D_{2\ell}^{\rm(com)}=0$, $2\ell>2$.   Concerning
Eq.\,(\ref{eq:sit2}), $D_2^{\rm(ncm)}$ and $\overline{D}_2$ can be expressed
in terms of $D_2^{\rm(com)}$ by solving related system.

On the other hand, the substitution of Eq.\,(\ref{eq:thp2}) into
(\ref{eq:tec12l}) yields:
\begin{lefteqnarray}
\label{eq:A2t2uc}
&& A_2\theta_{2,u}^{\rm(com)}(\xi_u)=D_2^{\rm(com)}\left[\hat{J}_{+5/2}
(B_u\xi_u)+\frac{\Gamma_{uv}\Lambda_v}{B_u^2D_2^{\rm(com)}}
\frac{\upsilon_v-\upsilon_u}6\xi_u^2\right]~~;
\end{lefteqnarray}
where, without loss of generality:
\begin{lefteqnarray}
\label{eq:t2ud}
&& \theta_{2,u}^{\rm(com)}(\xi_u)=\hat{J}_{+5/2}
(B_u\xi_u)+\frac{\Gamma_{uv}\Lambda_v}{B_u^2D_2^{\rm(com)}}
\frac{\upsilon_v-\upsilon_u}6\xi_u^2~~; \\
\label{eq:D2com}
&& D_2^{\rm(com)}=A_2~~;
\end{lefteqnarray}
and the substitution of Eqs.\,(\ref{eq:t2ud}) and (\ref{eq:D2com}) into
(\ref{eq:sit2}) produces:
\begin{lefteqnarray}
\label{eq:siu2}
&& \cases {
\hat{J}_{+5/2}(B_u\xi_u^\ast)+\frac
{\Gamma_{uv}\Lambda_v}{B_u^2A_2}\frac{\upsilon_v-\upsilon_u}6(\xi_u^\ast)^2
 & \cr
=D_2^\ast\hat{J}_{+5/2}(\Lambda_u^{1/2}\xi_u^\ast)+
\overline{D}_2^\ast\hat{J}_{-5/2}(\Lambda_u^{1/2}\xi_u^\ast)~~;
 & \cr
 & \cr
\hat{J}_{+5/2}^{\prime}(B_u\xi_u^\ast)+\frac{\Gamma_{uv}\Lambda_v}
{B_u^2A_2}\frac{\upsilon_v-\upsilon_u}3\xi_u^\ast
 & \cr
=D_2^\ast\hat{J}_{+5/2}^
{\prime}(\Lambda_u^{1/2}\xi_u^\ast)+\overline{D}_2^\ast\hat{J}_{-5/2}^{\prime}
(\Lambda_u^{1/2}\xi_u^\ast)~~; & \cr
}
\end{lefteqnarray}
where $D_2^\ast$, $\overline{D}_2^\ast$, are defined by Eq.\,(\ref{eq:D2la})
and $A_2$, in turn, depends on $D_2^{\rm(ncm)}$, $\overline{D}_2$, via
Eqs.\,(\ref{eq:tin12l})-(\ref{eq:din12l}) and (\ref{eq:ACsi}) written in
Appendix \ref{a:A2}.   Then the
system has to be solved through successive iterations in $A_2$, selecting an
appropriate initial value, up to convergence within an assigned
tolerance.

The EC2 associated functions, $\theta_{2\ell,u}(\xi_u)$, via
Eqs.\,(\ref{eq:thc1}), (\ref{eq:thc2}), (\ref{eq:D0uc}), after some algebra
take the explicit expression:
\begin{lefteqnarray}
\label{eq:ta0uc}
&& A_0\theta_{0,u}(\xi_u)=1+D_0\left[\hat{J}_{+1/2}(B_u\xi_u)-1\right]-
\frac{\Gamma_{uv}\Lambda_v}{B_u^2}\frac{\upsilon_v-\upsilon_u}6\xi_u^2~~; \\
\label{eq:ta02c}
&& A_{2}\theta_{2,u}(\xi_u)=D_{2}\hat{J}_{+5/2}(B_u\xi_u)+
\frac{\Gamma_{uv}\Lambda_v}{B_u^2}\frac{\upsilon_v-\upsilon_u}6\xi_u^2~~; \\
\label{eq:ta02lc}
&& A_{2\ell}\theta_{2\ell,u}(\xi_u)=D_{2\ell}\hat{J}_{+2\ell+1/2}
(B_u\xi_u)~~;\qquad2\ell>2~~;
\end{lefteqnarray}
where $A_{2\ell}=D_{2\ell}$,
$\theta_{2\ell,u}(\xi_u)=\hat{J}_{+2\ell+1/2}(B_u\xi_u)$, $2\ell>2$, with no
loss of generality.

The EC2 associated functions, $\theta_{2\ell,u}(\xi_u)$, can be expanded in
Taylor series
only in the special case of the singular starting point, $\xi_{0,u}=0$, where
the convergence radius is infinite.
Restricting to the cases of interest, $2\ell=0,2$, the trigonometric functions
appearing in Eqs.\,(\ref{eq:ta0uc})-(\ref{eq:ta02c}) via (\ref{eq:Jt}) and
(\ref{eq:Jt10})-(\ref{eq:Jt12p}) can be replaced by corresponding MacLaurin
series expansions and, after some algebra, the comparison with related
MacLaurin series expansions, Eq.\,(\ref{eq:sers0}), yields: 
\begin{leftsubeqnarray}
\slabel{eq:a0k11a}
&& A_0a_{0,2k+2}^{(u,u)}=(-1)^{k+2}\frac{D_0B_u^{2k+2}}{(2k+3)!}-\delta_{2k,2}
\frac{\Gamma_{uv}\Lambda_v}{B_u^2}\frac{\upsilon_v-\upsilon_u}6~~;\qquad  \\
\slabel{eq:a0k11b}
&& A_0a_{0,2k+1}^{(u,u)}=0~~;\qquad2k\ge0~~;\qquad A_0=1~~;
\label{seq:a0k11}
\end{leftsubeqnarray}
\begin{leftsubeqnarray}
\slabel{eq:a2k11a}
&& A_{2}a_{2,2k+2}^{(u,u)}=(-1)^{k+2}15\frac{(2k+2)(2k+4)}{(2k+5)!}B_u^{2k+2}+
\delta_{2k,2}\frac{\Gamma_{uv}\Lambda_v}{B_u^2}\frac{\upsilon_v-\upsilon_u}6
~~;\qquad \\
\slabel{eq:a2k11b}
&& A_{2}a_{2,2k+1}^{(u,u)}=0~~;\qquad2k\ge0~~;
\label{seq:a2k11}
\end{leftsubeqnarray}
where, unfortunately, the above results cannot be extended to the general case
of starting point, $\xi_{0,u}>0$, and 
the coefficients of Taylor series expansions, Eq.\,(\ref{eq:sersc}), cannot be
expressed in simpler form with respect to the regression formulae,
Eqs.\,(\ref{seq:a0kc}) and (\ref{seq:a2kc}).   For further details, an
interested reader is addressed to the parent paper [6].

Let $\theta_{0,v}(\Xi_{{\rm ex},v})=\theta_v(\Xi_v,\mu)=\theta_{{\rm b},v}$ be
the (fictitious) spherical isopycnic surface of the expanded sphere, related
to the interface.   By use of Eqs.\,(\ref{eq:Jt10}) and (\ref{eq:ta0uc}),
an explicit expression reads:
\begin{equation}
\label{eq:v10ex}
1+D_0\left[\frac{\sin(B_v\xi_v)}{B_v\xi_v}-1\right]-
\frac{\Gamma_{vu}\Lambda_u}{B_v^2}\frac{\upsilon_u-\upsilon_v}6\xi_v^2=
\theta_{{\rm b},v}~~;
\end{equation}
which is a transcendental equation where the scaled radius,
$\Xi_{{\rm ex},v}$, is the lowest positive solution.

Concerning the special case, $(n_v,n_u)=(1,0)$, the EC2 equation,
Eq.\,(\ref{eq:ECc0}), reduces to:
\begin{lefteqnarray}
\label{eq:ED10}
&& \frac1{\xi^2}\frac\partial{\partial\xi}\left(\xi^2
\frac{\partial\theta}{\partial\xi}\right)+
\frac1{\xi^2}\frac\partial{\partial\mu}\left[(1-\mu^2)\frac{\partial
\theta}{\partial\mu}\right]-\upsilon^\prime=-\theta~~;
\end{lefteqnarray}
where $\theta=\theta_v$, $\xi=\Lambda_v^{1/2}\xi_v$,
$\upsilon^\prime=\upsilon-\Lambda_{uv}$, $\upsilon=\upsilon_v/\Lambda_v$, via
Eqs.\,(\ref{eq:dimq}), (\ref{eq:snnr}), (\ref{eq:upp}), and
the parameter, $\upsilon^\prime$, can be conceived as a generalized distortion
including both centrifugal and tidal effects.

Keeping in mind the boundary conditions, Eq.\,(\ref{eq:EC2b}), which implies
absence of divergent terms at the origin, the solution of Eq.\,(\ref{eq:ED10})
via (\ref{eq:Ltot}) and (\ref{eq:thn2}) reads:
\begin{lefteqnarray}
\label{eq:thv12}
&& \theta(\xi,\mu)=\sum_{\ell=0}^{+\infty}D_{2\ell}\hat{J}_{+2\ell+1/2}(\xi)
P_{2\ell}(\mu)+\upsilon^\prime~~; \\
\label{eq:Jtp12}
&& \hat{J}_{+2\ell+1/2}(0)=\delta_{2\ell,0}~~; \\
\label{eq:D0v12}
&& D_0=1-\upsilon^\prime=\frac{1-\upsilon_v}{\Lambda_v}~~;
\end{lefteqnarray}
according to the properties of the Bessel functions of half-integer degree
e.g., [37], Chap.\,24.

In terms of the subsystem, $u$,
Eq.\,(\ref{eq:thv12}) via (\ref{eq:ispuv}) translates into:
\begin{lefteqnarray}
\label{eq:thu12}
&& \theta_u(\xi_u,\mu)=\Gamma_{vu}\sum_{\ell=0}^{+\infty}D_{2\ell}
\hat{J}_{+2\ell+1/2}(\Gamma_{uv}^{1/2}\Lambda_v^{1/2}\xi_u)P_{2\ell}(\mu)+
\Gamma_{vu}\frac{\upsilon_v-\Lambda_u}{\Lambda_v}+1-\Gamma_{vu} \nonumber \\
&& \phantom{\theta_u(\xi_u,\mu)=}
+\frac{\upsilon_u-\upsilon_v}6\xi_u^2\left[1-P_2(\mu)\right]~~;
\end{lefteqnarray}
and related expressions of the gravitational potential and radial component of the
gravitational force, via Eqs.\,(\ref{eq:csir}), (\ref{eq:VGd}),
(\ref{eq:Gauv}), and (\ref{eq:cccC}) written in Appendix \ref{a:A2}, read:
\begin{lefteqnarray}
\label{eq:VG10}
&& {\cal V}_G(\xi_u,\mu)=4\pi G\sum(\lambda_w)\alpha_u^2\left\{\Gamma_{vu}
\sum_{\ell=0}^{+\infty}D_{2\ell}\hat{J}_{+2\ell+1/2}(\Gamma_{uv}^{1/2}
\Lambda_v^{1/2}\xi_u)P_{2\ell}(\mu)
\right.\nonumber \\
&& \phantom{{\cal V}_G(\xi_u,\mu)=}\left.+
1-\Gamma_{vu}+\Gamma_{vu}\frac{\upsilon_v-\Lambda_u}{\Lambda_v}
-\frac16\upsilon_v\xi_u^2[1-P_2(\mu)]+c_{{\rm b},u}^\dagger\right\};\qquad \\
\label{eq:FG10}
&& \frac{\partial{\cal V}_G}{\partial r}=\frac1{\alpha_u}\frac
{\partial{\cal V}_G}{\partial\xi_u}=4\pi G\sum(\lambda_w)\alpha_u\left\{
\Gamma_{vu}\sum_{\ell=0}^{+\infty}D_{2\ell}\hat{J}_{+2\ell+1/2}^\prime
(\Gamma_{uv}^{1/2}\Lambda_v^{1/2}\xi_u)P_{2\ell}(\mu)\right.\nonumber \\
&& \phantom{\frac{\partial{\cal V}_G}{\partial r}=\frac1{\alpha_u}\frac
{\partial{\cal V}_G}{\partial\xi_u}=}\left.-
\frac13\upsilon_v\xi_u\left[1-P_2(\mu)\right]\right\};\qquad
\end{lefteqnarray}
where the prime denotes derivation with respect to $\xi_u$.   The counterparts
of Eqs.\,(\ref{eq:VG10}) and (\ref{eq:FG10}), related to the noncommon
region, are expressed by Eqs.\,(\ref{eq:VGn0}) and (\ref{eq:FGn0}),
respectively.

The continuity of the gravitational potential and the radial component of the
gravitational force on
the interface along a selected direction, $\xi_u^\ast=\xi_u^\ast(\mu)$,
via Eqs.\,(\ref{eq:VGn0}), (\ref{eq:FGn0}), implies the following systems of
equations:
\begin{lefteqnarray}
\label{eq:tit0}
&& \cases {
\Gamma_{vu}D_0^{\rm(com)}\hat{J}_{+1/2}(\Gamma_{uv}^{1/2}\Lambda_v^{1/2}
\xi_u^\ast)+\Gamma_{vu}\frac{\upsilon_v-\Lambda_u}{\Lambda_v}+1-\Gamma_{vu}-
\frac16\upsilon_v(\xi_u^\ast)^2+c_{{\rm b},u}^\dagger
& \cr
=D_0^{\rm(ncm)}+\frac{\overline{D}_0}{\xi_u^\ast}-\frac16\Lambda_u
(\xi_u^\ast)^2+c_{{\rm b},u}^\dagger~~;
 & \cr
 & \cr
\Gamma_{vu}D_0^{\rm(com)}\hat{J}_{+1/2}^{\prime}(\Gamma_{uv}^{1/2}\Lambda_v^
{1/2}\xi_u^\ast)-\frac13\upsilon_v\xi_u^\ast
=-\frac{\overline{D}_0}{(\xi_u^\ast)^2}-\frac13\Lambda_u\xi_u^\ast~~;
& \cr
} \\
\label{eq:tit2}
&& \cases {
\Gamma_{vu}D_2^{\rm(com)}\hat{J}_{+5/2}(\Gamma_{uv}^{1/2}\Lambda_v^{1/2}
\xi_u^\ast)+\frac16\upsilon_v(\xi_u^\ast)^2
 & \cr
=D_2^{\rm(ncm)}(\xi_u^\ast)^2+\frac{\overline{D}_2}{(\xi_u^\ast)^3}+\frac16
\Lambda_u(\xi_u^\ast)^2~~;
 & \cr
 & \cr
\Gamma_{vu}D_2^{\rm(com)}\hat{J}_{+5/2}^{\prime}(\Gamma_{uv}^{1/2}
\Lambda_v^{1/2}\xi_u^\ast)+\frac13\upsilon_v\xi_u^\ast
=2D_2^{\rm(ncm)}\xi_u^\ast-\frac{3\overline{D}_2}{(\xi_u^\ast)^4}+\frac13
\Lambda_u\xi_u^\ast~~; & \cr
} \\
\label{eq:tit2l}
&& \cases {
\Gamma_{vu}D_{2\ell}^{\rm(com)}\hat{J}_{+2\ell+1/2}(\Gamma_{uv}^{1/2}\Lambda_v
^{1/2}\xi_u^\ast)
=D_{2\ell}^{\rm(ncm)}(\xi_u^\ast)^{2\ell}+\frac{\overline{D}_{2\ell}}
{(\xi_u^\ast)^{2\ell+1}}~~;
 & \cr
 & \cr
\Gamma_{vu}2\ell D_{2\ell}^{\rm(com)}\hat{J}_{+2\ell+1/2}^{\prime}
(\Gamma_{uv}^{1/2}\Lambda_v^{1/2}\xi_u^\ast)
=2\ell D_{2\ell}^{\rm(ncm)}(\xi_u^\ast)^{2\ell-1}-
\frac{(2\ell+1)\overline{D}_{2\ell}}{(\xi_u^\ast)^{2\ell+2}}~~;
& \cr
}
\end{lefteqnarray}
in three unknowns, $D_{2\ell}^{\rm(com)}$, $D_{2\ell}^{\rm(ncm)}$,
$\overline{D}_{2\ell}$.

The substitution of Eq.\,(\ref{eq:D0v12}) into (\ref{eq:tit0}) yields an
explicit expression of $D_0^{\rm(ncm)}$, $\overline{D}_0$, as:
\begin{lefteqnarray}
\label{eq:Db010}
&& \overline{D}_0=(\xi_u^\ast)^2\left[-\Gamma_{vu}\frac{1-\upsilon_v}
{\Lambda_v}\hat{J}_{+1/2}^\prime(\Gamma_{uv}^{1/2}\Lambda_v^{1/2}\xi_u^\ast)+
\frac{\upsilon_v-\Lambda_u}3\xi_u^\ast\right]~~; \\
\label{eq:D010}
&& D_0^{\rm(ncm)}=\Gamma_{vu}\frac{1-\upsilon_v}{\Lambda_v}\left[
\hat{J}_{+1/2}(\Gamma_{uv}^{1/2}\Lambda_v^{1/2}\xi_u^\ast)+\xi_u^\ast
\hat{J}_{+1/2}^\prime(\Gamma_{uv}^{1/2}\Lambda_v^{1/2}\xi_u^\ast)\right]+
\Gamma_{vu}\frac{\upsilon_v-\Lambda_u}{\Lambda_v} \nonumber \\
&& \phantom{D_0^{\rm(ncm)}=}
+1-\Gamma_{vu}-\frac{\upsilon_v-\Lambda_u}2(\xi_u^\ast)^2~~;
\end{lefteqnarray}
which can be determined via Eq.\,(\ref{eq:Jt10}).  With regard to
Eq.\,(\ref{eq:tit2l}),
$D_{2\ell}^{\rm(ncm)}=0$, $\overline{D}_{2\ell}=0$, $2\ell>2$, via
Eqs.\,(\ref{eq:Dw2l0}) and (\ref{eq:ECn0u}), respectively, which implies
$D_{2\ell}^{\rm(com)}=0$, $2\ell>2$.   Concerning 
Eq.\,(\ref{eq:tit2}), $D_2^{\rm(ncm)}$ and $\overline{D}_2$ can be expressed
in terms of $D_2^{\rm(com)}$ by solving related system.

On the other hand, the comparison of Eq.\,(\ref{eq:thv12}) with
(\ref{eq:thc}), with regard to the terms of second order in Legendre
polynomials, yields:
\begin{lefteqnarray}
\label{eq:A2t2vo}
&& A_2\theta_{2,v}^{\rm(com)}(\xi_v)=D_2^{\rm(com)}\hat{J}_{+5/2}
(\Lambda_v^{1/2}\xi_v)~~;
\end{lefteqnarray}
and the substitution of Eq.\,(\ref{eq:A2t2vo}) into (\ref{eq:isq2l}) written
in Appendix \ref{a:seco} after some algebra produces:
\begin{lefteqnarray}
\label{eq:A2t2ul}
&& A_2\theta_{2,u}^{\rm(com)}(\xi_u)=\Gamma_{vu}D_2^{\rm(com)}\hat{J}_{+5/2}
(\Gamma_{uv}^{1/2}\Lambda_u^{1/2}\xi_u)+D_2^{\rm(com)}\frac
{\upsilon_v-\upsilon_u}{6D_2^{\rm(com)}}\xi_u^2~~;
\end{lefteqnarray}
where, without loss of generality:
\begin{lefteqnarray}
\label{eq:t2ue}
&& \theta_{2,u}^{\rm(com)}(\xi_u)=\Gamma_{vu}\hat{J}_{+5/2}
(\Gamma_{uv}^{1/2}\Lambda_u^{1/2}\xi_u)+\frac{\upsilon_v-\upsilon_u}{6A_2}
\xi_u^2~~; \\
\label{eq:D2dom}
&& D_2^{\rm(com)}=A_2~~;
\end{lefteqnarray}
accordingly, Eq.\,(\ref{eq:tit2}) may be cast into the equivalent form:
\begin{lefteqnarray}
\label{eq:tit3}
&& \cases {
D_2^\ast(\xi_u^\ast)^2+\frac{\overline{D}_2^\ast}{(\xi_u^\ast)^3}
=\Gamma_{vu}\hat{J}_{+5/2}(\Gamma_{uv}^{1/2}\Lambda_u^{1/2}\xi_u^\ast)+
\frac{\upsilon_v-\Lambda_u}{6A_2}(\xi_u^\ast)^2~~;
 & \cr
 & \cr
2D_2^\ast\xi_u^\ast-\frac{3\overline{D}_2^\ast}{(\xi_u^\ast)^4}
=\Gamma_{vu}\hat{J}_{+5/2}^{\prime}(\Gamma_{uv}^{1/2}\Lambda_u^{1/2}\xi_u^\ast)
+\frac{\upsilon_v-\Lambda_u}{3A_2}\xi_u^\ast~~;
 & \cr
}
\end{lefteqnarray}
where $D_2^\ast$, $\overline{D}_2^\ast$, are defined by Eq.\,(\ref{eq:D2la})
and $A_2$, in turn, depends on $D_2^{\rm(ncm)}$, $\overline{D}_2$, via
Eqs.\,(\ref{eq:A2t2ul}) and (\ref{eq:ACsi}) written in Appendix \ref{a:A2}.
Then the
system has to be solved through successive iterations in $A_2$, selecting an
appropriate initial value, up to convergence within an assigned tolerance.

Accordingly, Eq.\,(\ref{eq:h20}) reads:
\begin{lefteqnarray}
\label{eq:h20a}
&& \theta_{2,u}^{({\rm ncm})}(\xi_u)=D_2^\ast\xi_u^2+\frac{\overline{D}_2^
\ast}{\xi_u^3}-\frac{\upsilon_u-\Lambda_u}{6A_2}\xi_u^2~~;
\end{lefteqnarray}
which allows the determination of $\theta_{2,u}(\Xi_u)$,
$\theta_{2,u}^\prime(\Xi_u)$, on a selected point of the boundary,
$\Xi_u=\Xi_u(\mu)$, and the next iteration value of $A_2$ via
Eq.\,(\ref{eq:ACsi}) written in Appendix \ref{a:A2}.

The EC2 associated functions, $\theta_{2\ell,u}(\xi_u)$, via
Eqs.\,(\ref{eq:thc1}), (\ref{eq:thv12}), (\ref{eq:D0v12}), after some algebra
take the explicit expression:
\begin{lefteqnarray}
\label{eq:ta10vc}
&& A_0\theta_{0,v}(\xi_v)=1+\frac{1-\upsilon_v}{\Lambda_v}\left[\hat{J}_{+1/2}
(\Lambda_v^{1/2}\xi_v)-1\right]~~;\qquad A_0=1~~; \\
\label{eq:ta12lvc}
&& A_{2\ell}\theta_{2\ell,v}(\xi_v)=D_{2\ell}\hat{J}_{+2\ell+1/2}
(\Lambda_v^{1/2}\xi_v)~~;\qquad2\ell>0~~;
\end{lefteqnarray}
where $A_{2\ell}=D_{2\ell}$,
$\theta_{2\ell,v}(\xi_v)=\hat{J}_{+2\ell+1/2}(\Lambda_v^{1/2}\xi_v)$,
$2\ell>0$, with no loss of generality.   With regard to $u$ subsystem,
Eqs.\,(\ref{eq:ta10vc})-(\ref{eq:ta12lvc}) can be rewritten by use of
Eq.\,(\ref{eq:isp2l}) appearing in Appendix \ref{a:seco}.   In the limit of a
vanishing $u$
subsystem, $\Lambda_u\to0$, $\Lambda_v\to1$, the associated EC2 functions,
$\theta_{2\ell,v}(\xi_v)$, coincide with their counterparts related to EC1
polytropes [4], as expected.   

The EC2 associated functions, $\theta_{2\ell,v}(\xi_v)$, can be expanded in
Taylor series
only in the special case of the singular starting point, $\xi_{0,v}=0$, where
the convergence radius is infinite.
Restricting to the cases of interest, $2\ell=0,2$, the trigonometric functions
appearing in Eqs.\,(\ref{eq:ta10vc})-(\ref{eq:ta12lvc}) via (\ref{eq:Jt}) and
(\ref{eq:Jt10})-(\ref{eq:Jt12p}) can be replaced by corresponding MacLaurin
series expansions and, after some algebra, the comparison with related
MacLaurin series expansions, Eq.\,(\ref{eq:sers0}), yields: 
\begin{leftsubeqnarray}
\slabel{eq:a00k11a}
&& A_0a_{0,2k+2}^{(v,v)}=(-1)^{k+2}\frac{1-\upsilon_v}{\Lambda_v}\frac
{\Lambda_v^{k+1}}{(2k+3)!}~~; \\
\slabel{eq:a00k11b}
&& A_0a_{0,2k+1}^{(v,v)}=0~~;\qquad2k\ge0~~;\qquad A_0=1~~;
\label{seq:a00k11}
\end{leftsubeqnarray}
\begin{leftsubeqnarray}
\slabel{eq:a22k11a}
&& A_{2}a_{2,2k+2}^{(v,v)}=(-1)^{k+2}15\frac{(2k+2)(2k+4)}{(2k+5)!}
\Lambda_v^{k+1}~~; \\
\slabel{eq:a22k11b}
&& A_{2}a_{2,2k+1}^{(v,v)}=0~~;\qquad2k\ge0~~;
\label{seq:a22k11}
\end{leftsubeqnarray}
where, unfortunately, the above results cannot be extended to the general case
of starting point, $\xi_{0,v}>0$, and 
the coefficients of Taylor series expansions, Eq.\,(\ref{eq:sersc}), cannot be
expressed in simpler form with respect to the regression formulae,
Eqs.\,(\ref{seq:a0kc}) and (\ref{seq:a2kc}).   For further details, an
interested reader is addressed to the parent paper [6].

Let $\theta_{0,v}(\Xi_{{\rm ex},v})=\theta_v(\Xi_v,\mu)=\theta_{{\rm b},v}$ be
the (fictitious) spherical isopycnic surface of the expanded sphere, related
to the interface.   By use of Eqs.\,(\ref{eq:Jt10}) and (\ref{eq:ta10vc}),
an explicit expression reads:
\begin{equation}
\label{eq:v10e2}
1+\frac{1-\upsilon_v}{\Lambda_v}\left[\frac{\sin(\Lambda_v^{1/2}\xi_v)}
{\Lambda_v^{1/2}\xi_v}-1\right]=\theta_{{\rm b},v}~~;
\end{equation}
which is a transcendental equation where the dimensionless radius,
$\Xi_{{\rm ex},v}$, is the lowest positive solution.

Concerning the special case, $(n_v,n_u)=(0,1)$, 
Eq.\,(\ref{eq:ECc0}) reduces to (\ref{eq:ED10}) where $\theta=\theta_u$,
$\xi=\Lambda_u^{1/2}\xi_u$, $\upsilon^\prime=\upsilon-\Lambda_{vu}$,
$\upsilon=\upsilon_u/\Lambda_u$.   Then the solution of Eq.\,(\ref{eq:ED10})
is expressed by Eq.\,(\ref{eq:thv12}) where the constant, $D_0$, takes the
expression:
\begin{lefteqnarray}
\label{eq:D0c01}
&& D_0=1-\upsilon^\prime=\frac{1-\upsilon_u}{\Lambda_u}~~;
\end{lefteqnarray}
according to the properties of the Bessel functions of half-integer degree
e.g., [37], Chap.\,24.

Related expressions of the gravitational potential and radial component of the
gravitational force, via
Eqs.\,(\ref{eq:VGd}),
(\ref{eq:Gauv}), and (\ref{eq:cccC}) written in Appendix \ref{a:A2}, read:
\begin{lefteqnarray}
\label{eq:VG01}
&& {\cal V}_G(\xi_u,\mu)=4\pi G\sum(\lambda_w)\alpha_u^2\left\{
\sum_{\ell=0}^{+\infty}D_{2\ell}\hat{J}_{+2\ell+1/2}(\Lambda_u^{1/2}\xi_u)
P_{2\ell}(\mu)\right.\nonumber \\
&& \phantom{{\cal V}_G(\xi_u,\mu)=}\left.+
\frac{\upsilon_u-\Lambda_v}{\Lambda_u}-\frac16\upsilon_u\xi_u^2
[1-P_2(\mu)]+c_{{\rm b},u}^\dagger\right\};\qquad \\
\label{eq:FG01}
&& \frac{\partial{\cal V}_G}{\partial r}=\frac1{\alpha_u}\frac
{\partial{\cal V}_G}{\partial\xi_u}=4\pi G\sum(\lambda_w)\alpha_u\left\{
\sum_{\ell=0}^{+\infty}D_{2\ell}\hat{J}_{+2\ell+1/2}^\prime
(\Lambda_u^{1/2}\xi_u)P_{2\ell}(\mu)\right.\nonumber \\
&& \phantom{\frac{\partial{\cal V}_G}{\partial r}=\frac1{\alpha_u}\frac
{\partial{\cal V}_G}{\partial\xi_u}=}\left.-
\frac13\upsilon_u\xi_u\left[1-P_2(\mu)\right]\right\};\qquad
\end{lefteqnarray}
where the prime denotes derivation with respect to $\xi_u$.   The counterparts
of Eqs.\,(\ref{eq:VG01}) and (\ref{eq:FG01}), related to the noncommon
region, are expressed by Eqs.\,(\ref{eq:VGdn1}) and (\ref{eq:FGdn1}),
respectively.

The continuity of the gravitational potential and the gravitational force on
the interface along a selected direction, $\xi_u^\ast=\xi_u^\ast(\mu)$,
via Eqs.\,(\ref{eq:VGdn1}), (\ref{eq:FGdn1}), implies the following systems of
equations:
\begin{lefteqnarray}
\label{eq:tiu0}
&& \cases {
D_0^{\rm(com)}\hat{J}_{+1/2}(\Lambda_u^{1/2}\xi_u^\ast)+
\frac{\upsilon_u-\Lambda_v}{\Lambda_u}-
\frac16\upsilon_u(\xi_u^\ast)^2+c_{{\rm b},u}^\dagger
& \cr
=D_0^{\rm(ncm)}\hat{J}_{+1/2}(\Lambda_u^{1/2}\xi_u^\ast)
+\overline{D}_0\hat{J}_{-1/2}(\Lambda_u^{1/2}\xi_u^\ast)+\frac
{\upsilon_u}{\Lambda_u}-\frac16\upsilon_u(\xi_u^\ast)^2+
c_{{\rm b},u}^\dagger~~;
 & \cr
 & \cr
D_0^{\rm(com)}\hat{J}_{+1/2}^{\prime}(\Lambda_u^{1/2}\xi_u^\ast)-
\frac13\upsilon_u\xi_u^\ast
& \cr
=D_0^{\rm(ncm)}\hat{J}_{+1/2}^{\prime}(\Lambda_u^{1/2}\xi_u^\ast)+
\overline{D}_0\hat{J}_{-1/2}^{\prime}(\Lambda_u^{1/2}\xi_u^\ast)-
\frac13\upsilon_u\xi_u^\ast~~;
& \cr
} \\
\label{eq:tiu2l}
&& \cases {
D_{2\ell}^{\rm(com)}\hat{J}_{+2\ell+1/2}(\Lambda_u^{1/2}\xi_u^\ast)
+\delta_{2\ell,2}\frac16\upsilon_u(\xi_u^\ast)^2
 & \cr
=D_{2\ell}^{\rm(ncm)}\hat{J}_{+2\ell+1/2}(\Lambda_u^{1/2}\xi_u^\ast)+
\overline{D}_{2\ell}\hat{J}_{-2\ell-1/2}(\Lambda_u^{1/2}\xi_u^\ast)
+\delta_{2\ell,2}\frac16\upsilon_u(\xi_u^\ast)^2~~;
 & \cr
 & \cr
D_{2\ell}^{\rm(com)}\hat{J}_{+2\ell+1/2}^{\prime}(\Lambda_u^{1/2}\xi_u^\ast)
+\delta_{2\ell,2}\frac13\upsilon_u\xi_u^\ast
 & \cr
=D_{2\ell}^{\rm(ncm)}
\hat{J}_{+2\ell+1/2}^{\prime}(\Lambda_u^{1/2}\xi_u^\ast)+
\overline{D}_{2\ell}
\hat{J}_{-2\ell-1/2}^{\prime}(\Lambda_u^{1/2}\xi_u^\ast)
+\delta_{2\ell,2}\frac13\upsilon_u\xi_u^\ast~~;
& \cr
}
\end{lefteqnarray}
which, after some algebra, can be merged into a single system as:
\begin{lefteqnarray}
\label{eq:tiv2l}
&& \cases {
\left(D_{2\ell}^{\rm(ncm)}-D_{2\ell}^{\rm(com)}\right)
\hat{J}_{+2\ell+1/2}(\Lambda_u^{1/2}\xi_u^\ast)
+\overline{D}_{2\ell}\hat{J}_{-2\ell-1/2}(\Lambda_u^{1/2}\xi_u^\ast)
 & \cr
=-\delta_{2\ell,0}\Lambda_{vu}~~;
 & \cr
 & \cr
\left(D_{2\ell}^{\rm(ncm)}-D_{2\ell}^{\rm(com)}\right)
\hat{J}_{+2\ell+1/2}^{\prime}(\Lambda_u^{1/2}\xi_u^\ast)
+\overline{D}_{2\ell}
\hat{J}_{-2\ell-1/2}^{\prime}(\Lambda_u^{1/2}\xi_u^\ast)=0~~;
& \cr
}
\end{lefteqnarray}
in the two unknowns, $\left(D_{2\ell}^{\rm(ncm)}-D_{2\ell}^{\rm(com)}\right)$
and $\overline{D}_{2\ell}$.

In the special case, $2\ell=0$, the solution of Eq.\,(\ref{eq:tiv2l}) is
determined after some algebra as:
\begin{lefteqnarray}
\label{eq:seb0}
&& \overline{D}_0=\frac{\Lambda_{vu}\hat{J}_{+1/2}^\prime
(\Lambda_u^{1/2}\xi_u^\ast)}{{\cal D}_0^{(0)}}~~; \\
\label{eq:sed0}
&& D_0^{\rm(ncm)}-D_0^{\rm(com)}=-\frac{\Lambda_{vu}\hat{J}_{-1/2}^\prime
(\Lambda_u^{1/2}\xi_u^\ast)}{{\cal D}_0^{(0)}}~~; \\
\label{eq:det01b}
&& {\cal D}_0^{(0)}=\hat{J}_{+1/2}(\Lambda_u^{1/2}\xi_u^\ast)
\hat{J}_{-1/2}^\prime(\Lambda_u^{1/2}\xi_u^\ast)-
\hat{J}_{+1/2}^\prime(\Lambda_u^{1/2}\xi_u^\ast)
\hat{J}_{-1/2}(\Lambda_u^{1/2}\xi_u^\ast)~~;
\end{lefteqnarray}
according to standard methods.

Finally, the substitution of Eq.\,(\ref{eq:D0c01}) into (\ref{eq:sed0})
yields:
\begin{lefteqnarray}
\label{eq:sos0}
&& D_0^{\rm(ncm)}=-\frac{\Lambda_{vu}\hat{J}_{-1/2}^\prime(\Lambda_u^{1/2}
\xi_u^\ast)}{{\cal D}_0^{(0)}}+\frac{1-\upsilon_u}{\Lambda_u}~~;
\end{lefteqnarray}
and the EC2 associated function, $\theta_{0,u}$, can be explicitly expressed.
In the remaining cases, $2\ell>0$, related systems are made of linearly
independent equations.   Accordingly, only null solutions exist, hence
$D_{2\ell}^{\rm(ncm)}=D_{2\ell}^{\rm(com)}$, $\overline{D}_{2\ell}=0$.
Accordingly,the EC2 associated functions, $\theta_{2\ell,u}$,
$2\ell>0$, maintain their expression passing from the common to the noncommon
region, in the case under discussion.

Related explicit expressions, via Eqs.\,(\ref{eq:thc}), (\ref{eq:thv12}),
(\ref{eq:D0c01}), after some algebra read:
\begin{lefteqnarray}
\label{eq:ta1uc}
&& A_0\theta_{0,u}(\xi_u)=1+\frac{1-\upsilon_u}{\Lambda_u}\left[\hat{J}_{+1/2}
(\Lambda_u^{1/2}\xi_u)-1\right]~~;\qquad A_0=1~~; \\
\label{eq:ta12luc}
&& A_{2\ell}\theta_{2\ell,u}(\xi_u)=D_{2\ell}\hat{J}_{+2\ell+1/2}
(\Lambda_u^{1/2}\xi_u)~~;\qquad2\ell>0~~;
\end{lefteqnarray}
where $A_{2\ell}=D_{2\ell}$,
$\theta_{2\ell,u}(\xi_u)=\hat{J}_{+2\ell+1/2}(\Lambda_u^{1/2}\xi_u)$,
$2\ell>0$, with no loss of generality, via Eqs.\,(\ref{eq:D2dom}) and
(\ref{eq:ta12lvc}).   In the limit of a vanishing $v$
subsystem, $\Lambda_v\to0$, $\Lambda_u\to1$, the associated EC2 functions,
$\theta_{2\ell,u}(\xi_u)$, coincide with their counterparts related to EC1
polytropes [4], as expected.   

With regard to Taylor series expansions and dimensionless radius,
$\Xi_{{\rm ex},u}$, following the same procedure as in the case,
$(n_v,n_u)=(1,0)$, Eqs.\,(\ref{seq:a00k11})-(\ref{seq:a22k11}) and
(\ref{eq:v10e2}) hold provided the index, $v$, is replaced by $u$ therein.

Concerning the special case, $(n_v,n_u)=(5,0)$, the EC2 equation,
Eq.\,(\ref{eq:ECc0}), reduces to:
\begin{lefteqnarray}
\label{eq:ED50}
&& \frac1{\xi^2}\frac\partial{\partial\xi}\left(\xi^2
\frac{\partial\theta}{\partial\xi}\right)+
\frac1{\xi^2}\frac\partial{\partial\mu}\left[(1-\mu^2)\frac{\partial
\theta}{\partial\mu}\right]-\upsilon^\prime=-\theta^5~~;
\end{lefteqnarray}
where $\theta=\theta_v$, $\xi=\Lambda_v^{1/2}\xi_v$,
$\upsilon^\prime=\upsilon-\Lambda_{uv}$, $\upsilon=\upsilon_v/\Lambda_v$, via
Eqs.\,(\ref{eq:dimq}), (\ref{eq:snnr}), (\ref{eq:upp}), and
the parameter, $\upsilon^\prime$, can be conceived as a generalized distortion
including both centrifugal and tidal effects.

Similarly, the EC2 associated equations, Eqs.\,(\ref{eq:t0c})-(\ref{eq:t2lc}),
reduce to:
\begin{lefteqnarray}
\label{eq:EA205}
&& \frac1{\xi^2}\frac\diff{\diff\xi}\left(\xi^2\frac{\diff\theta_{0}}
{\diff\xi}\right)-\upsilon^\prime=-\theta_0^5~~; \\
\label{eq:EA225}
&& \frac1{\xi^2}\frac\diff{\diff\xi}\left(\xi^2\frac{\diff\theta_{2}}
{\diff\xi}\right)-\frac6{\xi^2}\theta_{2}=-5\theta_{0}^4\theta_2 \\
\label{eq:EA2l25}
&& \frac1{\xi^2}\frac\diff{\diff\xi}\left(\xi^2\frac{\diff\theta_{2\ell}}
{\diff\xi}\right)-\frac{2\ell(2\ell+1)}{\xi^2}\theta_{2\ell}=-5\theta_0^4
\theta_{2\ell}~~;
\end{lefteqnarray}
where $\theta_{2\ell}=\theta_{{2\ell},v}$ and the cases of interest, for
reasons mentioned above, are $2\ell=0,2$.

It can be seen Eqs.\,(\ref{eq:ED50})-(\ref{eq:EA2l25}) have the same formal
expression of their counterparts related to EC1 polytropes [8],
keeping in mind $\upsilon^\prime\to0$ does not
necessarily imply spherical isopycnic surfaces.   Accordingly,
Eqs.\,(\ref{eq:EA205}) and (\ref{eq:EA225}) can be integrated analytically
provided the generalized distortion, $\upsilon^\prime$,  is negligible with
respect to the other terms, which implies the inequality:
\begin{lefteqnarray}
\label{eq:upp5}
&& \vert\upsilon^\prime\vert=\left\vert\frac{\upsilon_v-\Lambda_u}{\Lambda_v}
\right\vert\ll1~~;
\end{lefteqnarray}
where, in particular, $\upsilon_v\to0$, $\Lambda_u\to0$, $\Lambda_v\to1$,
$\lambda_v\to+\infty$, that is a Roche model.   Related solutions are [8],
[20]:
\begin{lefteqnarray}
\label{eq:tha05}
&& \theta_0(\xi)=\cos\nu+\frac12\upsilon^\prime\tan^2\nu~~; \\
\label{eq:tha25}
&& \theta_2(\xi)=\frac{15}{128} \nonumber \\
&& \times\frac{3[\nu-\sin\nu\cos^3\nu+\sin^3\nu\cos\nu]
-8[\sin^3\nu\cos^5\nu-\sin^5\nu\cos^3\nu]}{\sin^3\nu\cos^2\nu}~~;\qquad \\
\label{eq:nu}
&& \nu=\arctan\frac\xi{\sqrt3}=\arctan\frac{\Lambda_v^{1/2}\xi_v}{\sqrt3}=
\arctan\frac{\Gamma_{uv}^{1/2}\Lambda_v^{1/2}\xi_u}{\sqrt3}~~;
\end{lefteqnarray}
where the second term on the right-hand side of Eq.\,(\ref{eq:tha05}) can be
neglected with respect to the first one via Eq.\,(\ref{eq:upp5}).

With regard to $u$ subsystem, the counterparts of Eqs.\,(\ref{eq:tha05}) and
(\ref{eq:tha25}) via Eq.\,(\ref{eq:isp2l}) written in Appendix \ref{a:seco},
after some algebra take the expression:
\begin{lefteqnarray}
\label{eq:thu05}
&& \theta_{0,u}(\xi_u)=\Gamma_{vu}\cos\nu+\frac12\Gamma_{vu}\frac{\upsilon_v-
\Lambda_u}{\Lambda_v}\tan^2\nu+1-\Gamma_{vu}+\frac{\upsilon_u-\upsilon_v}6
\xi_u^2~~;\qquad \\
\label{eq:thu25}
&& \theta_{2,u}(\xi_u)=\Gamma_{vu}\frac{15}{128}F(\nu)-
\frac{\upsilon_u-\upsilon_v}{6A_2}\xi_u^2~~; \\
\label{eq:Fnu}
&& F(\nu)=3F_1(\nu)-8F_2(\nu)=\frac{\Gamma_{uv}\Lambda_v}3\xi_u^2[3G_1(\nu)-8
G_2(\nu)]~~; \\
\label{eq:F1nu}
&& F_1(\nu)=\frac\nu{\sin^3\nu\cos^2\nu}-\frac{\cos\nu}{\sin^2\nu}+\frac1
{\cos\nu}
~~; \\
\label{eq:G1nu}
&& G_1(\nu)=\frac\nu{\sin^5\nu}-\frac{\cos^3\nu}{\sin^4\nu}+\frac{\cos\nu}
{\sin^2\nu}~~; \\
\label{eq:F2nu}
&& F_2(\nu)=\cos^3\nu-\sin^2\nu\cos\nu
~~; \\
\label{eq:G2nu}
&& G_2(\nu)=\frac{\cos^5\nu}{\sin^2\nu}-\cos^3\nu~~;
\end{lefteqnarray}
and related expressions of the gravitational potential and radial component of
the gravitational force, via Eqs.\,(\ref{eq:csir}), (\ref{eq:VGd}),
(\ref{eq:Gauv}), and (\ref{eq:cccC}) written in Appendix \ref{a:A2}, read:
\begin{lefteqnarray}
\label{eq:VG50}
&& {\cal V}_G(\xi_u,\mu)=4\pi G\sum(\lambda_w)\alpha_u^2\left\{\Gamma_{vu}
\left[\cos\nu+\frac12\frac{\upsilon_v-\Lambda_u}{\Lambda_v}\tan^2\nu
\right.\right.\nonumber \\
&& \phantom{{\cal V}_G(\xi_u,\mu)=}\left.\left.
+\frac{15}{128}A_2F(\nu)P_2(\mu)\right]
+1-\Gamma_{vu}-\frac16\upsilon_v\xi_u^2\left[1-P_2(\mu)\right]
+c_{{\rm b},u}^\dagger\right\};\qquad \\
\label{eq:FG50}
&& \frac{\partial{\cal V}_G}{\partial r}=\frac1{\alpha_u}\frac
{\partial{\cal V}_G}{\partial\xi_u}=4\pi G\sum(\lambda_w)\alpha_u\left\{
\Gamma_{vu}\left[\frac{\diff\cos\nu}{\diff\xi_u}+\frac{\upsilon_v-\Lambda_u}
{\Lambda_v}\tan\nu\frac{\diff\tan\nu}{\diff\xi_u}
\right.\right.\nonumber \\
&& \phantom{\frac{\partial{\cal V}_G}{\partial r}=\frac1{\alpha_u}\frac
{\partial{\cal V}_G}{\partial\xi_u}=}\left.\left.
+\frac{15}{128}A_2\frac{\diff F}{\diff\xi_u}P_2(\mu)\right]-
\frac13\upsilon_v\xi_u\left[1-P_2(\mu)\right]\right\};\qquad
\end{lefteqnarray}
where the derivatives via Eq.\,(\ref{eq:nu}) read:
\begin{lefteqnarray}
\label{eq:dtgnu}
&& \frac{\diff\tan\nu}{\diff\xi_u}=\frac{\tan\nu}{\xi_u}=
\frac{\Gamma_{uv}\Lambda_v}3\xi_u\frac{\cos\nu}{\sin\nu}~~; \\
\label{eq:dsnnu}
&& \frac{\diff\sin\nu}{\diff\xi_u}=\frac{\sin\nu\cos^2\nu}{\xi_u}=
\frac{\Gamma_{uv}\Lambda_v}3\xi_u\frac{\cos^4\nu}{\sin\nu}~~; \\
\label{eq:dcsnu}
&& \frac{\diff\cos\nu}{\diff\xi_u}=-\frac{\sin^2\nu\cos\nu}{\xi_u}=
-\frac{\Gamma_{uv}\Lambda_v}3\xi_u\cos^3\nu~~; \\
\label{eq:dnu}
&& \frac{\diff\nu}{\diff\xi_u}=\frac{\sin\nu\cos\nu}{\xi_u}=
\frac{\Gamma_{uv}\Lambda_v}3\xi_u\frac{\cos^3\nu}{\sin\nu}~~; \\
\label{eq:dFnu}
&& \frac{\diff F}{\diff\xi_u}=F^\prime(\xi_u)=\frac{\Gamma_{uv}\Lambda_v}3
\xi_u\left[\frac{3\cos\nu}{\sin^4\nu}-\frac{9\nu\cos^2\nu}{\sin^5\nu}+
\frac{6\nu}{\sin^3\nu}+\frac{3\cos^3\nu}{\sin^2\nu}+\frac{6\cos^5\nu}
{\sin^4\nu}\right. \nonumber \\
&& \phantom{\frac{\diff F}{\diff\xi_u}=F^\prime(\xi_u)=\frac{\Gamma_{uv}
\Lambda_v}3\xi_u\left[\right.}\left.
+3\cos\nu+40\cos^5\nu-8\sin^2\nu\cos^3\nu\right]~~;
\end{lefteqnarray}
which can be verified after a lot of algebra.   The counterparts
of Eqs.\,(\ref{eq:VG50}) and (\ref{eq:FG50}), related to the noncommon
region, are expressed by Eqs.\,(\ref{eq:VGn0}) and (\ref{eq:FGn0}),
respectively.

The continuity of the gravitational potential and the gravitational force on
the interface along a selected direction, $\xi_u^\ast=\xi_u^\ast(\mu)$,
via Eqs.\,(\ref{eq:VGn0}), (\ref{eq:FGn0}), implies the following systems of
equations:
\begin{lefteqnarray}
\label{eq:tit05}
&& \cases {
\Gamma_{vu}\left[\cos\nu^\ast+\frac12\frac{\upsilon_{v}-\Lambda_u}{\Lambda_v}
\tan^{2}\nu^\ast\right]+1-\Gamma_{vu}-\frac16\upsilon_v(\xi_u^\ast)^2+
c_{{\rm b},u}^\dagger
& \cr
=D_0+\frac{\overline{D}_0}{\xi_u^\ast}-\frac16\Lambda_u(\xi_u^\ast)^2+
c_{{\rm b},u}^\dagger~~;
 & \cr
 & \cr
\Gamma_{vu}\left[\left(\frac{\diff\cos\nu}{\diff\xi_u}\right)_{\xi_u^\ast}+
\frac{\upsilon_v-\Lambda_u}{\Lambda_v}\tan\nu^\ast\left(\frac{\diff\tan\nu}
{\diff\xi_u}\right)_{\xi_u^\ast}\right]-\frac13\upsilon_v\xi_u^\ast
=-\frac{\overline{D}_0}{(\xi_u^\ast)^2}-\frac13\Lambda_u\xi_u^\ast~~;
& \cr
} \\
\label{eq:tit25}
&& \cases {
\Gamma_{vu}\frac{15}{128}A_2F(\xi_u^\ast)+\frac16\upsilon_v(\xi_u^\ast)^2
=D_2(\xi_u^\ast)^2+\frac{\overline{D}_2}{(\xi_u^\ast)^3}+\frac16
\Lambda_u(\xi_u^\ast)^2~~;
 & \cr
 & \cr
\Gamma_{vu}\frac{15}{128}A_2F^\prime(\xi_u^\ast)+\frac13\upsilon_v\xi_u^\ast
=2D_2\xi_u^\ast-\frac{3\overline{D}_2}{(\xi_u^\ast)^4}+\frac13
\Lambda_u\xi_u^\ast~~; & \cr
}
\end{lefteqnarray}
where $\nu^\ast=\nu(\xi_u^\ast)$ and the following relations hold via 
Eq.\,(\ref{eq:nu}):
\begin{lefteqnarray}
\label{eq:t2gxi}
&& \tan^2\nu=\frac13\Gamma_{uv}\Lambda_v\xi_u^2~;\quad\sin^2\nu=\frac
{\Gamma_{uv}\Lambda_v\xi_u^2}{3+\Gamma_{uv}\Lambda_v\xi_u^2}~;\quad
\cos^2\nu=\frac3{3+\Gamma_{uv}\Lambda_v\xi_u^2}~;\qquad~
\end{lefteqnarray}
which define the trigonometric functions in terms of $\xi_u$.

The substitution of Eq.\,(\ref{eq:nu}) into (\ref{eq:tit05})
via (\ref{eq:dtgnu})-(\ref{eq:dFnu}), after long but stimulating algebra
yields an explicit expression of $D_0$, $\overline{D}_0$, as:
\begin{lefteqnarray}
\label{eq:Db050}
&& \overline{D}_0=\frac13\Lambda_v(\xi_u^\ast)^3\cos^3\nu^\ast~~; \\
\label{eq:D050}
&& D_0=\Gamma_{vu}\cos^3\nu^\ast+1-\Gamma_{vu}+\frac{\upsilon_u-\upsilon_v}6
(\xi_u^\ast)^2~~;
\end{lefteqnarray}
on the other hand, Eq.\,(\ref{eq:tit25}) may be cast under the equivalent
form:
\begin{lefteqnarray}
\label{eq:tit4}
&& \cases {
D_2^\ast(\xi_u^\ast)^2+\frac{\overline{D}_2^\ast}{(\xi_u^\ast)^3}
=\Gamma_{vu}\frac{15}{128}F(\xi_u^\ast)+\frac{\upsilon_v-\Lambda_u}{6A_2}
(\xi_u^\ast)^2~~;
 & \cr
 & \cr
2D_2^\ast\xi_u^\ast-\frac{3\overline{D}_2^\ast}{(\xi_u^\ast)^4}
=\Gamma_{vu}\frac{15}{128}F^\prime(\xi_u^\ast)+\frac{\upsilon_v-\Lambda_u}
{3A_2}\xi_u^\ast~~;
& \cr
}
\end{lefteqnarray}
where $D_2^\ast$, $\overline{D}_2^\ast$, are defined by
Eq.\,(\ref{eq:D2la}) and $A_2$, in turn, depends on $D_2$, $\overline{D}_2$,
via Eqs.\,(\ref{eq:h20}), (\ref{eq:e20}), and (\ref{eq:ACsi}) written in
Appendix \ref{a:A2}.

Then the system has to be solved through successive iterations in $A_2$,
selecting an appropriate initial value, up to convergence within an assigned
tolerance.   With regard to a generic step, after long but stimulating algebra
the result is:
\begin{lefteqnarray}
\label{eq:D2a50}
&& D_2^\ast=\frac15\frac{15}{128}\frac{\Lambda_v}3\left[\frac{15\nu^\ast}
{\sin^3\nu^\ast}-\frac{15\cos\nu^\ast}{\sin^2\nu^\ast}+30\cos\nu^\ast+40
\cos^3\nu^\ast+48\cos^5\nu^\ast\right] \nonumber \\
&& \phantom{D_2^\ast=}
-\frac16\frac{\upsilon_u-\upsilon_v}
{A_2}+\frac16\frac{\upsilon_v-\Lambda_u}{A_2}~~; \\
\label{eq:D2b50}
&& \overline{D}_2^\ast=\frac15\frac{15}{128}\frac{\Lambda_v}3(\xi_u^\ast)^5
\left[\frac{15\nu^\ast\cos^2\nu^\ast}{\sin^5\nu^\ast}
-\frac{15\cos\nu^\ast}{\sin^4\nu^\ast}+\frac{5\cos\nu^\ast}{\sin^2\nu^\ast}+
10\cos\nu^\ast\right. \nonumber \\
&& \phantom{\overline{D}_2^\ast=\frac15\frac{15}{128}\frac{\Lambda_v}3
(\xi_u^\ast)^5~}\left.
+40\cos^3\nu^\ast-48\cos^5\nu^\ast\right]~~;
\end{lefteqnarray}
where $\overline{D}_2^\ast$ shows no explicit dependence on $A_2$.
Accordingly, values of $\theta_2(\Xi_u)$, $\theta_2^\prime(\Xi_u)$, on a
selected point of the boundary, $\Xi_u=\Xi_u(\mu)$, can be determined via
Eq.\,(\ref{eq:h20a}), and the next iteration value of $A_2$ can be determined
via Eq.\,(\ref{eq:ACsi}) written in Appendix \ref{a:A2}.

The EC2 associated functions, $\theta_{0,v}(\xi_v)$, can be expanded in
Taylor series only in the special case of the singular starting point,
$\xi_{0,v}^\dagger=0$, restricted to $\upsilon^\prime\to0$.
Accordingly, Eq.\,(\ref{eq:tha05}) via (\ref{eq:nu}) and (\ref{eq:t2gxi})
reduces to:
\begin{lefteqnarray}
\label{eq:t0v50l}
&& A_0\theta_{0,v}(\xi_v)=\cos\nu=\left(1+\frac13\Lambda_v\xi_v^2\right)^{-1/2}
~~;\qquad A_0=1~~;
\end{lefteqnarray}
where the right-hand side can be expanded in binomial series.   The comparison
with related MacLaurin series counterpart,  Eq.\,(\ref{eq:sers0}), yields:
\begin{leftsubeqnarray}
\slabel{eq:a0k50a}
&& A_0a_{0,2k+2}^{(v,v)}=(-1)^k\left(\frac{\Lambda_v}3\right)^k\frac{1\cdot3\cdot
...\cdot(2k+1)}{2\cdot4\cdot...\cdot(2k+2)}~~; \\
\slabel{eq:a0k50b}
&& A_0a_{0,2k+1}^{(v,v)}=0~~;\qquad 2k\ge0~~;\quad A_0=1~~;\qquad
\label{seq:a0k50}
\end{leftsubeqnarray}
where the convergence radius is $\Delta_{\rm C}\xi_v=(3/\Lambda_v)^{1/2}$.
Unfortunately, the above results cannot be extended to the general case
of starting point, $\xi_{0,v}>0$, and 
the coefficients of Taylor series expansions, Eq.\,(\ref{eq:sersc}), cannot be
expressed in simpler form with respect to the regression formula,
Eq.\,(\ref{seq:a0kc}). 
For further details, an interested reader is addressed to the parent paper
[6].

Let $\theta_{0,v}(\Xi_{{\rm ex},v})=\theta_v(\Xi_v,\mu)=\theta_{{\rm b},v}$ be
the (fictitious) spherical isopycnic surface of the expanded sphere, related
to the interface.   By use of Eq.\,(\ref{eq:tha05}), an explicit expression
reads:
\begin{lefteqnarray}
\label{eq:v50e2}
&& \cos\nu+\frac12\upsilon^\prime\tan^2\nu=\theta_{{\rm b},v}~~;
\end{lefteqnarray}
which is a transcendental equation where the scaled radius,
$\Xi_{{\rm ex},v}$, is the lowest positive solution.   In the limit,
$\upsilon^\prime\to0$, Eq.\,(\ref{eq:v50e2}) via (\ref{eq:nu}) and
(\ref{eq:t2gxi}) reduces to:
\begin{lefteqnarray}
\label{eq:v50p2}
&& \left(\frac3{3+\Lambda_v\xi_v^2}\right)^{1/2}=\theta_{{\rm b},v}~~;
\end{lefteqnarray}
which has a unique (acceptable) solution as:
\begin{lefteqnarray}
\label{eq:Xiev50}
&& \Xi_{{\rm ex},v}=\left[\frac{3(1-\theta_{{\rm b},v}^2)}
{\Lambda_v\theta_{{\rm b},v}^2}\right]^{1/2}~~;
\end{lefteqnarray}
that in the special case, $\theta_{{\rm b},v}=0$, $\Lambda_v=1$, reduces to
its counterpart related to EC1 polytropes [4], [6], as expected.

\section{A guidance example}
\label{guex}

To the knowledge of the author, subsystems in hydrostatic equilibrium with
intersecting boundaries have never been considered in literature.   To this
respect, an application shall be presented below restricting to the simplest
case, $(n_i,n_j)=(0,0)$, $(\theta_{{\rm b},i},\theta_{{\rm b},j})=(0,0)$,
which via Eq.\,(\ref{eq:raij}) implies $\Gamma_{ji}\ge1$, to be intended as a
guidance example.

\subsection{Input and output parameters}
\label{iopa}

The equipotential surfaces within the common region are expressed by
Eq.\,(\ref{eq:ECc0t}) which, in the special case of the interface,
$\theta_{{\rm b},w}=0$, reduces to:
\begin{lefteqnarray}
\label{eq:ibc}
&& 1-\frac{1-\upsilon_w}6\Xi_w^2-\frac16\upsilon_w\Xi_w^2P_2(\mu)=0~~;
\end{lefteqnarray}
and the substitution of Eq.\,(\ref{eq:Lp}) into (\ref{eq:ibc}) after some
algebra yields:
\begin{lefteqnarray}
\label{eq:ibi}
&& \Xi_w=\sqrt6\left[1-\frac32\upsilon_w(1-\mu^2)\right]^{-1/2}~~;
\end{lefteqnarray}
which is the equation of a spheroid. In particular, the polar $(\mu=1)$ and
the equatorial $(\mu=0)$ scaled semiaxes are:
\begin{lefteqnarray}
\label{eq:ibpe}
&& \Xi_{{\rm p},w}=\sqrt6~~;\qquad\Xi_{{\rm e},w}=\sqrt6\left(1-\frac32
\upsilon_w\right)^{-1/2}~~;
\end{lefteqnarray}
where $\Xi_{{\rm p},j}$ and $\Xi_{{\rm e},i}$ are different from their
counterparts related to the boundary, in that the boundary is not spheroidal
via Eq.\,(\ref{eq:ECn00t}).

\begin{figure*}[t]  
\begin{center}      
\includegraphics[scale=0.8]{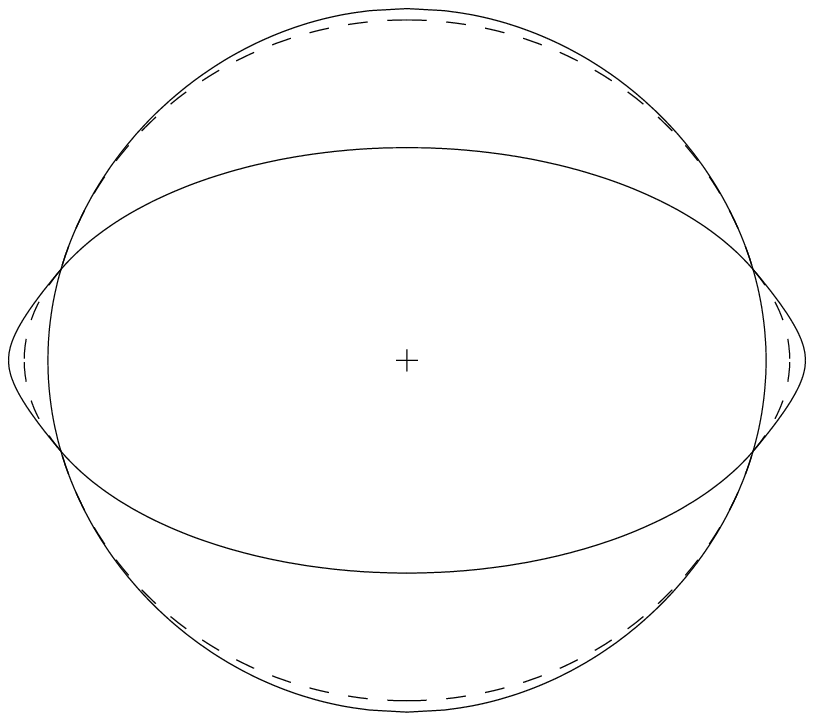}                      
\caption[ddbb]{Subsystems with intersecting boundaries, $(n_i,n_j)=(0,0)$
and $(\theta_{{\rm b},i},\theta_{{\rm b},j})=(0,0)$.   By definition, the
inner subsystem, $i$, has the pole closer to the centre, and the outer
subsystem, $j$, has the pole farther, hence
$\alpha_i\Xi_{{\rm p},i}\le\alpha_j\Xi_{{\rm p},j}$.   The interface lies on
two spheroids, where the locus of intersection points is related to
the polar angle, $\hat\delta=\arccos\hat\mu$ (not shown to save clarity).
The boundary of the system
cannot be fitted by the analytical continuation of the interface (dashed),
owing to the occurrence of
nonlinear terms  in related equations.   See text for further details.}
\label{f:ib00}     
\end{center}       
\end{figure*}

The substitution of Eq.\,(\ref{eq:ibpe}) into (\ref{eq:ibi}) yields:
\begin{lefteqnarray}
\label{eq:ibp}
&& \alpha_w\Xi_w=\alpha_w\Xi_{{\rm p},w}\left[1-\frac32\upsilon_w(1-\mu^2)
\right]^{-1/2}~~;
\end{lefteqnarray}
in terms of the radial coordinate, $R=\alpha_w\Xi_w$.  Owing to intersecting
boundaries, the spheroid is restricted to the polar region for $i$ subsystem
and to the equatorial region for $j$ subsystem, as shown in
Fig.\,\ref{f:ib00}.

The intersection of the two spheroids is defined as
$\alpha_i\Xi_i(\hat\mu)=\alpha_j\Xi_j(\hat\mu)$ which, via
Eqs.\,(\ref{eq:Gauv}) and (\ref{eq:ibp}), after some algebra yields:
\begin{lefteqnarray}
\label{eq:ibmt}
&& \hat\mu=\mp\left[1-\frac23\frac{1-\Gamma_{ij}}{\upsilon_i-\Gamma_{ij}
\upsilon_j}\right]^{1/2}~~;
\end{lefteqnarray}
accordingly, the interface depends on the fractional scaling radius,
$\Gamma_{ji}^{1/2}$, and the rotation parameters, $\upsilon_i$, $\upsilon_j$.

The dimensionless polar semiaxis, $\Xi_{{\rm p},i}$, and the dimensionless
equatorial semiaxis, $\Xi_{{\rm e},j}$, via Eq.\,(\ref{eq:ibpe}) read:
\begin{lefteqnarray}
\label{eq:Cspi}
&& \Xi_{{\rm p},i}=\sqrt6~~; \\
\label{eq:Csej}
&& \Xi_{{\rm e},j}=\sqrt6\left(1-\frac32\upsilon_j\right)^{-1/2}~~;
\end{lefteqnarray}
while the remaining ones, $\Xi_{{\rm p},j}$ and $\Xi_{{\rm e},i}$, cannot be
explicitly expressed in that they extend outside the interface.

The equipotential surfaces within the noncommon region are expressed by
Eq.\,(\ref{eq:ECn00t}) which, in the special case of the boundary,
$\theta_{{\rm b},w}=0$, reduces to:
\begin{lefteqnarray}
\label{eq:ibu}
&& D_{0,w}+\frac{\overline D_{0,w}}{\Xi_w}-\frac16\Lambda_w\Xi_w^2+\frac16
\upsilon_w\Xi_w^2[1-P_2(\mu)]=0~~;
\end{lefteqnarray}
where the constants, $D_{0,w}$, $\overline D_{0,w}$, via Eq.\,(\ref{eq:Du00})
read:
\begin{lefteqnarray}
\label{eq:D0w}
&& D_{0,w}=1-\frac{1-\Lambda_w}2(\xi_w^\ast)^2~~;\qquad\overline D_{0,w}=\frac
{1-\Lambda_w}3(\xi_w^\ast)^3~~;
\end{lefteqnarray}
where $\xi_w^\ast=\xi_w^\ast(\hat\mu)$ has to be selected in the case under
discussion to ensure continuity.   Accordingly, the boundary depends on the
fractional scaling radius, $\Gamma_{ji}^{1/2}$, the fractional density,
$\Lambda_{ji}=\lambda_j/\lambda_i$ via $\Lambda_u=1/(1+\Lambda_{vu})$, and the
rotation parameters, $\upsilon_i$, $\upsilon_j$.

With regard to a generic point on the equatorial plane,
$(\xi_w,0)$, the substitution of Eqs.\,(\ref{eq:ECc0t}) and (\ref{eq:ECn00t}),
respectively, into (\ref{eq:thw0}) via (\ref{eq:diEC}), (\ref{eq:csir}), after
little algebra yields:
\begin{lefteqnarray}
\label{eq:uecom}
&& \upsilon_{{\rm eq},w}^{(\rm com)}=\frac23~~;  \\
\label{eq:uencm}
&& \upsilon_{{\rm eq},w}^{(\rm ncm)}=\frac{2\Lambda_w\xi_w^3+6\overline
D_{0,w}}{3\xi_w^3}~~;
\end{lefteqnarray}
according if centrifugal support takes place on the common or noncommon
region, respectively.  Then
$\upsilon_w>\upsilon_{{\rm eq},w}$ implies instability.

The substitution of Eq.\,(\ref{eq:Lp}) into (\ref{eq:ibu}) after some algebra
yields:
\begin{lefteqnarray}
\label{eq:ibe}
&& \frac16\left[\Lambda_w-\frac32\upsilon_w(1-\mu^2)\right]\Xi_w^3-D_{0,w}
\Xi_w-\overline D_{0,w}=0~~;
\end{lefteqnarray}
and the boundary is defined by the (positive) solution of the above
third-degree equation, which matches the intersection between subsystem
surfaces.   More specifically, $\alpha_i\Xi_i(\mu)$,
$\mu\le\hat\mu$, and $\alpha_j\Xi_j(\mu)$, $\mu\ge\hat\mu$, have to be
considered, as shown in Fig.\,\ref{f:ib00}.

The subsystem volume can be inferred from the boundary as [7]:
\begin{lefteqnarray}
\label{eq:Sw}
&& S_w=4\pi\alpha_w^3I_{{\rm S},w}~~; \\
\label{eq:ISw}
&& I_{{\rm S},w}=\int_0^1\diff\mu\int_0^{\Xi_w(\mu)}\xi_w^2\diff\xi_w=
\frac13\int_0^{\hat\mu}\Xi_w^3(\mu)\diff\mu+\frac13\int_{\hat\mu}^1\Xi_w^3
(\mu)\diff\mu~~;
\end{lefteqnarray}
where $\hat\mu>0$ is expressed by Eq.\,(\ref{eq:ibmt}).

The subsystem mass and mass ratio are:
\begin{lefteqnarray}
\label{eq:Mw}
&& M_w=\lambda_wS_w~~;\qquad m=\frac{M_j}{M_i}=\frac{\lambda_j}{\lambda_i}
\frac{S_j}{S_i}=\Lambda_{ji}\Gamma_{ji}^{3/2}\frac{I_{{\rm S},j}}
{I_{{\rm S},i}}~~;
\end{lefteqnarray}
in the case under discussion of homogeneous configurations, $(n_i,n_j)=(0,0)$.

In conclusion, the input: fractional scaling radius,
$\Gamma_{ji}^{1/2}=\alpha_j/\alpha_i$; fractional density,
$\Lambda_{ji}=\lambda_j/\lambda_i$; rotation
parameters, $\upsilon_i$, $\upsilon_j$; implies the output: scaled radius,
$\Xi_w(\mu)$; homologous axis ratio, $\eta_r=a_{r,j}/a_{r,i}$; subsystem axis
ratio, $\epsilon_w=a_{{\rm p},w}/a_{{\rm e},w}$; interface, $\Xi_i(\mu)$,
$1\ge\mu\ge\hat\mu$, $\Xi_j(\mu)$, $0\le\mu\le\hat\mu$; boundary,
$\Xi_i(\mu)$, $0\le\mu\le\hat\mu$, $\Xi_j(\mu)$, $1\ge\mu\ge\hat\mu$; volume
ratio, $s=S_j/S_i$; mass ratio, $m=M_j/M_i$.

\subsection{Numerical values}
\label{nuva}

\begin{figure*}[t]  
\begin{center}      
\includegraphics[scale=0.8]{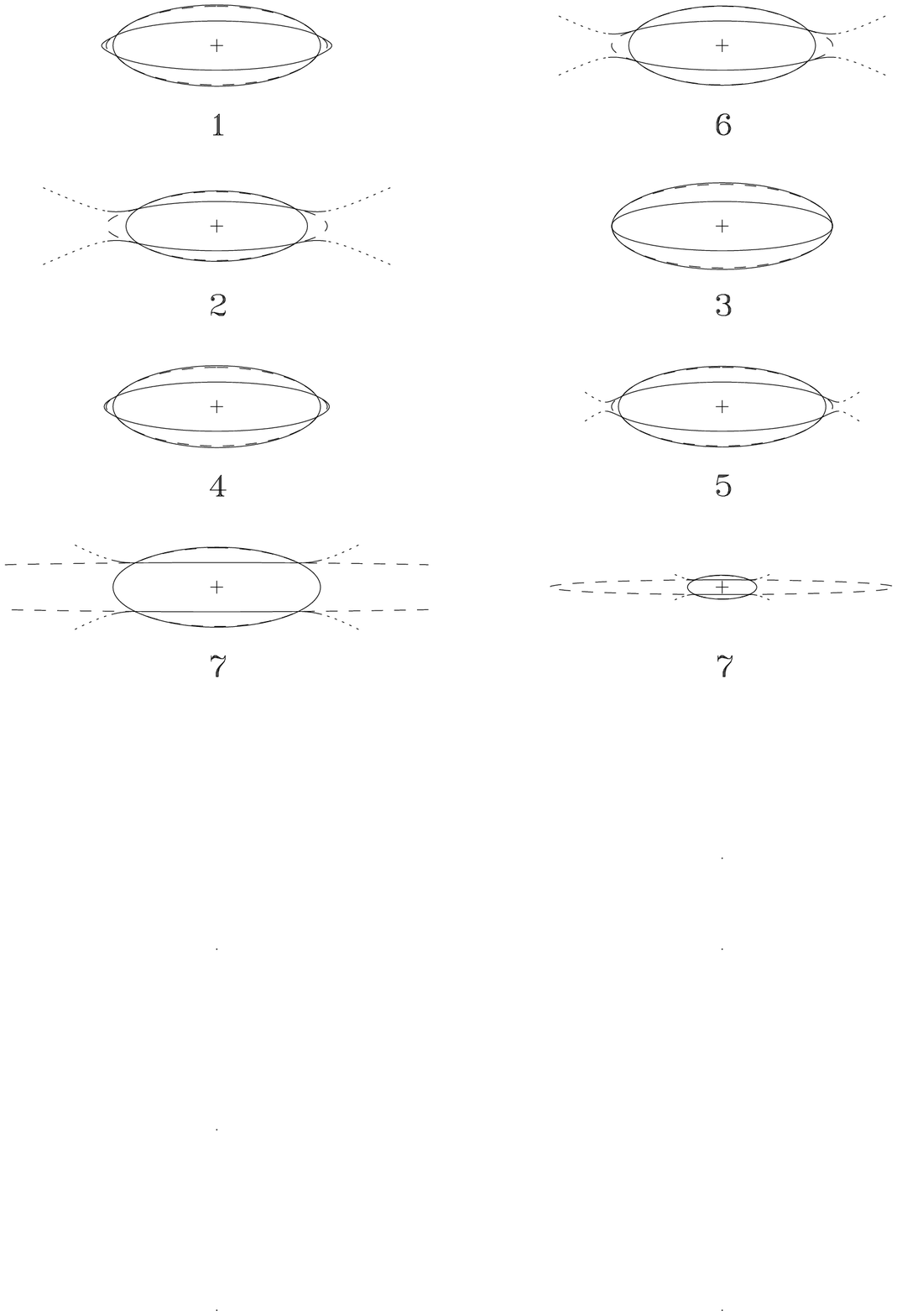}                      
\caption[ddbb]{Boundaries and interfaces  of EC2 polytropes where
$(n_i,n_j)=(0,0)$ and $(\theta_{{\rm b},i},\theta_{{\rm b},j})=(0,0)$,
related to cases 1-7 listed in Tables \ref{t:diss}-\ref{t:dist}.   The top
left panel is
the source of Fig.\,\ref{f:ib00}.   Case 7 is also shown in reduced scale on
the bottom right panel.
The analytical continuation of the interface within the noncommon region is
shown by dashed curves.   With regard to the inner subsystem, the analytical
continuation of open boundaries (where centrifugal support is exceeded) is
shown by dotted lines.   See text for further details.}
\label{f:ec6}     
\end{center}       
\end{figure*}

Let the substantially flattened, inner subsystem represent a disk, and the
slightly flattened, outer subsystem represent a stellar cluster.   Values of
output parameters for different choices of input parameters are listed in
Tables \ref{t:diss} and \ref{t:dist}.   Interfaces and boundaries of related
configurations are shown in Fig.\,\ref{f:ec6}, where the top left panel is
the source of Fig.\,\ref{f:ib00} and the bottom right panel replots case 7 in
reduced scale.
\begin{table*}
\caption[par]{Input and output parameters (prm) for different configurations of EC2
polytropes where $(n_i,n_j)=(0,0)$ and 
$(\theta_{{\rm b},i},\theta_{{\rm b},j})=(0,0)$.   Input: fractional scaling
radius, $\Gamma_{ji}^{1/2}=\alpha_j/\alpha_i$; fractional density, 
$\Lambda_{ji}=\lambda_j/\lambda_i$; rotation parameters, 
$\upsilon_i$, $\upsilon_j$.   Output:  scaled equatorial semiaxis,
$\Xi_{{\rm e},w}=a_{{\rm e},w}/\alpha_w$; scaled polar semiaxis,
$\Xi_{{\rm p},w}=a_{{\rm p},w}/\alpha_w$
($\Xi_{{\rm p},i}=\sqrt6\approx2.4495$ in all
cases); cosine of polar angle at surface intersection,
$\hat\mu=\cos\hat\delta$;
subsystem axis ratio, $\epsilon_w=a_{{\rm p},w}/a_{{\rm e},w}$;
homologous axis ratio, 
$\eta_r=a_{r,j}/a_{r,i}$;
volume integral, $I_{{\rm S},w}$; fractional volume, $s=S_j/S_i$; fractional
mass, $m=M_j/M_i$.
For open boundaries (marked by asterisks on the case number),
$\Xi_{{\rm e},i}$ and related quantities, $\epsilon_i$,
$\eta_{\rm e}$, $I_{{\rm S},i}$, $s$, $m$, are determined on the point which
is nearer to the equatorial plane i.e. where $\mu$ attains the minimum value.
See text for further details.}
\label{t:diss}
\begin{center}
\begin{tabular}{lllll} \hline
\multicolumn{1}{l}{case:} &
\multicolumn{1}{c}{1} &
\multicolumn{1}{c}{2$^\ast$} &
\multicolumn{1}{c}{3} &
\multicolumn{1}{c}{4} 
\\
prm                 &             &             &             &             \\
$\Gamma_{ji}^{1/2}$ & 1.6000E$+$0 & 1.4000E$+$0 & 1.7000E$+$0 & 1.6000E$+$0 \\
$\Lambda_{ji}$      & 1.0000E$+$0 & 1.0000E$+$0 & 1.0000E$+$0 & 5.0000E$-$1 \\
$\upsilon_j$        & 2.4000E$-$1 & 2.4000E$-$1 & 2.4000E$-$1 & 2.4000E$-$1 \\
$\upsilon_i$        & 5.2000E$-$1 & 5.2000E$-$1 & 5.2000E$-$1 & 5.2000E$-$1 \\
$\Xi_{{\rm e},j}$   & 3.0619E$+$0 & 3.0619E$+$0 & 3.0619E$+$0 & 3.0619E$+$0 \\
$\hat\mu$           & 2.1661E$-$1 & 4.2266E$-$1 & 4.7088E$-$2 & 2.1661E$-$1 \\ 
$\Xi_{{\rm p},j}$   & 2.5314E$+$0 & 2.5065E$+$0 & 2.5413E$+$0 & 2.5498E$+$0 \\
$\Xi_{{\rm e},i}$   & 5.4359E$+$0 & 5.3735E$+$0 & 5.2226E$+$0 & 5.3158E$+$0 \\
$\epsilon_j$        & 8.2676E$-$1 & 8.1863E$-$1 & 8.2999E$-$1 & 8.3276E$-$1 \\
$\epsilon_i$        & 4.5061E$-$1 & 4.5585E$-$1 & 4.6902E$-$1 & 4.6079E$-$1 \\
$\eta_{\rm e}$      & 9.0123E$-$1 & 7.9773E$-$1 & 9.9666E$-$1 & 9.2159E$-$1 \\
$\eta_{\rm p}$      & 1.6535E$+$0 & 1.4326E$+$0 & 1.7637E$+$0 & 1.6655E$+$0 \\
$I_{{\rm S},j}$     & 1.9293E$+$1 & 1.5479E$+$1 & 2.2580E$+$1 & 1.9399E$+$1 \\
$I_{{\rm S},i}$     & 4.2964E$+$1 & 1.9719E$+$1 & 2.6722E$+$1 & 4.2100E$+$1 \\
$s$                 & 1.8393E$+$0 & 2.1539E$+$0 & 4.1515E$+$0 & 1.8874E$+$0 \\
$m$                 & 1.8393E$+$0 & 2.1539E$+$0 & 4.1515E$+$0 & 9.4370E$-$1 \\
\noalign{\smallskip}
\hline
\end{tabular}                     
\end{center}                      
\end{table*}                       
\begin{table*}
\caption{Continuation of Table \ref{t:diss}.   The reference case, 1, has been
repeated to facilitate comparison.}
\label{t:dist}
\begin{center}
\begin{tabular}{lllll} \hline
\multicolumn{1}{l}{case:} &
\multicolumn{1}{c}{1} &
\multicolumn{1}{c}{5$^\ast$} &
\multicolumn{1}{c}{6$^\ast$} &
\multicolumn{1}{c}{7$^\ast$} \\
prm                 &             &             &             &             \\
$\Gamma_{ji}^{1/2}$ & 1.6000E$+$0 & 1.6000E$+$0 & 1.6000E$+$0 & 1.6000E$+$0 \\
$\Lambda_{ji}$      & 1.0000E$+$0 & 2.0000E$+$0 & 1.0000E$+$0 & 1.0000E$+$0 \\
$\upsilon_j$        & 2.4000E$-$1 & 2.4000E$-$1 & 1.4000E$-$1 & 2.4000E$-$1 \\
$\upsilon_i$        & 5.2000E$-$1 & 5.2000E$-$1 & 5.2000E$-$1 & 6.6000E$-$1 \\
$\Xi_{{\rm e},j}$   & 3.0619E$+$0 & 3.0619E$+$0 & 2.7559E$+$0 & 3.0619E$+$0 \\
$\hat\mu$           & 2.1661E$-$1 & 2.1661E$-$1 & 3.5627E$-$1 & 5.3156E$-$1 \\ 
$\Xi_{{\rm p},j}$   & 2.5314E$+$0 & 2.5097E$+$0 & 2.4685E$+$0 & 2.4910E$+$0 \\
$\Xi_{{\rm e},i}$   & 5.4359E$+$0 & 5.5473E$+$0 & 5.5083E$+$0 & 5.6401E$+$0 \\
$\epsilon_j$        & 8.2676E$-$1 & 8.1966E$-$1 & 8.9573E$-$1 & 8.1354E$-$1 \\
$\epsilon_i$        & 4.5061E$-$1 & 4.4156E$-$1 & 4.4469E$-$1 & 4.3430E$-$1 \\
$\eta_{\rm e}$      & 9.0123E$-$1 & 8.8313E$-$1 & 8.0050E$-$1 & 8.6860E$-$1 \\
$\eta_{\rm p}$      & 1.6535E$+$0 & 1.6393E$+$0 & 1.6124E$+$0 & 1.6271E$+$0 \\
$I_{{\rm S},j}$     & 1.9293E$+$1 & 1.9173E$+$1 & 1.3784E$+$1 & 1.3647E$+$1 \\
$I_{{\rm S},i}$     & 4.2964E$+$1 & 3.0798E$+$1 & 2.3106E$+$1 & 1.2196E$+$1 \\
$s$                 & 1.8393E$+$0 & 2.5499E$+$0 & 2.4436E$+$0 & 4.5835E$+$0 \\
$m$                 & 1.8393E$+$0 & 5.0998E$+$0 & 2.4436E$+$0 & 4.5835E$+$0 \\
\noalign{\smallskip}
\hline
\end{tabular}                     
\end{center}                      
\end{table*}                       
In particular, cases 1, 3, 4, are characterized by hydrostatic equilibrium,
while the contrary holds for cases 2, 5, 6, 7, where centrifugal support is
exceeded within the noncommon region related to $i$ subsystem.  From a
mathematical standpoint, the reason is the following.

In presence of hydrostatic equilibrium, the third-degree equation which
describes the boundary, Eq.\,(\ref{eq:ibu}) via (\ref{eq:ibe}), has a single
real solution
within the range, $\Xi_i(\hat\mu)\le\Xi_i(\mu)\le\Xi_i(0)=\Xi_{{\rm e},i}$,
and the surface can be defined as shown in panels 1, 3, 4, of
Fig.\,\ref{f:ec6}.
Conversely, when centrifugal support is exceeded, there are three real
solutions among which at least two are positive, getting closer and closer as
$\mu$ decreases, up to coincide (via a null discriminant) at $\mu=\mu_0>0$,
and the surface can be defined up to this point (transition from full to
dotted lines) as shown in panels 2, 5, 6, 7, of Fig.\,\ref{f:ec6}.  The
analytical continuation of the surface can be traced via the second positive
solution up to $\mu=\hat\mu$, as shown by dotted lines.    Scaled equatorial
semiaxes, $\Xi_{{\rm e},i}$, subsystem axis ratios, $\epsilon_i$, homologous
axis ratios, $\eta_{\rm e}$, volume integrals, $I_{{\rm S},i}$, fractional
volumes, $s$, and fractional masses, $m$, are calculated up to $\mu=\mu_0$
when centrifugal support is exceeded.

Further inspection of
Tables \ref{t:diss}-\ref{t:dist} and Fig.\,\ref{f:ec6} discloses the following
main features.
\begin{description}
\item[$\bullet$]
With respect to the analytical continuation of the interface (dashed curves),
the boundary appears less flattened in the polar region and more flattened in
the equatorial region.
\item[$\bullet$]
Decreasing the fractional scaling radius,
$\Gamma_{ji}^{1/2}=\alpha_j/\alpha_i$, yields more flattened configurations,
lower fractional volume, $s=S_j/S_i$, lower fractional mass, $m=M_j/M_i$, and
vice versa (cases 3 and 1).
\item[$\bullet$]
Increasing the fractional central density, $\Lambda_{ji}=\lambda_j/\lambda_i$,
yields more flattened configurations, lower $s$, larger $m$, and vice versa
(cases 4 and 1).
\item[$\bullet$]
Increasing the rotation parameter, $\upsilon_j$, yields more flattened $j$
configurations, less flattened $i$ configurations, lower $s$, lower $m$, and
vice versa (cases 6 and 1).
\item[$\bullet$]
Increasing the rotation parameter, $\upsilon_i$, yields more flattened
configurations, larger $s$, larger $m$, and vice versa (cases 1 and 7).
\end{description}
Caution is needed for configurations where centrifugal support is exceeded
(case number with asterisk in Tables \ref{t:diss}-\ref{t:dist}), in that the
boundary has been truncated at $\mu=\mu_0$, where the discriminant of related
third-degree equation is null (transition from full to dotted lines in
Fig.\,\ref{f:ec6}).

\section{Discussion}
\label{disc}

Similarly to distorted EC1 polytropes [13], [15],
distorted EC2 polytropes [10], [5], [7] can be
described in terms of solutions of associated EC2 equations,
Eqs.\,(\ref{eq:t0n})-(\ref{eq:t2ln}) and (\ref{eq:t0c})-(\ref{eq:t2lc}) for
the noncommon and the common region, respectively.   The above mentioned
solutions can be expanded in MacLaurin series via
recursion formulae expressed by Eqs.\,(\ref{seq:a0kc0}) and (\ref{seq:a2kc0}),
up to a convenient point within the convergence radius.   Taking that point as
starting point, Taylor series expansions can be performed via
Eqs.\,(\ref{seq:a0kc}),
(\ref{seq:a2kc}), and (\ref{seq:a0k}), (\ref{seq:a2k}), for the common and the
noncommon region, respectively, up to a convenient point within the
convergence radius.   Then the procedure can be repeated up to the first zero,
$\Xi_{{\rm ex},w}$, of the associated EC2 function, $\theta_{0,w}$, keeping in mind
the convergence radius is steadily decreasing down to zero as
$\xi_{0,w}\to\Xi_{{\rm ex},w}$.   Accordingly, starting points of series
expansions involving $\theta_{0,w}<0$ should imply
$\xi_{0,w}>\Xi_{{\rm ex},w}$. In the limit of a vanishing subsystem (other
than $w$), the above
results reduce to their counterparts related to EC1 polytropes [6].

In general, the knowledge of the convergence radius, $\Delta_{\rm C}\xi_w$,
related to
a Taylor series expansion of starting point, $\xi_{0,w}$, is very important
owing not only to convergence in itself but, in addition, to uniform
convergence wherein $\vert\xi_w-\xi_{0,w}\vert<\Delta_{\rm C}\xi_w$, which
implies the
series can be differentiated or integrated term-by-term inside the convergence
radius, behaving almost like an analytical function e.g., [36], [29].

EC2 polytropes are a useful tool for the description of large-scale celestial
bodies, such as galaxies or galaxy clusters, where a visible baryonic
(including leptons) subsystem is embedded within a dark nonbaryonic halo.   In
most applications, each component is treated separately and the model is a
simple superposition of the two matter distributions e.g., [12].
To this respect, EC2 polytropes make a further step in that each subsystem
readjusts itself in presence of tidal interaction and, in
addition, both collisional and collisionless fluids can be represented within
the range of polytropic index, $1/2\le n\le5$ [39].   Finally, a specific
model [34], widely used for the description of galaxies and globular
clusters, is a different formulation of EC1 polytropes where $n=5$ e.g., [11].

With regard to the parent paper [10], the current investigation includes
additional points, namely (i) boundaries where the density is not vanishing
e.g., [24], Chap.\,IX, \S235, which could be useful in some
applications e.g., modelling strange quark stars [16], [25] and
computing the total mass of a rotating configuration as a function of the
central density [35]; (ii) subsystems with intersecting
boundaries, where both fill (different volumes of) the noncommon region, which
could be useful for the description of special configurations e.g., a
substantially flattened subsystem extending outside a slightly flattened one,
mimicing a
bulge-disk galaxy; (iii) detailed analysis of a few particular cases,
$(n_v,n_u)=(0,0), (0,1), (1,0), (1,1), (5,0)$, which could be useful as
guidance examples for better understanding a real situation.

To this respect, it is worth emphasizing the current investigation on the
special case, $(n_v,n_u)=(0,0)$, is restricted to the same approximation used
for the general case, aiming to improve comparison.   An outline of exact
solution can be found in an earlier paper [7].   More specifically, the
formal expression of the EC2 function remains the same, but the constant,
$A_2$, is expressed using a different approximation in comparison to the
present paper i.e. spheroidal equipotential surfaces within the noncommon
region, which implies the nonlinear term on the right-hand side of 
Eq.\,(\ref{eq:ECn00t}) is neglegible with respect to the others, or
$1-\Lambda_u\ll1$ via Eq.\,(\ref{eq:Du00}) i.e. a nearly vanishing $v$
subsystem.

Homogeneous, concentric and copolar spheroids with intersecting boundaries,
regardless of hydrostatic equilibrium, were considered in an earlier
investigation [31].   Due attention should be devoted to equilibrium
configurations with intersecting boundaries, even if restricted to toy models,
to gain more insight on astrophysical systems such as bulge-disk galaxies and
star cluster-accretion disk (including the supermassive black hole) within the
central parcec.   The guidance example shown in Section \ref{guex} has been
restricted to $(n_i,n_j)=(0,0)$ for simplicity, which rules out stellar
systems in that collisionless equilibrium fluids need $n_w\ge1/2$, but
$(n_i,n_j)=(1,1)$, well holds to this respect.

With regard to results listed in Tables \ref{t:diss}-\ref{t:dist} and plotted
in Fig.\,\ref{f:ec6}, configurations in hydrostatic equilibrium (cases 1, 3,
4) show the inner subsystem slightly exceeds the outer one along the
equatorial plane, the volume of the outer subsystem exceeds the inner one by
a factor up to about four, and the mass of the outer subsystem is slightly
lower than or exceeds the outer one by a factor up to about four. On the other
hand, configurations where hydrostatic equilibrium is broken by centrifugal
forces (cases 2, 5, 6, 7), disclose similar results (implying a factor up to
about five) but with respect to boundaries truncated at
$\mu=\mu_0$ (transition point from full to dotted lines in Fig.\,\ref{f:ec6}),
where the discriminant of the boundary third-degree equation is null.
Accordingly, values of equatorial semiaxis, volume and mass of the inner
subsystem have to be considered as lower limits.

Though the model under discussion fails in quantitative predictions, still a
correct trend could be expected for resembling astrophysical systems, such as
early-type disk galaxies and nuclear star clusters embedding a supermassive
black hole and the related inner accretion disk.

\section{Conclusion}
\label{conc}

The theory of EC2 polytropes has been reformulated.   The method
used in earlier investigations [36], [29],
[6] for series expansion of the solution of EC1 equation and
related associated equations has been extended to the solution of EC2 equation
and related associated equations.
In addition, special cases where the solution can be expressed analytically
have been considered in detail: $(n_i,n_j)=(0,0), (0,1), (1,0),(1,1), (5,0)$.
Subsystems with nonvanishing density on the boundary and subsystems
with intersecting boundaries have also been included in the general theory,
with regard to both collisional and collisionless fluids.

A selected class of configurations has been defined in terms of central
densities, $\lambda_w$, scaling radii, $\alpha_w$, rotation parameters,
$\upsilon_w$, where $w=i,j$.   Accordingly, it has been realized subsystems
belong to one among the following states: (1) rotating to a different extent
and showing different scaling radius; (2) rotating to a different extent but
showing equal scaling radius; (3) rotating to the same extent but showing
different scaling radius; (4) rotating to the same extent and showing equal
scaling radius.

With regard to the noncommon region, it has been realized the solution of the
EC2 equation and associated EC2 equations can be expressed similarly to its
counterpart related to EC1 polytropes.   It has been recognized the contrary
holds for the common region, where the solution of EC2 equation and associated
EC2 equations involves both subsystems instead of only one, leaving aside a
few special cases.

The main results of the current paper may be summarized in the following
points.
\begin{description}
\item[(i)]
Subsystems with nonvanishing density on the boundary are included in the
description, which could be useful in e.g., modelling neutron stars and
strange quark stars [16], [25] and
computing the total mass of a rotating configuration as a function of the
central density [35].
\item[(ii)]
Subsystems with intersecting boundaries are included in the description, which
could be useful for representing special configurations e.g., a substantially
flattened subsystem extending outside a slightly flattened one.
\item[(iii)]
Special cases where the results can be expressed analytically are analysed in
detail with the addition of a guidance example involving homogeneous
configurations for simplicity, but with intersecting boundaries.
\end{description}


\appendix
\section*{Appendix}

\section{Series expansions within the common region}
\label{a:seco}

With regard to the common region, the series expansion of the solution of the
EC2 equation, Eq.\,(\ref{eq:thc}), may be inserted into Eq.\,(\ref{eq:ispuv})
and the terms of different degree in Legendre polynomials may be equated
separately.   After some algebra, the result is:
\begin{lefteqnarray}
\label{eq:isp2l}
&& A_{2\ell,v}\theta_{2\ell,v}(\xi_v)=A_{2\ell,u}\theta_{2\ell,u}(\xi_u)+(1-
\Gamma_{uv})[\delta_{2\ell,0}-A_{2\ell,u}\theta_{2\ell,u}(\xi_u)] \nonumber \\
&& \phantom{A_{2\ell,v}\theta_{2\ell,v}(\xi_v)=A_{2\ell,u}\theta_{2\ell,u}
(\xi_u)}+(\delta_{2\ell,0}-\delta_{2\ell,2})\Gamma_{uv}\frac{\upsilon_v-
\upsilon_u}6\xi_u^2~;~~\xi_w\le\xi_w^\ast~;\qquad
\end{lefteqnarray}
which, related to the EC2 associated functions, is the counterpart of
Eq.\,(\ref{eq:ispuv}), related to the EC2 functions.

The right-hand side of the EC2 associated equations,
Eqs.\,(\ref{eq:At0c})-(\ref{eq:At2lc}), is independent of the subsystem under
consideration, which implies the same holds for the left-hand side, as it can
be verified after a long algebra using Eqs.\,(\ref{eq:dimq}) and
(\ref{eq:isp2l}).   The result is:
\begin{lefteqnarray}
\label{eq:EC2uv}
&& \frac1{\xi_u^2}\frac\diff{\diff\xi_u}\left[\xi_u^2\frac{\diff(A_{2\ell,u}
\theta_{2\ell,u})}{\diff\xi_u}\right]-\frac{2\ell(2\ell+1)}{\xi_u^2}
A_{2\ell,u}\theta_{2\ell,u}-\delta_{2\ell,0}\upsilon_u \nonumber \\
&& =\frac1{\xi_v^2}\frac\diff{\diff\xi_v}\left[\xi_v^2\frac{\diff(A_{2\ell,v}
\theta_{2\ell,v})}{\diff\xi_v}\right]-\frac{2\ell(2\ell+1)}{\xi_v^2}
A_{2\ell,v}\theta_{2\ell,v}-\delta_{2\ell,0}\upsilon_v~~;
\end{lefteqnarray}
where $A_{0,w}=1$ owing to the boundary conditions, Eqs.\,(\ref{eq:EC2b}) and
(\ref{eq:t02ci}).   Without loss of generality, the EC2 associated functions,
$\theta_{2\ell,w}$, can be normalized to yield coinciding coefficients,
$A_{2\ell,w}$, $2\ell>0$.  Accordingly, the following relations hold:
\begin{equation}
\label{eq:Auv2l}
A_{0,u}=A_{0,v}=A_0=1~~;\quad A_{2\ell,u}(\upsilon_u,\upsilon_v)=
A_{2\ell,v}(\upsilon_u,\upsilon_v)=A_{2\ell}(\upsilon_u,\upsilon_v)~~;
\end{equation}
regardless of the subsystem under consideration.   Finally, the substitution
of Eq.\,(\ref{eq:Auv2l}) into (\ref{eq:isp2l}) yields:
\begin{lefteqnarray}
\label{eq:isq2l}
&& \theta_{2\ell,v}(\xi_v)=\theta_{2\ell,u}(\xi_u)+(1-\Gamma_{uv})\left[\frac
{\delta_{2\ell,0}}{A_{2\ell}}-\theta_{2\ell,u}(\xi_u)\right] \nonumber \\
&& \phantom{A_{2\ell,v}\theta_{2\ell,v}(\xi_v)=A_{2\ell,u}\theta_{2\ell,u}
(\xi_u)}+
\Gamma_{uv}\frac{\delta_{2\ell,0}-\delta_{2\ell,2}}{A_{2\ell}}
\frac{\upsilon_v-\upsilon_u}6\xi_u^2~;~~\xi_w\le\xi_w^\ast~;\qquad
\end{lefteqnarray}
where the EC2 associated functions, $\theta_{2\ell,v}$, may be expanded in
Taylor series as:
\begin{lefteqnarray}
\label{eq:sersv}
&& \theta_{2\ell,v}(\xi_v)=\sum_{k=0}^{+\infty}a_{2\ell,k}^{(v,v)}(\xi_{0,v})
(\xi_v-\xi_{0,v})^k=\sum_{k=0}^{+\infty}a_{2\ell,k}^{(v,u)}(\xi_{0,u})
(\xi_u-\xi_{0,u})^k~~;~~\quad \\
\label{eq:cersv}
&& a_{2\ell,0}^{(v,u)}(\xi_{0,u})=\theta_{2\ell,v}(\xi_{0,u});~~
a_{2\ell,k}^{(v,u)}(\xi_{0,u})=\frac1{k!}\left(\frac{\diff^k\theta_{2\ell,v}}
{\diff\xi_u^k}\right)_{\xi_{0,u}};
\end{lefteqnarray}
and, in particular, $a_{2\ell,1}^{(v,u)}(\xi_{0,u})=\theta_{2\ell,v}^\prime
(\xi_{0,u})$ and $a_{2\ell,2}^{(v,u)}(\xi_{0,u})=\theta_{2\ell,v}^\pprime
(\xi_{0,u})/2$.   The variables, $\xi_u$ and $\xi_v$, are linked via
Eq.\,(\ref{eq:csir}).   The coefficients belonging to different series
expansions appearing in Eq.\,(\ref{eq:sersv}), via Eqs.\,(\ref{eq:csir}),
(\ref{eq:Gauv}) and (\ref{eq:sersc}), are related as:
\begin{equation}
\label{eq:a2lG}
a_{2\ell,k}^{(v,v)}=\Gamma_{vu}^{k/2}a_{2\ell,k}^{(v,u)}~~;
\end{equation}
with regard to the common region.

The substitution of Eqs.\,(\ref{eq:sersc}) and (\ref{eq:sersv}) into
(\ref{eq:isq2l}) by use of the identity:
\begin{equation}
\label{eq:idcs}
\xi_w^2=\xi_{0,w}^2+2\xi_{0,w}(\xi_w-\xi_{0,w})+(\xi_w-\xi_{0,w})^2~~;
\end{equation}
after equating the terms of same degree in $(\xi_w-\xi_{0,w})^k$ yields:
\begin{lefteqnarray}
\label{eq:alkuv}
&& a_{2\ell,k}^{(v,u)}(\xi_{0,u})=\Gamma_{uv}\left\{a_{2\ell,k}^{(u,u)}
(\xi_{0,u})-\frac{\delta_{0k}\delta_{2\ell,0}}{A_{2\ell}}(1-\Gamma_{vu})
\right. \nonumber \\
&& \phantom{a_{2\ell,k}^{(v,u)}(\xi_{0,u})=\Gamma_{uv}^k\left\{\right.}-\left.
\frac{\delta_{2\ell,0}-\delta_{2\ell,2}}{A_{2\ell}}
\frac{\upsilon_u-\upsilon_v}6
[\delta_{0k}\xi_{0,u}^2+2\delta_{1k}\xi_{0,u}+\delta_{2k}]\right\};\quad
\end{lefteqnarray}
provided both series are convergent.

In the special case, $k=0$, Eq.\,(\ref{eq:alkuv}) via (\ref{eq:cersc}) and
(\ref{eq:cersv}) reduces to Eq.\,(\ref{eq:isq2l}) particularized to the
starting point, $\xi_w=\xi_{0,w}$.
In the special case, $k=1$, Eq.\,(\ref{eq:alkuv}) via (\ref{eq:cersc}) and
(\ref{eq:cersv}) reduces to Eq.\,(\ref{eq:isq2l}) particularized to the
starting point, $\xi_w=\xi_{0,w}$, after derivation on both sides with respect
to $\xi_w$.   Similar considerations hold for generic $k$.

In the special case of the singular starting point, $\xi_{0,w}=0$, 
Eq.\,(\ref{eq:alkuv}) reduces to:
\begin{lefteqnarray}
\label{eq:lkuv0}
&& a_{2\ell,k}^{(v,u)}(0)=\Gamma_{uv}\left[a_{2\ell,k}^{(u,u)}(0)-\frac
{\delta_{0k}\delta_{2\ell,0}}{A_{2\ell}}(1-\Gamma_{vu})-\delta_{2k}\frac
{\delta_{2\ell,0}-\delta_{2\ell,2}}{A_{2\ell}}\frac{\upsilon_u-\upsilon_v}6
\right];\qquad
\end{lefteqnarray}
provided both series are convergent.

Finally, Taylor series expansions expressed by Eqs.\,(\ref{eq:cn1s}) and
(\ref{eq:cn2s}) are special cases of the power, $1/\xi_w^m$.
Aiming to write a general formula, the $k$-th derivative is:
\begin{equation}
\label{eq:dkcsi}
\frac{\diff^k}{\diff\xi_w^k}\frac1{\xi_{w}^m}=\frac{(-1)^km(m+1)...(m+k-1)}
{\xi_w^{m+k}}~~;
\end{equation}
and related Taylor series expansion after some algebra reads:
\begin{equation}
\label{eq:Tfm}
\frac1{\xi_{w}^m}=\frac1{\xi_{0,w}^m}\sum_{k=0}^{+\infty}(-1)^k
{k+m-1 \choose m-1}\left(\frac{\xi_{w}-\xi_{0,w}}{\xi_{0,w}}\right)^k~~;\quad
\vert\xi_{w}-\xi_{0,w}\vert<\xi_{0,w}~~;
\end{equation}
where $\xi_{0,w}$ is the starting point.  In the special cases, $m=1,2$,
Eq.\,(\ref{eq:Tfm}) reduces to (\ref{eq:cn1s}), (\ref{eq:cn2s}), respectively.

\section{Determination of the coefficients, $A_{2\ell}$}
\label{a:A2}

With regard to the outer (along the direction considered) subsystem, $u$, the
substitution of Eqs.\,(\ref{eq:thn}) and (\ref{eq:thc}) into
(\ref{eq:VGd}) changes the expression of the gravitational potential and the
radial component of the gravitational force as:
\begin{lefteqnarray}
\label{eq:grap}
&& {\cal V}_G=4\pi G\sum(\lambda_w)\alpha_u^2\sum_{\ell=0}^{+\infty}\left[
A_{2\ell}^{(\rm xxx)}\theta_{2\ell,u}^{(\rm xxx)}(\xi_u)-\frac
{\delta_{2\ell,0}-\delta_{2\ell,2}}6\upsilon_u\xi_u^2\right]P_{2\ell}(\mu)
\nonumber \\
&& \phantom{{\cal V}_G=}+
{\cal V}_{{\rm b},u}^\dagger~~; \\
\label{eq:graf}
&& \frac{\diff{\cal V}_G}{\diff r}=\frac1{\alpha_u}\frac{\diff{\cal V}_G}
{\diff\xi_u}=4\pi G\sum(\lambda_w)\alpha_u \nonumber \\
&& \phantom{\frac{\diff{\cal V}_G}{\diff r}=}\times
\sum_{\ell=0}^{+\infty}\left[
A_{2\ell}^{(\rm xxx)}\theta_{2\ell,u}^{\prime(\rm xxx)}(\xi_u)-\frac
{\delta_{2\ell,0}-\delta_{2\ell,2}}3\upsilon_u\xi_u\right]P_{2\ell}(\mu)~~; \\
\label{eq:grac}
&& {\cal V}_{{\rm b},u}^\dagger={\cal V}_{{\rm b},u}-4\pi G\sum(\lambda_w)
\alpha_u^2\theta_{{\rm b},u}~~;
\end{lefteqnarray}
where $A_0^{(xxx)}=1$ according to Eqs.\,(\ref{eq:A2ln}) and (\ref{eq:A2lc}),
Eq.\,(\ref{eq:thn}) has been expressed in terms of $\xi_u$ instead of
$\xi_{u1}=\Lambda_u^{1/2}\xi_u$,
and xxx = com, ncm, according if $(\xi_u,\mu)$    
belongs to the common region, $0\le\xi_u\le\xi_u^\ast$, or the noncommon
region, $\xi_u^\ast\le\xi_u\le\Xi_u$.

The continuity of the gravitational potential and the radial component of the
gravitational force on the interface, $\xi_u=\xi_u^\ast$, implies the validity
of the following relations:
\begin{lefteqnarray}
\label{eq:copi}
&& A_{2\ell}^{(\rm com)}\theta_{2\ell,u}^{(\rm com)}(\xi_u^\ast)=
   A_{2\ell}^{(\rm ncm)}\theta_{2\ell,u}^{(\rm ncm)}(\xi_u^\ast)~~; \\
\label{eq:cofi}
&& A_{2\ell}^{(\rm com)}\theta_{2\ell,u}^{\prime(\rm com)}(\xi_u^\ast)=
   A_{2\ell}^{(\rm ncm)}\theta_{2\ell,u}^{\prime(\rm ncm)}(\xi_u^\ast)~~;
\end{lefteqnarray}
where, in general, $\theta_{2\ell,w}^\prime=\diff\theta_{2\ell,w}/\diff\xi_w$.  On
the other hand, the continuity of the first latitudinal component of the
gravitational force,
$\partial{\cal V}_G/\partial\mu$, on the interface, reduces to
Eq.\,(\ref{eq:copi}).   In fractional form, Eqs.\,(\ref{eq:copi}) and
(\ref{eq:cofi}) translate into:
\begin{lefteqnarray}
\label{eq:cpfi}
&& \frac{A_{2\ell}^{(\rm ncm)}}{A_{2\ell}^{(\rm com)}}=\frac{\theta_{2\ell,u}^
{(\rm com)}(\xi_u^\ast)}{\theta_{2\ell,u}^{(\rm ncm)}(\xi_u^\ast)}=
\frac{\theta_{2\ell,u}^{\prime(\rm com)}(\xi_u^\ast)}{\theta_{2\ell,u}^
{\prime(\rm ncm)}(\xi_u^\ast)}~~;
\end{lefteqnarray}
or, in other words, $\theta_{2\ell,u}^{(\rm com)}$ and
$\theta_{2\ell,u}^{(\rm ncm)}$ on the interface differ by a constant factor at
most.

Outside the boundary of a body of revolution, the gravitational potential and
the radial component of the gravitational force, via
Eqs.\,(\ref{eq:diEC})-(\ref{eq:csir}), at
sufficiently large distances can be expressed as e.g., [27], Chap.\,VII,
\S193:
\begin{lefteqnarray}
\label{eq:grpe}
&& {\cal V}_G(\xi_u,\mu)=4\pi G\sum(\lambda_w)\alpha_u^2\sum_{\ell=0}^
{+\infty}\frac{c_{2\ell,u}}{\xi_u^{2\ell+1}}P_{2\ell}(\mu)~~;\qquad\xi_u\gg
\Xi_u~~; \\
\label{eq:frpe}
&& \frac{\partial{\cal V}_G}{\partial r}=\frac1{\alpha_u}\frac
{\partial{\cal V}_G}{\partial\xi_u}=4\pi G\sum(\lambda_w)\alpha_u
\sum_{\ell=0}^{+\infty}\frac{-(2\ell+1)c_{2\ell,u}}{\xi_u^{2\ell+2}}
P_{2\ell}(\mu)~~;
\end{lefteqnarray}
where $c_{{2\ell},u}$ are dimensionless coefficients and odds terms are
ruled out by symmetry with respect to the equatorial plane.   Strictly
speaking, Eqs.\,(\ref{eq:grpe})-(\ref{eq:frpe}) are exact for
spherical-symmetric matter
distributions and Roche models [24], Chap.\,IX \S\S229-232, but hold
to an acceptable extent provided
oblateness remains sufficiently small and/or concentration maintains
sufficiently high.   For further details, an interested reader is addressed to
an earlier investigation [9].

The continuity of the gravitational potential and the radial component of the
gravitational force on
a selected point of the boundary, $\Xi_u=\Xi_u(\mu)$, implies
Eqs.\,(\ref{eq:grap})-(\ref{eq:graf}) and (\ref{eq:grpe})-(\ref{eq:frpe}),
respectively, match at $(\Xi_u,\mu)$ for the terms of same degree in Legendre
polynomials.   The result is e.g., [9]:
\begin{lefteqnarray}
\label{eq:ACsi}
&& A_{2\ell}^{(\rm ncm)}=-\delta_{2\ell,2}\frac56\frac{\upsilon_u\Xi_u^2}
{3\theta_{2\ell,u}^{(\rm ncm)}(\Xi_u)+\Xi_u\theta_{2\ell,u}^
{\prime(\rm ncm)}(\Xi_u)}~~;\qquad2\ell>0~~; \\
\label{eq:cCsi}
&& c_{2\ell,u}=\frac15A_{2\ell}^{(\rm ncm)}\Xi_u^3\left[
2\theta_{2\ell,u}^{(\rm ncm)}(\Xi_u)-\Xi_u\theta_{2\ell,u}^
{\prime(\rm ncm)}(\Xi_u)\right]~~;\qquad2\ell>0~~; \\
\label{eq:AcC0}
&& A_0^{(\rm ncm)}=1~~;\qquad c_{0,u}=-\Xi_u^2\theta_{0,u}^{\prime(\rm ncm)}
(\Xi_u)+\frac13\upsilon_u\Xi_u^3~~; \\
\label{eq:ccCs}
&& c_{{\rm b},u}^\dagger=\frac{{\cal V}_{{\rm b},u}^\dagger}
{4\pi G\sum(\lambda_w)\alpha_u^2}=-\theta_{0,u}^{(\rm ncm)}(\Xi_u)-\Xi_u
\theta_{0,u}^{\prime(\rm ncm)}(\Xi_u)+\frac12\upsilon_u\Xi_u^2~~; \\
\label{eq:cccC}
&& c_{{\rm b},u}=\frac{{\cal V}_{{\rm b},u}}{4\pi G\sum(\lambda_w)\alpha_u^2}=
c_{{\rm b},u}^\dagger+\theta_{{\rm b},u}~~;
\end{lefteqnarray}
where a value of $\Xi_u$ on the boundary has necessarily to be fixed for
defining the approximation used and calculating the coefficients,
$A_{2\ell}^{(\rm ncm)}$, among others.   Viable alternatives could be
$(\Xi_u,\mu)=(\Xi_{{\rm E},u},1/\sqrt3)$ [13] and
$(\Xi_u,\mu)=(\Xi_{{\rm p},u},1)$ [9], where $\Xi_{{\rm E},u}$
relates to the nonrotating sphere and $\Xi_{{\rm p},u}$ to the rotating
configuration, respectively.

With regard to the common region, the sole restriction is from
Eq.\,(\ref{eq:cpfi}).   Accordingly,
$A_{2\ell}^{(\rm ncm)}=A_{2\ell}^{(\rm com)}=A_{2\ell}$ without loss of
generality, and using Eqs.\,(\ref{eq:Auv2l}), (\ref{eq:ACsi}),
yields:
\begin{lefteqnarray}
\label{eq:AC2l}
&& A_{2\ell}=\delta_{2\ell,0}-\delta_{2\ell,2}\frac56\frac{\upsilon_u\Xi_u^2}
{3\theta_{2\ell,u}^{(\rm ncm)}(\Xi_u)+\Xi_u\theta_{2\ell,u}^
{\prime(\rm ncm)}(\Xi_u)}~~;
\end{lefteqnarray}
for a selected point of the boundary, $(\Xi_u,\mu)$.

For specified configurations where $\theta_{2\ell,w}^{(\rm com)}(\xi_w)$ is
known via numerical integrations, the particularization of
Eq.\,(\ref{eq:isq2l}) to $2\ell=2$ after some algebra yields:
\begin{lefteqnarray}
\label{eq:A2uv}
&& A_2=-\frac16\frac{\Gamma_{uv}(\upsilon_v-\upsilon_u)\xi_u^2}
{\theta_{2,v}^{(\rm com)}(\xi_v)-\Gamma_{uv}\theta_{2,u}^{(\rm com)}(\xi_u)}
~~;
\end{lefteqnarray}
where no approximation is involved.

\section{Corrigendum}
\label{a:C14c}

With regard to a quoted reference [9], Eq.\,(76) therein is affected
by a printing error, which is propagated to the next Eq.\,(77).   The
correct formulation reads:
\begin{lefteqnarray*}
\label{eq:76C14}
&& \theta_2^\prime(\xi)=\psi_2^\prime(\xi)=15\left[\left(-\frac9{\xi^3}+\frac4
\xi\right)\frac{\sin\xi}\xi+\left(\frac9{\xi^2}-1\right)\frac{\cos\xi}\xi
\right]~~;\hspace{23.5mm}(76) \\
\label{eq:77C14}
&& A_2=-\frac{\upsilon\Xi^2}{18}\left[\frac{\sin\Xi}\Xi-\cos\Xi\right]^{-1}~~;
\hspace{71.5mm}(77)
\end{lefteqnarray*}
respectively.   On the other hand, the above mentioned equations were not used
in computations, which implies all results reported therein remain unchanged.

{\bf Received: Month xx, 20xx}

\end{document}